\documentclass[10pt,journal,compsoc]{IEEEtran}

\usepackage{graphicx}
\usepackage{multirow}
\usepackage{ulem}
\usepackage{color}
\usepackage{cite}

\usepackage[breaklinks=true,colorlinks,bookmarks=false,citecolor=green,linkcolor=green]{hyperref}
\usepackage[activate]{microtype}
\usepackage{svg}

\usepackage{authblk}
\usepackage{booktabs}

\usepackage{url}
\usepackage{colortbl}
\usepackage[separate-uncertainty = true,multi-part-units = repeat]{siunitx}
\usepackage{array,amsfonts,amsmath}
\usepackage{caption}
\usepackage{threeparttable}
\usepackage{rotating}
\usepackage{tablefootnote}
\usepackage[english]{babel}
\usepackage[utf8]{inputenc}
\usepackage{algorithm}
\usepackage{adjustbox}
\usepackage{pdflscape}
\usepackage{afterpage}
\usepackage{sidecap}
\PassOptionsToPackage{noend}{algpseudocode}
\usepackage{algpseudocode}
\usepackage{pdflscape}
\usepackage{lscape}
\usepackage{epstopdf}

\usepackage[most]{tcolorbox}
\graphicspath{{./figures/}} 

\usepackage{boldline} 
\usepackage{makecell}
\usepackage{footnotehyper}
\definecolor{mygray}{gray}{.9}

\begin{document}
\title{Transformers in Medical Imaging: A Survey}

\author{\textsf{ Fahad~Shamshad,~%
        Salman~Khan,~
        Syed~Waqas~Zamir,~%
        Muhammad~Haris~Khan,~\linebreak%
        Munawar~Hayat,~%
        Fahad~Shahbaz~Khan,~%
        and~Huazhu~Fu}
\thanks{F. Shamshad, S. Khan, M. H. Khan and F. S. Khan are with the MBZ University of Artificial Intelligence, Abu Dhabi, UAE.\\
E-mail: firstname.lastname@mbzuai.ac.ae \\
\textit{S. W. Zamir is with the Inception Institute of Artificial Intelligence, Abu Dhabi, UAE.} \\
\textit{M. Hayat is with the Faculty of IT, Monash University, Clayton VIC 3800, Australia.}\\
\textit{S. Khan is also with the CECS, Australian National University, Canberra ACT 0200, Australia.} \\
\textit{F. S. Khan is also with the Computer Vision Laboratory, Linköping University, Sweden. }\\
\textit{H. Fu is with the Institute of High Performance Computing, Agency for Science, Technology and Research (A*STAR), Singapore.}
}
}

\IEEEtitleabstractindextext{%
\begin{abstract}
Following unprecedented success on the natural language tasks, Transformers have been successfully applied to several computer vision problems, achieving state-of-the-art results and prompting researchers to reconsider the supremacy of convolutional neural networks (CNNs) as {de facto} operators.
Capitalizing on these advances in computer vision, the medical imaging field has also witnessed growing interest for Transformers that can capture global context compared to CNNs with local receptive fields. Inspired from this transition, in this survey, we attempt to provide a comprehensive review of the applications of Transformers in medical imaging covering various aspects, ranging from recently proposed architectural designs to unsolved issues. Specifically, we survey the use of Transformers in medical image segmentation, detection, classification, reconstruction, synthesis, registration, clinical report generation, and other tasks. In particular, for each of these applications, we develop taxonomy, identify application-specific challenges as well as provide insights to solve them, and highlight recent trends. Further, we provide a critical discussion of the field's current state as a whole, including the identification of key challenges, open problems, and outlining promising future directions. We hope this survey will ignite further interest in the community and provide researchers with an up-to-date reference regarding applications of Transformer models in medical imaging. Finally, to cope with the rapid development in this field, we intend to regularly update the relevant latest papers and their open-source implementations at \textbf{{\fontfamily{qcr}\selectfont{\color{magenta}\href{https://github.com/fahadshamshad/awesome-transformers-in-medical-imaging}{\nolinkurl{https://github.com/fahadshamshad/awesome-transformers-in-medical-imaging}}}}}.
\end{abstract}
\begin{IEEEkeywords}
 Transformers, medical image analysis, vision transformers, deep neural networks, clinical report generation
\end{IEEEkeywords}}
\maketitle

\IEEEdisplaynontitleabstractindextext

\IEEEpeerreviewmaketitle
\IEEEraisesectionheading{\section{Introduction}\label{sec:introduction}}

\IEEEPARstart{C}{onvolutional} Neural Networks (CNNs)~\cite{goodfellow2016deep,lecun1989backpropagation,krizhevsky2012imagenet,liu2022convnet} have significantly impacted the field of medical imaging
due to their ability to learn highly complex representations in a data-driven manner. Since their renaissance, CNNs have 
demonstrated remarkable improvements on numerous medical imaging modalities, including Radiography \cite{lakhani2017deep}, Endoscopy \cite{min2019overview}, Computed Tomography (CT) \cite{wurfl2016deep,lell2020recent}, Mammography Images (MG) \cite{hamidinekoo2018deep}, Ultrasound Images \cite{liu2019deep}, Magnetic Resonance Imaging (MRI) \cite{lundervold2019overview,akkus2017deep}, and Positron Emission Tomography (PET) \cite{reader2020deep}, to name a few.
The workhorse in CNNs is the \textit{convolution} operator, which operates locally and provides translational equivariance. While these properties help in developing efficient and generalizable medical imaging solutions, the local receptive field in convolution operation limits capturing long-range pixel relationships. Furthermore, the convolutional filters have stationary weights that are not adapted for the given input image content at inference time. 

Meanwhile, significant research effort has been made by the vision community to integrate the attention mechanisms~\cite{vaswani2017attention,devlin2018bert,fedus2021switch} in CNN-inspired architectures \cite{wang2018non,yin2020disentangled,ramachandran2019stand,bello2019attention,vaswani2021scaling,dosovitskiy2020image}. 
These attention-based `Transformer' models have become an attractive solution 
due to their ability to encode long-range dependencies and learn highly effective feature representations \cite{chaudhari2019attentive}. 
Recent works have shown that these Transformer modules can fully replace the standard convolutions in deep neural networks by operating on a sequence of image patches, giving rise to Vision Transformers (ViTs)~\cite{dosovitskiy2020image}.
Since their inception, ViT models have been shown to push the state-of-the-art in numerous vision tasks, including image classification~\cite{dosovitskiy2020image}, object detection~\cite{zhu2020deformable}, semantic segmentation~\cite{zheng2021rethinking}, image colorization~\cite{kumar2021colorization}, low-level vision~\cite{chen2021pre}, and video understanding~\cite{arnab2021vivit} to name a few. Furthermore, recent research indicate that the prediction errors of ViTs are more consistent with those of humans than CNNs \cite{naseer2021intriguing,portelance2021emergence,geirhos2021partial,tuli2021convolutional}. These desirable properties of ViTs have sparked great interest in the medical community to adapt them for medical imaging applications, thereby mitigating the inherent inductive biases of CNNs \cite{matsoukas2021time}.

\begin{figure*}[t]
\centering
\includegraphics[trim={0.45cm 7.6cm 0.8cm 7cm},clip,width = \textwidth]{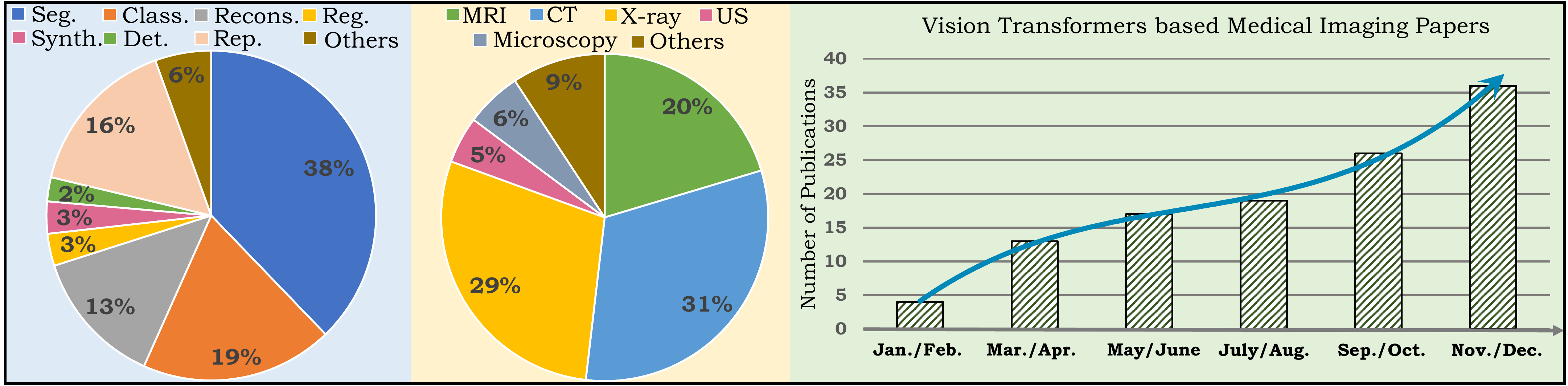} 
\caption{(Left) The pie-charts show statistics of the papers included in this survey according to medical imaging problem settings and data modalities. The rightmost figure shows consistent growth in the recent literature (for year 2021). Seg: segmentation, Class: classification, Recons: reconstruction, Reg: registration, Synth: synthesis, Det: detection, Rep: report generation, US: ultrasound.}
\label{fig:appvsnumbers}
\end{figure*}

\textbf{Motivation and Contributions:} Recently, medical imaging community has witnessed an exponential growth in the number of Transformer based techniques, especially after the inception of ViTs (see Fig. \ref{fig:appvsnumbers}).
The topic is now dominant in prestigious medical imaging conferences and journals, and it is getting increasingly difficult to keep pace with the recent progress due to the rapid influx of papers. 
As such, a survey of the existing relevant works is timely 
to provide a comprehensive account of new methods 
in this emerging field. To this end, we provide a holistic overview of the applications of Transformer models in medical imaging. We hope this work can provide a roadmap for the researchers to explore the field further. Our major contributions include:

\begin{figure}[t]
\centering
\includegraphics[width = 0.45\textwidth]{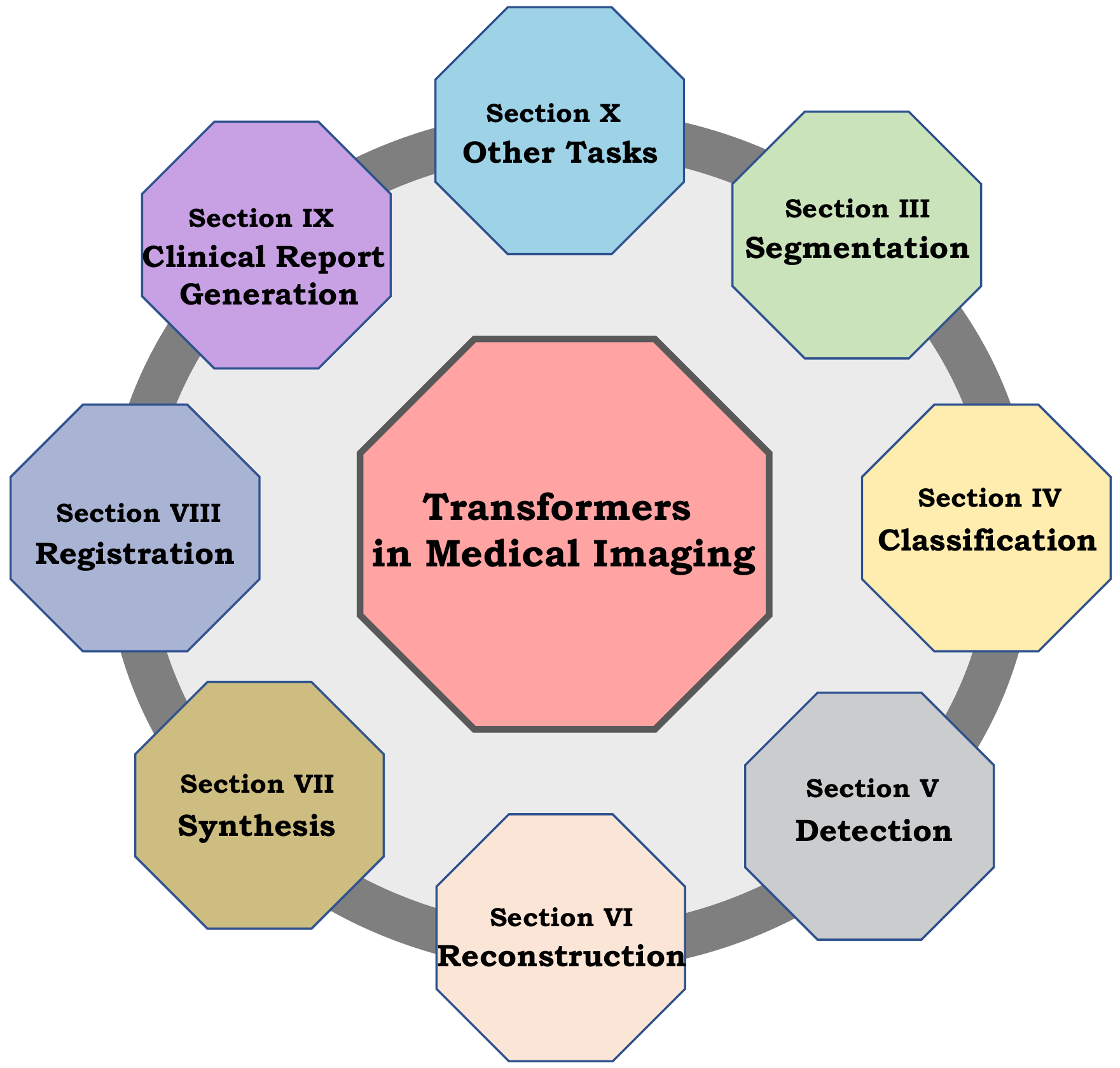}
\caption{A diverse set of application areas of Transformers in medical imaging covered in this survey.}
\label{fig:applications_transformers}
\end{figure}
\begin{itemize}
\item This is the first survey paper that comprehensively covers applications of Transformers in the medical imaging domain, thereby bridging the gap between the vision and medical imaging community in this rapidly evolving area. Specifically, we present a comprehensive overview of more than 125 relevant papers to cover the recent progress.

\item  We  provide a detailed coverage of the field by categorizing the papers based on their applications in medical imaging as depicted in Fig. \ref{fig:applications_transformers}. 
For each of these applications, we develop a taxonomy, highlight task-specific challenges, and provide insights about solving them based on the literature reviewed. 

\item Finally, we provide a critical discussion of the field’s current state as a whole, including identifying key challenges, highlighting open problems, and outlining promising future  directions.  

\item Although the main focus of this survey is on Vision Transformers, we are also the first since the inception of the \text{original} Transformer, about half a decade ago, to extensively cover its language modeling capabilities in the clinical report generation task (see Sec.~\ref{sec:cli}).
\end{itemize}
\textit{\textbf{Paper Organization.}} The rest of the paper is organized as follows. In Sec. \ref{sec:background}, we provide background of the field with a focus on salient concepts underlying Transformers. From Sec. \ref{sec:seg} to Sec. \ref{sec:other_app}, we comprehensively cover applications of Transformers in several medical imaging tasks as shown in Fig. \ref{fig:applications_transformers}. In particular, for each of these tasks, we develop a taxonomy and identify task-specific challenges. 
Sec. \ref{sec:openproblems} presents open problems and future directions about the field as a whole. Finally, in Sec. \ref{sec:conc}, we give recommendations to cope with the rapid development of the field and conclude the paper.

\begin{figure*}[]
\centering
\includegraphics[width = \textwidth]{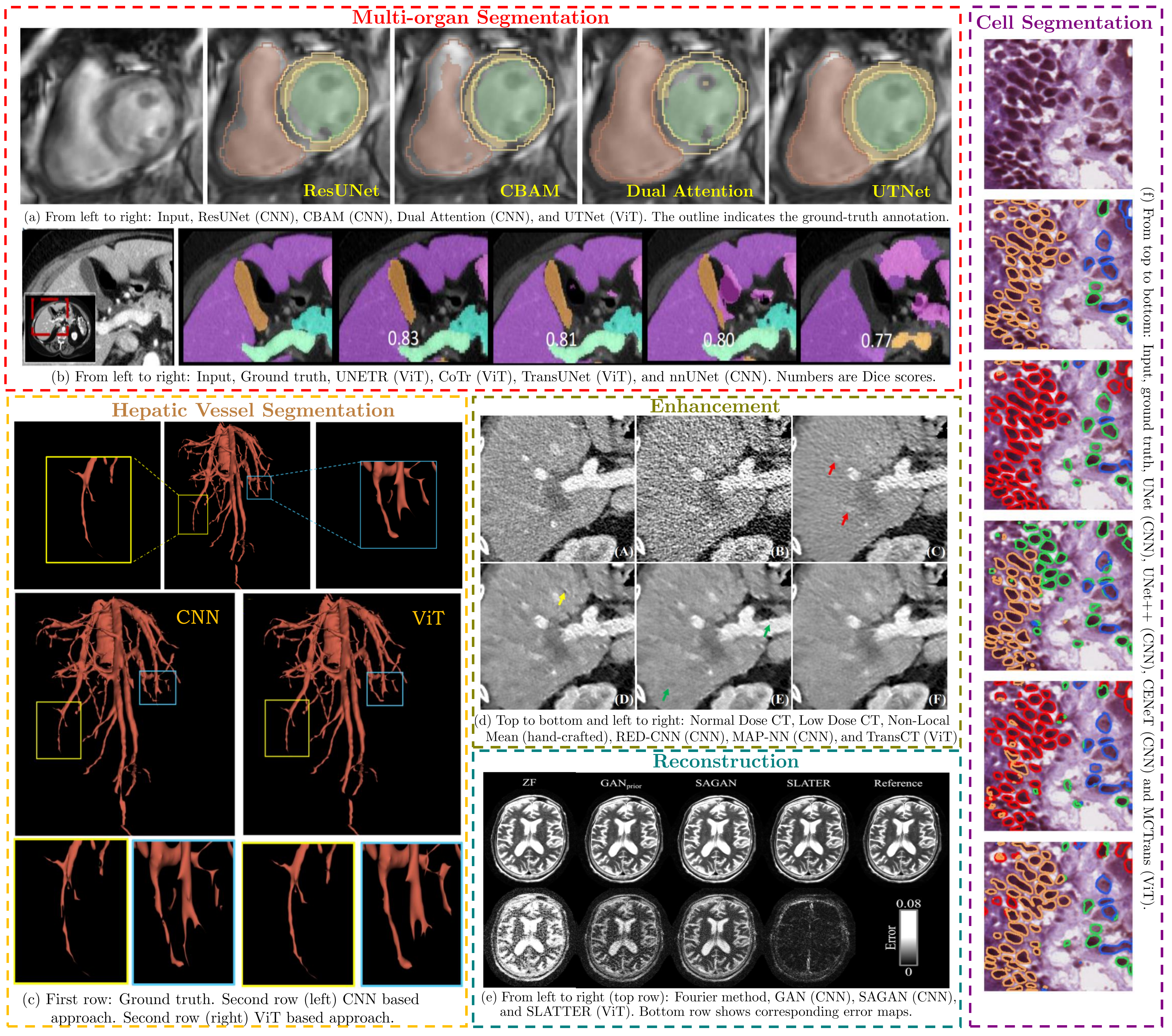} 
\caption{Applications of ViTs in various medical imaging problems along with the baseline CNN-based approaches. ViT-based approaches give superior
performance as compared to CNN-based methods due to their ability to model the global context. Figure sources: (a) \cite{gao2021utnet}, (b) \cite{hatamizadeh2021unetr}, (c) \cite{wu2021hepatic}, (d) \cite{zhang2021transct}, (e) \cite{korkmaz2021unsupervised}, (f) \cite{ji2021multi}. }
\label{fig:applications}
\end{figure*}

\section{Background} \label{sec:background}

Medical imaging approaches have undergone significant advances over the past few decades.
In this section, we briefly provide a background of these advancements and broadly group them into three 
categories: hand-crafted, CNN-based, and ViT-based. For the hand-crafted and CNN-based methods, we describe the underlying working principles along with their major strengths and shortcomings in the context of medical imaging. For the ViT-based methods, we highlight core concepts behind their success and defer further details to later sections.

\subsection{Hand-Crafted Approaches}
Conventional algorithms to solve medical imaging tasks are based on hand-crafted mathematical models designed by field experts using domain knowledge.
The development of these hand-crafted models with a focus on refining discriminative features and efficient optimization algorithms for a range of medical imaging problems has been the central research topic in the past \cite{suetens2017fundamentals,zhang2008medical}. Successful hand-crafted models in medical imaging include total-variation \cite{rudin1992nonlinear}, non-local self-similarity \cite{xu2015patch}, sparsity/structured sparsity \cite{eldar2012compressed}, Markov-tree models on wavelet coefficients \cite{choi2000hidden}, and untrained neural networks \cite{ulyanov2018deep,qayyum2021untrained,qayyum2020single}.
These models have been extensively leveraged in medical domain for image segmentation \cite{ramesh2021review}, reconstruction \cite{ravishankar2019image}, disease classification \cite{miranda2016survey}, enhancement \cite{jawdekar2021review}, and anomaly detection \cite{tschuchnig2021anomaly} due to their interpretability with solid mathematical foundations and theoretical supports on the robustness, recovery, and complexity \cite{ruping2006learning,ahmad2018interpretable}. Further, unlike deep learning-based approaches, they do not require large annotated medical imaging datasets for training. This reduced reliance on labeled datasets 
is crucial to the medical research community as collecting voluminous, reliable, and labeled data in the medical domain is difficult due to the lack of expert annotators, high time consumption, ethical considerations, and financial costs. 
However, due to inadequacy to leverage the expressive power of large medical imaging data sets, these hand-crafted models often suffer from poor discriminative capability \cite{zhou2021review}. Consequently, these models often fail to represent nuances of high-dimensional complex medical imaging data that can hamper the performance of the medical imaging diagnosis systems \cite{hegde2018algorithmic,lin2020comparison}. To circumvent the poor discriminability and generalization issue, learned hand-crafted models have been proposed to exploit data better.
The representative approaches include 
optimal directions \cite{engan1999method}, K-SVD \cite{aharon2006k}, data-driven tight frame~\cite{cai2014data}, low-rank models \cite{zhou2014low}, and piece-wise smooth image model\cite{an2007gamma}. Next, we explain the popular data-driven approaches explored in the literature.

\begin{figure*}[t]
\centering
\includegraphics[width = \textwidth]{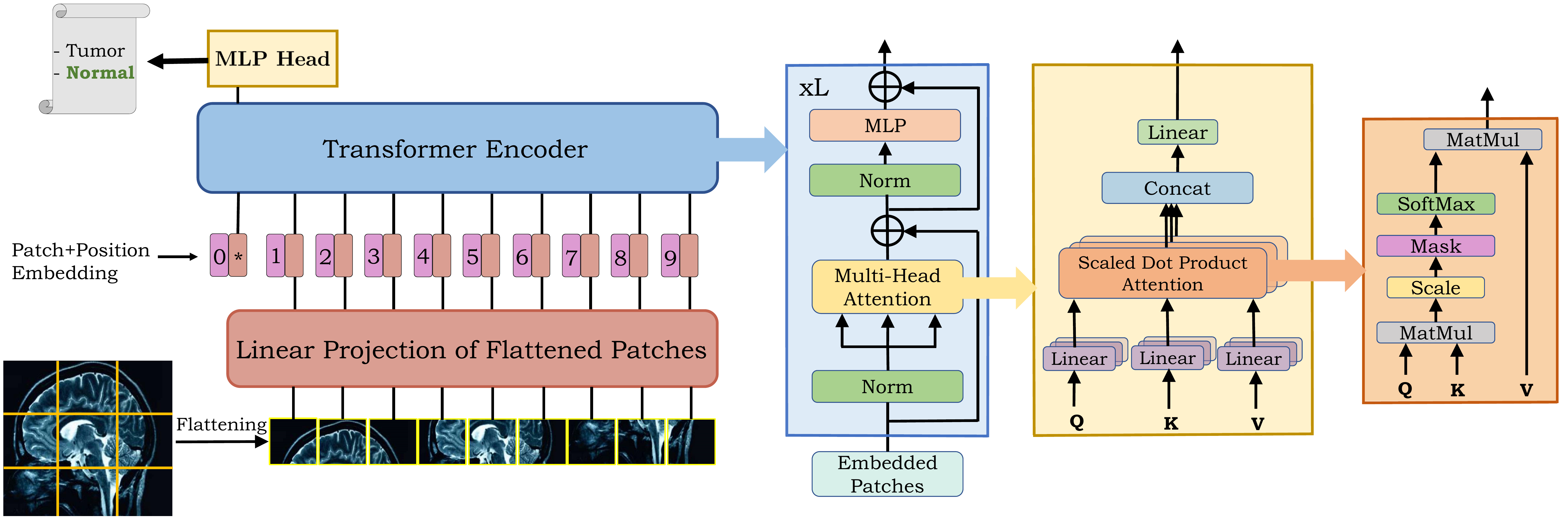} 
\caption{Architecture of the Vision Transformer (\textit{on the left}) and details of the Vision Transformer encoder block (\textit{on the right}). Vision transformer first
splits the input image into patches and projects them (after flattening) into a feature space where a transformer encoder processes them to produce the final classification output.}
\label{fig:vit}
\end{figure*}
\subsection{CNN-based methods}

CNNs are effective at learning discriminative features and extracting generalizable priors from large-scale medical datasets, thus providing excellent performance on medical imaging tasks, 
making them an integral component of modern AI-based medical imaging systems.
The advancements in CNNs have been mainly fueled by novel architectural designs, better optimization procedures, availability of special hardware (\textit{e.g.,} GPUs) and purpose-built open source software libraries \cite{gibson2018niftynet,perez2021torchio,beers2021deepneuro}. 
We refer interested readers to comprehensive survey papers related to CNNs applications in medical imaging \cite{yi2019generative,litjens2017survey,greenspan2016guest
,zhou2017deep,shen2017deep,cheplygina2019not,hesamian2019deep
,duncan2019biomedical,haskins2020deep,zhou2021review}. Despite considerable performance gains, the reliance of CNNs on large labeled datasets limits their applicability over the full spectrum of medical imaging tasks. Furthermore, CNNs-based approaches are generally more challenging to interpret and often act as black box solutions. 
Therefore, there has been an increasing effort in the medical imaging community to amalgamate the strengths of hand-crafted and CNNs based methods resulting in the prior information-guided CNNs models \cite{shlezinger2020model}. These hybrid methods contain special domain-specific layers, and include unrolled optimization \cite{monga2021algorithm}, generative models \cite{ongie2020deep}, and learned denoiser-based approaches \cite{ahmad2020plug}. 
Despite these architectural and algorithmic advancements, the decisive factor behind CNNs success has been primarily attributed to their image-specific inductive bias in dealing with scale invariance and modeling local visual structures. While this intrinsic locality (limited receptive field) brings efficiency to CNNs, it impairs their ability to capture long-range spatial dependencies in an input image, thereby stagnating performance~\cite{matsoukas2021time} (see Fig.~\ref{fig:applications}).
This demands an alternative architectural design capable of modeling long-range pixel relationships for better representation learning.

\subsection{Transformers}
Transformers were introduced by Vaswani \textit{et al.} \cite{vaswani2017attention} as a new attention-driven building block for machine translation. Specifically, these attention blocks are neural network layers that aggregate information from the entire input sequence \cite{bahdanau2014neural}. Since their inception, these models have demonstrated state-of-the-art performance on several Natural Language Processing (NLP) tasks, thereby becoming the default choice over recurrent models.
In this section, we will focus on Vision Transformers (ViTs) \cite{dosovitskiy2020image} that are built on vanilla Transformer model~\cite{vaswani2017attention} by cascading multiple transformer layers to capture the global context of an input image. Specifically, Dosovitskiy \textit{et al.} \cite{dosovitskiy2020image} interpret an image as a sequence of patches and process it by a standard transformer encoder as used in NLP. These ViT models continue the long-lasting trend of removing hand-crafted visual features and inductive biases from models in an effort to leverage the availability of larger datasets coupled with increased computational capacity. ViTs have garnered immense interest in the medical imaging community, and a number of recent approaches have been proposed which build upon ViTs. We highlight the working principle of ViT in a step-by-step manner in Algorithm \ref{alg:vit} for medical image classification.

\makeatletter
\def\BState{\State\hskip-\ALG@thistlm}
\makeatother
\algnewcommand\algorithmicinput{\textbf{Input:}}
\algnewcommand\algorithmicoutput{\textbf{Output:}}
\algnewcommand\Input{\item[\algorithmicinput]}%
\algnewcommand\Output{\item[\algorithmicoutput]}%

\begin{algorithm}[t]
    \caption{ViT Working Principle}\label{alg:vit}
    \begin{algorithmic}[1]
    \item Split a medical image into patches of fixed sizes 
    \item Vectorize image patches via flattening operation
    \item Create lower-dimensional linear embedding from vectorized patches via trainable linear layer
    \item Add positional encoding to lower dimensional linear embeddings
    \item Feed the sequence to ViT encoder as shown in Figure \ref{fig:vit}
    \item Pre-train the ViT model on a large-scale image dataset
    \item Fine-tune on the down stream medical image classification task
    \end{algorithmic}
\end{algorithm}

Below, we briefly describe the core components behind the success of ViTs that are \textit{self-attention} and \textit{multi-head self-attention}. For a more in-depth analysis of numerous ViT architectures and applications, we refer interesting readers to the recent relevant survey papers \cite{chaudhari2019attentive,han2020survey,khan2021transformers,tay2020efficient,lin2021survey}.

\begin{figure*}[t]
\centering
\includegraphics[trim = {5cm 14.5cm 4cm 4cm},clip,width = \textwidth]{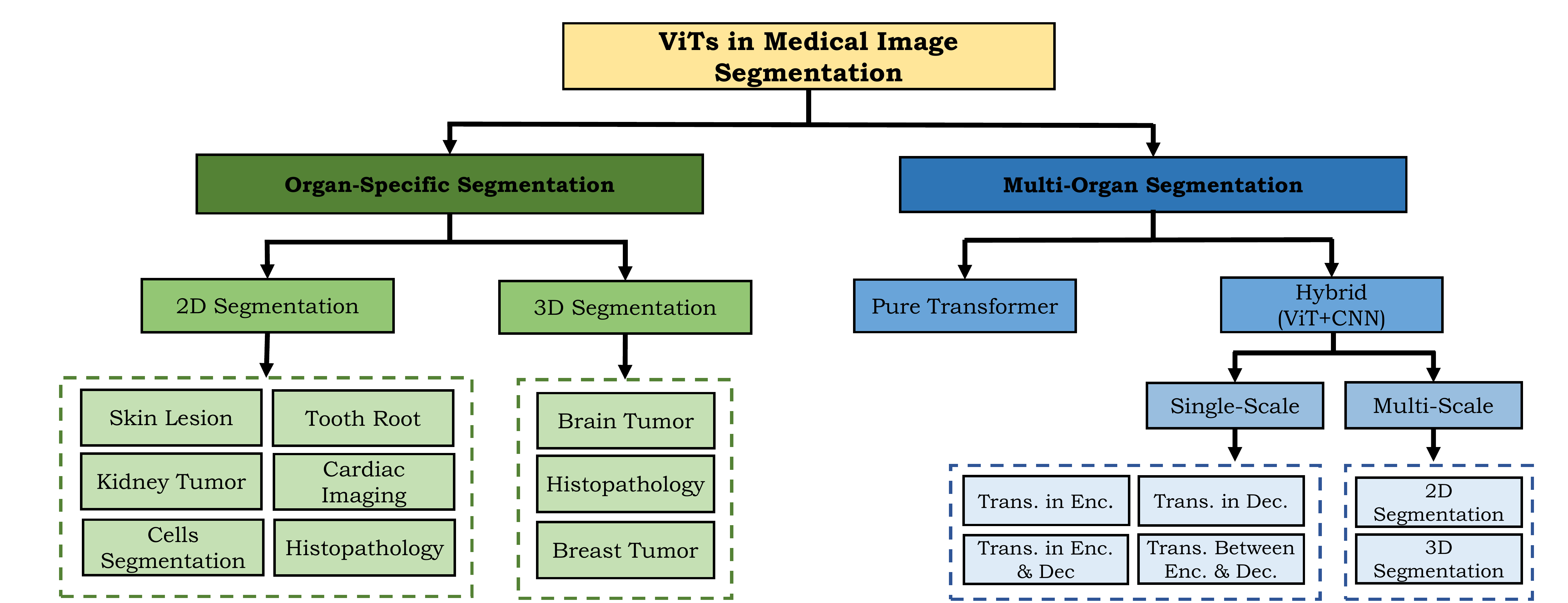}
\caption{Taxonomy of ViT-based medical image segmentation approaches.}
\label{fig:seg_taxonomy}
\end{figure*}
\subsubsection{\textbf{Self-Attention}}The success of the Transformer models has been widely attributed to the self-attention (SA) mechanism due to its ability to model long-range dependencies. The key idea behind the SA mechanism is to learn self-alignment, that is, to determine the relative importance of a single token (patch embedding) with respect to all other tokens in the sequence \cite{bahdanau2014neural}.
For 2D images, we first reshape the image $\mathbf{x} \in \mathbb{R}^{H \times W \times C}$ into a sequence of flattened 2D patches $\mathbf{x}_p \in \mathbb{R}^{N\times(P^2·C)}$, where $H$ and $W$ denotes height and width of the original image respectively, $C$ is the number of channels, $P \times P$ is the resolution of each image patch, and $N = HW/P^2$ is the resulting number of patches. These flattened patches are projected to $D$ dimension via trainable linear projection layer and can be represented in matrix form as $\mathbf{X} \in \mathbb{R}^{N \times D}$. The goal of self-attention is to capture the interaction amongst all these $N$ embeddings, that is done by defining three learnable weight matrices to transform input $\mathbf{X}$ into Queries (via $\mathbf{W}^Q \in \mathbb{R}^{D \times D_q}$), Keys (via $\mathbf{W}^K \in \mathbb{R}^{D \times D_k}$) and Values (via $\mathbf{W}^V \in \mathbb{R}^{D \times D_v}$), where $D_q = D_k$. The input sequence $\mathbf{X}$ is first projected onto these weight matrices to get $\mathbf{Q}=\mathbf{X}\mathbf{W}^Q$, $\mathbf{K}=\mathbf{X}\mathbf{W}^K$ and $\mathbf{V}=\mathbf{X}\mathbf{W}^V$.
The corresponding attention matrix $\mathbf{A} \in \mathbb{R}^{N \times N}$ can be written as,
$$\mathbf{A} = \mathbf{softmax}\left (\frac{\mathbf{Q}\mathbf{K}^T}{\sqrt{D_q}}\right ).$$
The output $\mathbf{Z}\in\mathbb{R}^{N \times D_v}$ of the SA layer is then given by,
$$ \mathbf{Z} = \text{SA}(\mathbf{X}) =  \mathbf{A}\mathbf{V}.$$

\subsubsection{\textbf{Multi-Head Self-Attention}} Multi-Head Self Attention (MHSA) consists of multiple SA blocks (heads) concatenated together channel-wise to model complex dependencies between different elements in the input sequence. Each head has its own learnable weight matrices denoted by  $\{\mathbf{W}^{Q_i},\mathbf{W}^{K_i},\mathbf{W}^{V_i} \}$, where $i=0 \cdots (h{-}1)$ and $h$ denotes total number of heads in MHSA block. Specifically, we can write,
$$\text{MHSA}(\mathbf{Q},\mathbf{K},\mathbf{V}) = [\mathbf{Z}_0,\mathbf{Z}_1,...,\mathbf{Z}_{h-1}]\mathbf{W}^{O},$$
whereas $\mathbf{W}^{O} \in \mathbb{R}^{h.D_v \times N}$ computes linear transformation of heads and $\mathbf{Z}_i$ can be written as,
$$ \mathbf{Z}_i = \mathbf{softmax}\left (\frac{\mathbf{QW}^{Q_i}(\mathbf{KW}^{K_i})^{T}}{\sqrt{D_q/h}}\right ) \mathbf{VW}^{V_i}.$$

Note that the complexity of computing the softmax for SA block is quadratic with respect to the length of the input sequence that can limit its applicability to high-resolution medical images. Recently, numerous efforts have been made to reduce complexity, including sparse attention \cite{rao2021dynamicvit}, linearization attention \cite{katharopoulos2020transformers}, low-rank attention \cite{xiong2021nystr}, memory compression based approaches \cite{choromanski2020rethinking}, and improved MHSA \cite{shazeer2020talking}. We will discuss the efficient SA in the context of medical imaging in the relevant sections.

Further, we find it important to clarify that several alternate attention approaches \cite{jin2020ra,schlemper2019attention,maji2022attention,guo2021attention} have been explored in the literature based on convolutional architectures. In this survey, we focus on the specific attention used in transformer blocks (MHSA) which has recently gained significant research attention in medical image analysis. Next, we outline these methods categorized according to specific application domains.

\section{Medical Image Segmentation} \label{sec:seg}

Accurate medical image segmentation is a crucial step in computer-aided diagnosis, image-guided surgery, and treatment planning. 
The global context modeling capability of Transformers is crucial for accurate medical image segmentation because the organs spread over a large receptive field can be effectively encoded by modeling the relationships between spatially distant pixels (e.g., lungs segmentation). Furthermore,  the background in medical scans is generally scattered (e.g., in ultrasound scan \cite{avola2021ultrasound}); therefore, learning global context between the pixels corresponding to the background can help the model in preventing misclassification.

Below, we highlight various attempts to integrate ViT-based models for medical image segmentation. We broadly classify the ViT-based segmentation approaches into \textit{organ-specific} and \textit{multi-organ} categories, as depicted in Fig. \ref{fig:seg_taxonomy}, due to the varying levels of context modeling required in both sets of methods.

\subsection{Organ-Specific Segmentation}

ViT-based organ-specific approaches generally consider a specific aspect of the underlying organ to design architectural components or loss functions. We mention specific examples of such design choices in this section. We have further categorized organ-specific categories into 2D and 3D-based approaches depending on the input type. 
\subsubsection{\textbf{2D Segmentation}} Here, we describe the organ-specific ViT-based segmentation approaches for 2D medical scans.

 {{\textbf{Skin Lesion Segmentation.}}}
\begin{figure}[t]
\centering
\includegraphics[width = 0.48\textwidth]{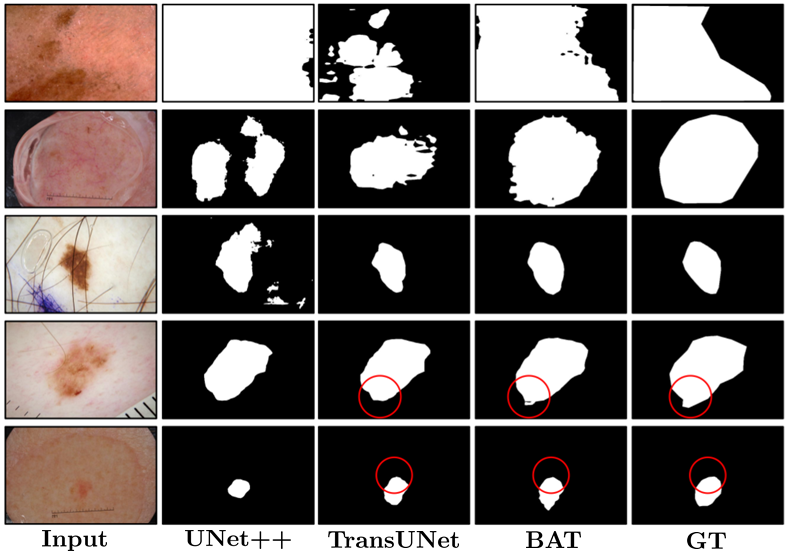}
\caption{Comparison of different skin lesion segmentation approaches. From left to right: Input image, CNN based UNet++ \cite{zhou2019unet++}, ViT-based TransUNet \cite{chen2021transunet}, Boundary aware transformer (BAT) \cite{wang2021boundary}, and the ground truth (GT) image. Red circles highlight small regions with an ambiguous boundary where BAT can perform well due to the use of boundary-wise prior knowledge. Image taken from \cite{wang2021boundary}.}
\label{fig:lesion}
\end{figure}
Accurate skin lesion segmentation for identifying melanoma (cancer cells) is crucial for cancer diagnosis and subsequent treatment planning. However, it remains a challenging task due to significant variations in color, size, occlusions, and contrast of skin lesion areas, resulting in ambiguous boundaries~\cite{yuan2017automatic} and consequently deterioration in segmentation performance. 
To address the issue of ambiguous boundaries, Wang \textit{et al.}~\cite{wang2021boundary} propose a novel Boundary-Aware Transformer (BAT).
Specifically, they design a boundary-wise attention gate in Transformer architecture to exploit the prior knowledge about boundaries. The auxiliary supervision of the boundary-wise attention gate provides feedback to train BAT effectively. 
Extensive experiments on ISIC 2016+PH2 \cite{gutman2016skin,mendoncca2013ph} and ISIC 2018 \cite{codella2019skin} validate the efficacy of their boundary-wise prior, as shown in Fig. \ref{fig:lesion}. 
Similarly, Wu \textit{et al.}~\cite{wu2021fat} propose a dual encoder-based feature adaptive transformer network (FAT-Net) that consists of CNN and transformer branches in the encoder. To effectively fuse the features from these two branches, a memory-efficient decoder and feature adaptation module have been designed. Experiments on ISIC 2016-2018~\cite{gutman2016skin,berseth2017isic,codella2019skin}, and PH2~\cite{mendoncca2013ph} datasets demonstrate the effectiveness of FAT-Net fusion modules.

 {{\textbf{Tooth Root Segmentation.}}}
Tooth root segmentation is one of the critical steps in
root canal therapy to treat periodontitis (gum infection) \cite{gao2010individual}. However, it is challenging due to blurry boundaries and overexposed and underexposed images. To address these challenges, Li \textit{et al.}~\cite{li2021gt} propose Group Transformer U-Net (GT U-Net) that consists of transformer and convolutional layers to encode global and local context, respectively. A shape-sensitive Fourier Descriptor loss function \cite{zahn1972fourier} has been proposed to deal with the fuzzy tooth boundaries. Furthermore, grouping and bottleneck structure has been introduced in the GT U-Net to significantly reduce the computational cost. Experiments on their in-house Tooth Root segmentation dataset with six evaluation metrics demonstrate the effectiveness of GT U-Net architectural components and Fourier-based loss function. In another work, Li \textit{et al.}~\cite{li2021agmb} propose anatomy-guided multibranch Transformer (AGMB-Transformer) to incorporate the strengths of group convolutions~\cite{chollet2017xception} and progressive Transformer network. Experiments on their self-collected dataset of 245 tooth root X-ray images show the effectiveness of AGMB-Transformer.

 {{\textbf{Cardiac Image Segmentation.}}}
Despite their impressive performance in medical image segmentation, Transformers are computationally demanding to train and come with a high parameter budget. To handle these challenges for cardiac image segmentation task, Deng \textit{et al.}\cite{deng2021transbridge} propose TransBridge, a lightweight parameter-efficient hybrid model.
TransBridge consists of Transformers and CNNs based encoder-decoder structure for left ventricle segmentation in echocardiography. Specifically, the patch embedding layer of the Transformer has been re-designed using the shuffling layer \cite{zhang2021sa} and group convolutions to significantly reduce the number of parameters. 
Extensive experiments on large-scale left ventricle segmentation dataset, echo-cardiographs~\cite{ouyang2020video} demonstrate the benefit of TransBridge over CNNs and Transformer-based baseline approaches~\cite{xie2021cotr}.

\begin{figure*}[t]
\centering
\includegraphics[width = 1\textwidth]{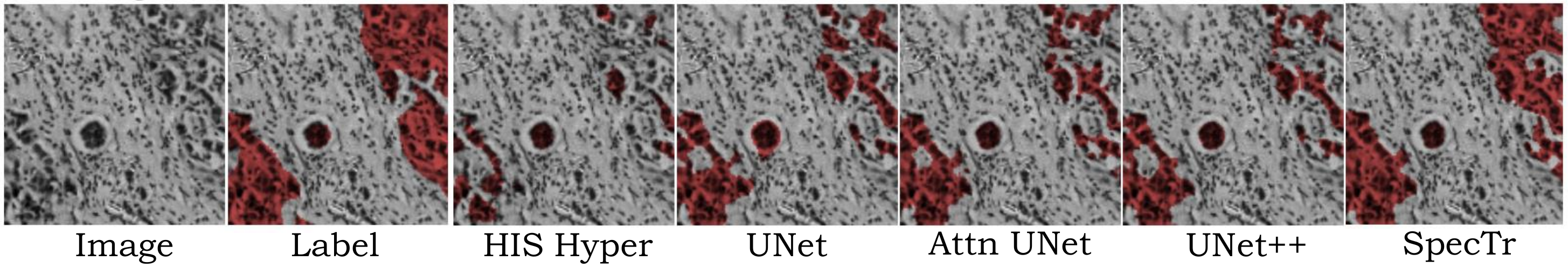}
\caption{Segmentation results for hyperspectral pathology dataset using Spectral Transformer (SpecTr). From left to right: Input image, Ground truth label, HIS Hyper (CNN-based)~\cite{wang2020identification}, UNet (CNN-based)~\cite{ronneberger2015u}, Attn UNet (CNN-based)~\cite{oktay2018attention}, UNet++ (CNN-based)~\cite{zhou2019unet++}, and SpecTr (ViT-based)~\cite{yun2021spectr}. Image adapted from \cite{yun2021spectr}.}
\label{fig:corneal}
\end{figure*}

 {{\textbf{Kidney Tumor Segmentation.}}}
Accurate segmentation of kidney tumors via computer diagnosis systems can reduce the effort of radiologists and is a critical step in related surgical procedures. However, it is challenging due to varying kidney tumor sizes and the contrast between tumors and their anatomical surroundings. To address these challenges, Shen \textit{et al.}~\cite{shen2021automated} propose a hybrid encoder-decoder architecture, COTR-Net, that consists of convolution and transformer layers for end-to-end kidney, kidney cyst, and kidney tumor segmentation. Specifically, the encoder of COTR-Net consists of several convolution-transformer blocks, and the decoder comprises several up-sampling layers with skip connections from the encoder. The encoder weights have been initialized using a pre-trained ResNet~\cite{he2016deep} architecture to accelerate convergence, and deep supervision has been exploited in the decoder layers to boost segmentation performance. Furthermore, the segmentation masks are refined using morphological operations as a post-processing step. Extensive experiments on the Kidney Tumor Segmentation dataset (KiTS21)~\cite{kits2021} demonstrate the effectiveness COTR-Net.

\begin{table}[]
\centering
\resizebox{0.48\textwidth}{!}{%
\begin{tabular}{V{3}l|l|lV{3}}
\hlineB{3}
\rowcolor{mygray} \textbf{CNN-based Models}         & \textbf{40x Magnification} & \textbf{20x Magnification} \\ \hline
PSPNet      \cite{zhao2017pyramid}              & 0.58$\pm$0.33         & 0.49$\pm$0.27         \\ \hline
U-Net \cite{ronneberger2015u}                    & 0.65$\pm$0.24         & 0.60$\pm$0.31         \\ \hline
DeepLabV3 \cite{chen2017deeplab}              & 0.63$\pm$0.28         & 0.67$\pm$0.24         \\ \hline
FPN  \cite{lin2017feature}                    & 0.64$\pm$0.20         & 0.72$\pm$0.22         \\ \hline
PAN     \cite{li2018pyramid}                 & 0.63$\pm$0.24         & 0.69$\pm$0.23         \\ \hline
LinkNet \cite{chaurasia2017linknet}                 & 0.35$\pm$0.33         & 0.54$\pm$0.25         \\ \hline 
\rowcolor{mygray} \textbf{Transformer-based Models} & \textbf{40x Magnification} & \textbf{20x Magnification} \\ \hline
TransUNet  \cite{chen2021transunet}              & 0.77$\pm$0.12         & 0.77$\pm$0.13         \\ \hline
Swin-UNet  \cite{cao2021swin}               & 0.53$\pm$0.23         & 0.42+0.23         \\ \hline
Swin Transformer (Base) \cite{liu2021swin}  & 0.79$\pm$0.14         & 0.71$\pm$0.26         \\ \hline
Segmenter  \cite{strudel2021segmenter}              & 0.80$\pm$0.14         & 0.82$\pm$0.11         \\ \hline
Medical Transformer \cite{valanarasu2021medical}     & 0.71$\pm$0.14         & 0.62$\pm$0.17         \\ \hline
BEiT  \cite{bao2021beit}                   & 0.72$\pm$0.21         & 0.66$\pm$0.28         \\ \hlineB{3}
\end{tabular}
}
\caption{Evaluation of Transformer-based semantic segmentation methods for pathological image segmentation in terms of average jaccard index on PAIP liver histopathological dataset \cite{kim2021paip}. It can be seen that transformer-based models tend to outperform CNNs with the exception of Swin-UNet. Results are from \cite{nguyen2021evaluating}, which is one of the first study to systematically evaluate the performance of transformers on pathological image segmentation task. }
\label{tab:histopathology}
\end{table}
 {{\textbf{Cell Segmentation.}}}
Inspired from the Detection Transformers (DETR)~\cite{carion2020end}, Zhang \textit{et al.} proposed Cell-DETR~\cite{prangemeier2020attention}, a Transformer-based framework for instance segmentation of biological cells. Specifically, they integrate a dedicated attention branch to the DETR framework to obtain instance-wise segmentation masks in addition to box predictions. During training, focal loss \cite{lin2017focal} and Sorenson dice loss \cite{carion2020end} are used for the segmentation branch. To enhance performance, they integrate three residual decoder blocks \cite{he2016deep} in Cell-DETR to generate accurate instance masks. Experiments on their in-house yeast cells dataset demonstrate the effectiveness of Cell-DETR relative to U-Net based baselines \cite{ronneberger2015u}. 
Similarly, existing medical imaging segmentation approaches generally struggle for Corneal endothelial cells due to blurry edges caused by the subject's movement \cite{van2019corneal}. This demands preserving more local details and making full use of the global context. Considering these attributes, Zhang \textit{et al.} \cite{zhang2021multi} propose a Multi-Branch hybrid Transformer Network (MBT-Net) consisting of convolutional and transformer layers. Specifically, they propose a body-edge branch that provides precise edge location information and promotes local consistency. Extensive ablation studies  on their self-collected TM-EM3000 and public Alisarine dataset \cite{ruggeri2010system} of Corneal Endothelial Cells show the effectiveness of MBT-Net architectural components.

 \textbf{Histopathology.}
Histopathology refers to the diagnosis and study of the diseases of tissues under a microscope and is the
gold standard for cancer recognition. Therefore accurate automatic segmentation of histopathology images can substantially alleviate the workload of pathologists. Recently, Nguyen \textit{et al.} \cite{nguyen2021evaluating}
systematically evaluate the performance of six latest ViTs, and CNNs-based approaches on whole slide images of the PAIP liver histopathological dataset \cite{kim2021paip}. Their results (shown in Table \ref{tab:histopathology})  demonstrate that almost all Transformer-based models indeed exhibit superior performance as compared to CNN-based approaches due to their ability to encode the global context. 

\subsubsection{\textbf{3D Medical Segmentation}} Here, we describe ViT-based segmentation approaches for volumetric medical data.

 {{\textbf{Brain Tumor Segmentation.}}}
An automatic and accurate brain tumor segmentation approach can lead to the timely diagnosis of neurological disorders such as Alzheimer’s disease. Recently, ViT-based models have been proposed to segment brain tumors effectively. Wang \textit{et al.}~\cite{wang2021transbts} have made the first attempt to leverage Transformers for 3D multimodal brain tumor segmentation by effectively modeling local and global features in both spatial and depth dimensions. Specifically, their encoder-decoder architecture, TransBTS, employs a 3D CNN to extract local 3D volumetric spatial features and Transformers to encode global features. Progressive upsampling in the 3D CNN-based decoder has been used to predict the final segmentation map.
To further boost the performance, they make use of test-time augmentation. Extensive experimentation on BraTS 2019\footnote{https://www.med.upenn.edu/cbica/brats2019} and BraTS 2020\footnote{https://www.med.upenn.edu/cbica/brats2020} datasets show the effectiveness of their proposed approach compared to CNN-based methods. 
Unlike most of the ViT-based image segmentation approaches, TransBTS does not require pre-training on large datasets and has been trained from scratch. 
In another work, inspired from the architectural design of TransBTS \cite{wang2021transbts}, Jia \textit{et al.} \cite{jia2021bitr} propose Bi-Transformer U-Net (BiTr-UNet) that performs relatively better on BraTS 2021~\cite{baid2021rsna} segmentation challenge. Different from TransBTS \cite{wang2021transbts}, BiTr-UNet consists of an attention module to refine encoder and decoder features and has two ViT layers (instead of one as in TransBTS).
Furthermore, BiTr-UNet adopts a post-processing strategy to eliminate a volume of predicted segmentation if the volume is smaller than a threshold \cite{isensee2018nnu} followed by model ensemble via majority voting \cite{lam1997application}.
Similarly, Peiris \textit{et al.}~\cite{peiris2021volumetric} propose a light-weight UNet shaped volumetric transformer, VT-UNet, to segment 3D medical image modalities in a hierarchical manner. Specifically, two self-attention layers have been introduced in the encoder of VT-UNet to capture both global and local contexts. Furthermore, the introduction of window-based self-attention and cross-attention modules and Fourier positional encoding in the decoder significantly improve the accuracy and efficiency of VT-UNet. Experiments on BraTs 2021~\cite{baid2021rsna} show that VT-UNet is robust to data artifacts and exhibits strong generalization ability.
\begin{table}
\centering
\resizebox{0.48\textwidth}{!}{%
\begin{tabular}{V{3}lcccV{3}} 
\hlineB{3}
\rowcolor{mygray} \textbf{Method} & \textbf{\#params}      & \textbf{Flops}        & \textbf{Dice Score (Avg.)}  \\ 
\hline
TransBTS~\cite{wang2021transbts}                                 & 33 M           & 333 G         & 84.99              \\
BiTr-UNet~\cite{jia2021bitr}                                 & -           & -         &    86.20           \\
UNETR~\cite{hatamizadeh2021unetr}                                    & 102.5 M        & 193.5 G       & 84.51              \\
nnFormer~\cite{zhou2021nnformer}                                 & 39.7 M         & 110.7 G       & 86.56              \\
Swin UNETR~\cite{hatamizadeh2022swin}                               & 61.98 M        & ~394.84 G     & \textbf{88.97}     \\
VT-UNET-T~\cite{peiris2021volumetric}                                & \textbf{5.4 M} & \textbf{52 G} & 86.82              \\
VT-UNET-S~\cite{peiris2021volumetric}                                & 11.8 M         & 100.8 G       & 87.00              \\
VT-UNET-B~\cite{peiris2021volumetric}                                & 20.8 M         & 165 G         & 88.07              \\
\hlineB{3}
\end{tabular}
}
\caption{Segmentation results and parameters of various Transformer-based models on 3D Multimodal Brain Tumor BraTS 2021 dataset~\cite{baid2021rsna}.}
\label{tab:brats21}
\end{table}
In another similar work, Hatamizadeh \textit{et al.}~\cite{hatamizadeh2022swin} propose Swin UNet based architecture, Swin UNETR, that consists of Swin transformer as the encoder and a CNN-based decoder. Specifically,  Swin UNETR computes self-attention in an efficient shifted window partitioning scheme and is a top-performing model on BraTs 2021~\cite{baid2021rsna} validation set. In Table \ref{tab:brats21}, we provide dice score and other parameters of various Transformer based models for the 3D multimodal BraTs 2021 dataset~\cite{baid2021rsna}.

 {{\textbf{Histopathology.}}}
Boxiang \textit{et. al} \cite{yun2021spectr} propose Spectral Transformer (SpecTr) for hyperspectral pathology image segmentation, which employs transformers to learn the contextual feature across the spectral dimension. To discard the irrelevant spectral bands, they introduce a sparsity-based scheme \cite{correia2019adaptively}. Furthermore, they employ separate group normalization for each band to eliminate the interference caused by distribution mismatch among spectral images. Extensive experimentation on the hyperspectral pathology dataset, Cholangiocarcinoma~\cite{zhang2019multidimensional}, shows the effectiveness of SpecTr as also shown in Fig. \ref{fig:corneal}.

 {{\textbf{Breast Tumor Segmentation.}}} Detection of breast cancer in the early stages can reduce the fatality rate by more than 40$\%$~\cite{huang2017breast}. Therefore, automatic breast tumor detection is of immense importance to doctors. Recently, Zhu \textit{et al.}~\cite{zhu2021region} propose a region aware transformer network (RAT-Net) to effectively fuse the Breast tumor region information into multiple scales to obtain precise segmentation. Extensive experiments on a large ultrasound breast tumor segmentation dataset show that RAT-Net outperforms CNN and transformer-based baselines.
Similarly, Liu \textit{et al.}~\cite{liu20213d} also propose a hybrid architecture consisting of transformer layers in the decoder part of 3D UNet~\cite{cciccek20163d} to effectively segment tumors from volumetric breast data.

\subsection{Multi-organ Segmentation}

Multi-organ segmentation aims to segment several organs simultaneously and is challenging due to inter-class imbalance and varying sizes, shapes, and contrast of different organs. ViT models are particularly suitable for the multi-organ segmentation due to their ability to effectively model global relations and differentiate multiple organs. We have categorized multi-organ segmentation approaches based on the architectural design, as these approaches do not consider any organ-specific aspect and generally focus on boosting performance by designing effective and efficient architectural modules~\cite{lei2020deep}. We categorize multi-organ segmentation approaches into \textit{Pure Transformer} (only ViT layers) and \textit{Hybrid Architectures} (both CNNs and ViTs layers).

\begin{figure}[t]
\centering
\includegraphics[width = \columnwidth]{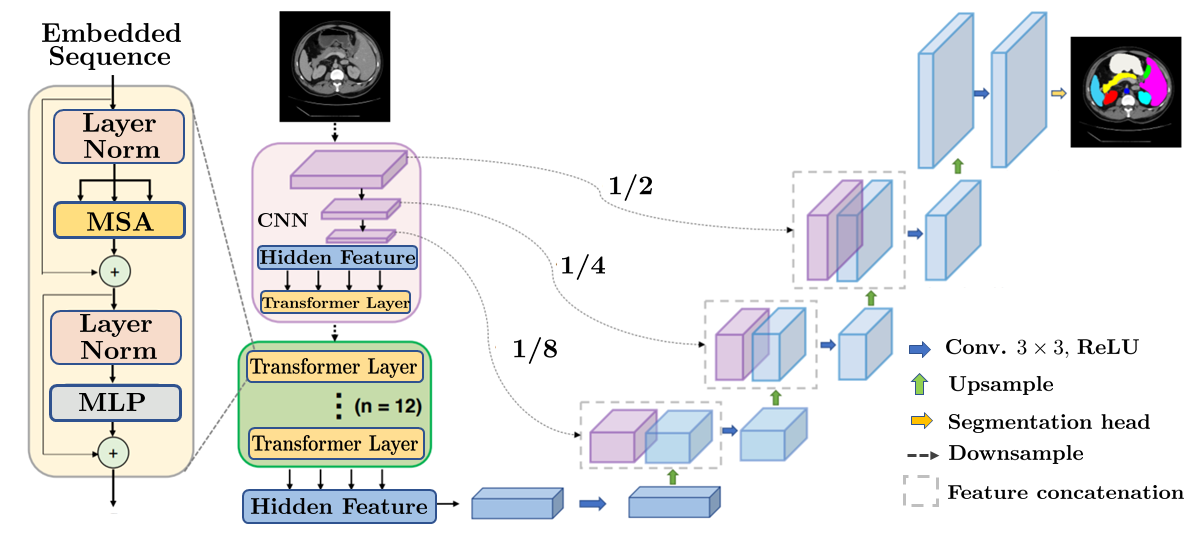}
\caption{Overview of TransUNet architecture \cite{chen2021transunet} proposed for multi-organ segmentation. It is one of the first transformer-based architecture proposed for medical image segmentation and merits both transformer and UNet. It employs a hybrid CNN-Transformer architecture for encoder, followed by multiple upsampling layers in decoder to output final segmentation mask. Image adapted from \cite{chen2021transunet}.}
\label{fig:transunet}
\end{figure}

\subsubsection{\textbf{Pure Transformers}} 
Pure Transformer based architectures consist of only ViT layers and have seen fewer applications in medical image segmentation compared to hybrid architectures as both global and local information is crucial for dense prediction tasks like segmentation~\cite{chen2021transunet}.
Recently, Karimi \textit{et. al} \cite{karimi2021convolution}  propose a pure Transformer-based model for 3D medical image segmentation by leveraging self-attention \cite{wang2018non} between neighboring linear embedding of 3D medical image patches. They also propose a method to effectively pre-train their model when only a few labeled images are available. Extensive experiments show the effectiveness of their convolution-free network on three benchmark 3D medical imaging datasets related to brain cortical plate~\cite{dou2020deep}, pancreas, and hippocampus. 
One of the drawbacks of using Pure Transformer-based models in segmentation is the quadratic complexity of self-attention with respect to the input image dimensions. This can hinder the ViTs applicability in the segmentation of high-resolution medical images. To mitigate this issue, Cao \textit{et al.} \cite{cao2021swin} propose Swin-UNet that, like Swin Transformer~\cite{liu2021swin}, computes self-attention within a local window and has linear computational complexity with respect to the input image. 
Swin-UNet also contains a patch expanding layer for upsampling decoder's feature maps and shows superior performance in recovering fine details compared to bilinear upsampling.
Experiments on Synapse and ACDC~\cite{bernard2018deep} dataset demonstrate the effectiveness of the Swin-UNet architectural design. 

\subsubsection{\textbf{Hybrid Architectures}} 

Hybrid architecture-based approaches combine the complementary strengths of Transformers and CNNs to effectively model global context and capture local features for accurate segmentation. We have further categorized these hybrid models into single and multi-scale approaches.

\paragraph{{\textbf{Single-Scale Architectures}}}

These methods process the input image information at one scale only and have seen widespread applications in medical image segmentation due to their low computational complexity compared to multi-scale architectures. 
We can sub-categorized single-scale architectures based on the position of the Transformer layers in the model. These sub-categories include \textit{Transformer in Encoder}, \textit{Transformer between Encoder and Decoder}, \textit{Transformer in Encoder and Decoder}, and \textit{Transformer in Decoder}.

{{\textbf{Transformer in Encoder.}}} Most initially developed Transformer-based medical image segmentation approaches have Transformer layers in the model's encoder.
The first work in this category is TransUNet~\cite{chen2021transunet} that consists of 12 Transformer layers in the encoder as shown in Figure \ref{fig:transunet}. These Transformer layers encode the tokenized image patches from the CNN layers.
The resulting encoded features are upsampled via up-sampling layers in the decoder to output the final segmentation map. With skip-connection incorporated, TransUnet sets new records (at the time of publication) on synapse multi-organ segmentation dataset~\cite{synapse} and automated cardiac diagnosis challenge (ACDC)~\cite{bernard2018deep}.
In other work, Zhang \textit{et al.} propose TransFuse~\cite{zhang2021transfuse} to effectively fuse features from the Transformer and CNN layers via BiFusion module.
The BiFusion module leverages the self-attention and multi-modal fusion mechanism to selectively fuse the features.
Extensive evaluation of TransFuse on multiple modalities (2D and 3D), including Polyp segmentation, skin lesion segmentation, Hip segmentation, and prostate segmentation, demonstrate its efficacy. 
Both TransUNet~\cite{chen2021transunet} and TransFuse~\cite{zhang2021transfuse} require pre-training on ImageNet dataset~\cite{deng2009imagenet} to effectively learn the positional encoding of the images. To learn this positional bias without any pre-training, Valanarasu \textit{et al.} \cite{valanarasu2021medical} propose a modified gated axial attention layer~\cite{wang2020axial} that works well on small medical image segmentation datasets. Furthermore, to boost segmentation performance, they propose a Local-Global training scheme to focus on the fine details of input images.
Extensive experimentation on brain anatomy segmentation \cite{wang2018automatic}, gland segmentation \cite{sirinukunwattana2017gland}, and MoNuSeg (microscopy) \cite{kumar2019multi} demonstrate the effectiveness of their proposed gated axial attention module.

In another work, Tang \textit{et al.}~\cite{tang2021self} introduce Swin UNETR, a novel self-supervised learning framework with proxy tasks to pre-train Transformer encoder on 5,050 images of CT dataset. They validate the effectiveness of pre-training by fine-tuning the Transformer encoder with a CNN-based decoder on the downstream task of MSD and BTCV segmentation datasets.
Similarly, Sobirov \textit{et al.}~\cite{sobirov2022automatic} show that transformer-based models can achieve comparable results to state-of-the-art CNN-based approaches on the task of head and neck tumor segmentation. 
Few works have also investigated the effectiveness of Transformer layers by integrating them into the encoder of UNet-based architectures in a plug-and-play manner. For instance, Cheng \textit{et al.} \cite{chang2021transclaw} propose TransClaw UNet by integrating Transformer layers in the encoding part of the Claw UNet~\cite{yao2020claw} to exploit multi-scale information.
TransClaw-UNet achieves an absolute gain of 0.6 in dice score compared to Claw-UNet on Synapse multi-organ segmentation dataset and shows excellent generalization.
Similarly, inspired from the LeViT \cite{graham2021levit}, Xu \textit{et al.}~\cite{xu2021levit} propose LeViT-UNet which aims to optimize the trade-off between accuracy and efficiency. 
LeViT-UNet is a multi-stage architecture 
that demonstrates good performance and generalization ability on Synapse and ACDC 
benchmarks.

{{\textbf{Transformer between Encoder and Decoder.}}} In this category, Transformer layers are between the encoder and decoder of a U-Shape architecture. These architectures are more suitable to avoid the loss of details during down-sampling in the encoder layers. The first work in this category is TransAttUNet~\cite{chen2021transattunet}
that leverages guided attention and multi-scale skip connection to enhance the flexibility of traditional UNet. Specifically, a robust self-aware attention module has been embedded between the encoder and decoder of UNet to concurrently exploit the expressive abilities of global spatial attention and transformer self-attention. Extensive experiments on five benchmark medical imaging segmentation datasets demonstrate the effectiveness of TransAttUNet architecture. Similarly, Yan \textit{et al.}~\cite{yan2021after} propose Axial Fusion Transformer UNet (AFTer-UNet) that contains a computationally efficient axial fusion layer between encoder and decoder to effectively fuse inter and intra-slice 
information for 3D medical image segmentation. Experimentation on BCV~\cite{simpson2019large}, Thorax-85~\cite{chen2021deep}, and SegTHOR~\cite{lambert2020segthor} datasets demonstrate the effectiveness of their proposed fusion layer.

{{\textbf{Transformer in Encoder and Decoder.}}} Few works integrate Transformer layers in both encoder and decoder of a U-shape architecture to better exploit the global context for medical image segmentation. The first work in this category is UTNet that efficiently reduces the complexity of the self-attention mechanism from quadratic to linear~\cite{wang2020linformer}.
Furthermore, to model the image content effectively, UTNet exploits the two-dimensional relative position encoding~\cite{bello2019attention}. Experiments show strong generalization ability of UTNet on multi-label and multi-vendor cardiac MRI challenge dataset cohort~\cite{campello2021multi}.
\begin{figure}[t]
\centering
\includegraphics[width = 0.48\textwidth]{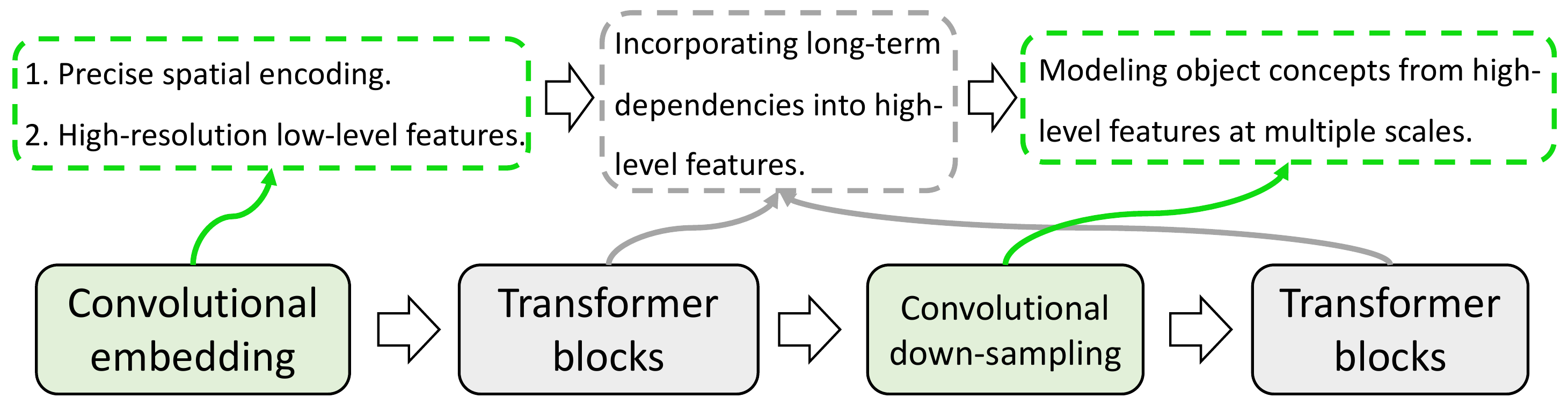} 
\caption{Overview of the interleaved encoder not-another transFormer (nnFormer) \cite{zhou2021nnformer} for volumetric medical image segmentation. Note that  convolution and transformer layers are interleaved to give full play to their strengths. Image taken from \cite{zhou2021nnformer}.}
\label{fig:nnformer}
\end{figure}
Similarly, to optimally combine convolution and transformer layers for medical image segmentation, Zhou \textit{et al.}~\cite{zhou2021nnformer} propose nnFormer, an interleave encoder-decoder based architecture, where convolution layer encodes precise spatial information and Transformer layer encodes global context as shown in Fig. \ref{fig:nnformer}. Like Swin Transformers \cite{liu2021swin}, the self-attention in nnFormer has been computed within a local window to reduce the computational complexity. Moreover, deep supervision in the decoder layers has been employed to enhance performance.
Experiments on ACDC and Synapse datasets show that
nnFormer surpass Swin-UNet~\cite{cao2021swin} (transformer-based medical segmentation approach) by over 7$\%$ (dice score) on Synapse dataset.
In other work, Lin \textit{et al.} propose  Dual Swin Transformer UNet (DS-TransUNet) \cite{lin2021ds} to incorporate the advantages of Swin Transformer in U-shaped architecture for medical image segmentation. They split the input image into non-overlapping patches at two scales and feed them into the two Swin Transformer-based branches of the encoder. A novel Transformer Interactive Fusion module has been proposed to build long-range dependencies between different scale features in encoder. DS-TransUNet outperforms CNN-based methods on four standard datasets related to Polyp segmentation, ISIC 2018, GLAS, and Datascience bowl 2018. 

{{\textbf{Transformer in Decoder.}}} Li \textit{et al.}~\cite{li2021more} investigate the use of Transformer as an upsampling block in the decoder of the UNet for medical image segmentation. Specifically, they adopt a window-based self-attention mechanism to better complement the upsampled feature maps while maintaining 

\begin{table*}[]
\rotatebox{90}{
\centering
\begin{adjustbox}{max width=1.2\textwidth}
\begin{tabular}{V{3}c|c|p{2.2cm}|c|p{3.5cm}|p{3cm}|c|c|p{11.5cm}V{3}}
\hlineB{3}
\rowcolor{mygray} \textbf{Method} & \textbf{Organ} & \textbf{Modality}        & \textbf{Type} & \textbf{Datasets}                                                                  & \textbf{Metrics}                                             & \textbf{Arch.} & \textbf{P.T.} & {\textbf{Highlights}}                                                                                                                                                                                     \\ \hline
TransUNet\cite{chen2021transunet}       &     Multi-organ           & CT, MRI                & 2D            & Synapse~\cite{synapse} , ACDC~\cite{bernard2018deep}                                              &                     Dice, Hausdorff distance                                         &         Hybrid             &  Yes    &  Encodes strong global
context by treating the image features as sequences but also well utilizes the low-level CNN features via a u-shaped hybrid architectural design.                                                                                                                   \\ \hline
TransFuse\cite{zhang2021transfuse}       &      Multi-organ          & Colonoscopy                        & 2D, 3D        &                                        KVASIR~\cite{jha2020kvasir}, Clinic DB~\cite{bernal2015wm}, Colon DB~\cite{tajbakhsh2015automated}, EndoScene~\cite{vazquez2017benchmark}, ETIS~\cite{silva2014toward}, ISIC 2017~\cite{berseth2017isic}, MSD~\cite{simpson2019large}                                 &         Dice, Jaccard index                                                      &         Hybrid             &   Yes     & Leverages the inductive bias of CNNs on modeling spatial correlation and the powerful capability of Transformers on modelling global relationship. Novel Bi-Fusion module to fuse CNN and  Transformers features for segmentation.                                                                                                                                          \\ \hline
MedT\cite{zhang2021pyramid}            &     Multi-organ           & Ultrasound, Microscopy   & 2D            & Brain US (Private), GLAS\cite{sirinukunwattana2017gland}, MoNuSeg~\cite{kumar2017dataset}                                                     & F1                                                             &        Hybrid               &    No  & Gated axial attention layer for positional encoding. Local global training for training on both full resolution images as well in patches.                                                                        \\ \hline
Conv. Free\cite{karimi2021convolution}      &      Multi-organ          & MRI, CT                  & 3D            & Brain crotial, (private)                                               &                                         Dice, Hausdorff Distance, Average Symmetric Surface Distance                  &          Pure             &   Yes   & Convolutional free medical segmentation model. Based on self-attention between neighboring 3D patches.                                                                                                                                                          \\ \hline
CoTr\cite{xie2021cotr}            &      Multi-organ          & CT                       & 3D            & BCV~\cite{simpson2019large}                                                                       &                                                  Dice           &          Hybrid            &   Yes    &  Deformable self-attention mechanism to reduce the computational and spatial complexities of modelling the long range dependency.                                                                                                                                                                    \\ \hline
SpecTr\cite{yun2021spectr}          &      Bile-duct          & Hyperspectral                      & 3D            & Choledock~\cite{zhang2019multidimensional} &                                              Dice, IoU, Hausdorff distance                &           Hybrid            &   No   & First application to hyperspectral and learn contextual feature across spectral dimension.                                                                                                              \\ \hline
TransBTS\cite{wang2021transbts}        &        Brain        & MRI                      & 3D            & BraTS 19~\cite{brats19}, BraTS 20~\cite{brats20}                                                          &        Dice, Hausdorff distance                                                      &           Hybrid            &   Yes   & 3D CNN for capturing local volumetric features and transformers for encoding global features.                                                                                                                    \\ \hline
U-Transformer\cite{petit2021u}   &      Multi-organ          & CT                       & 2D            & TCIA~\cite{clark2013cancer}, Private multi-organ &                                      Dice                        &           Hybrid            &    No  & Propose self and cross-attention modules
to model long-range interactions and spatial dependencies.                                                                                       \\ \hline
UNETR\cite{hatamizadeh2021unetr}           &      Brain, Spleen          & MRI, CT                  & 3D            & BTCV, MSD~\cite{simpson2019large}  &    Dice, Hausdorff distance  &            Hybrid           & No     &  Transformer as the encoder to learn sequence representations of the input volume and effectively capture the global multi-scale information                                                                                                        \\ \hline
PMTrans\cite{zhang2021pyramid}         &       Multi-organ         & Microscopy, CT, PET      & 2D            & GLAS~\cite{sirinukunwattana2017gland}, MoNuSeg~\cite{kumar2017dataset}, HECKTOR~\cite{andrearczyk2020overview}                                                &      Dice                                                        &            Hybrid           &   No   & Integrate multi-scale attention and CNN feature extraction using a pyramidal network architecture. An adaptive partitioning scheme was implemented to retain informative relations and to access different receptive fields efficiently.                                                                                                                               \\ \hline
Swin-UNet\cite{cao2021swin}       &       Multi-organ         & CT                       & 2D            & Synapse~\cite{synapse}, ACDC~\cite{bernard2018deep}                                                                      &       Dice                                                       &       Pure                &   Yes   & Swin transformer based pure transformer architecture to segmentation with patch expanding layer design in decoder.                                                                                      \\ \hline
Segtran\cite{li2021medical}         & Multi-organ    & Fundus, Colonoscopy, MRI & 2D, 3D        & REFUGE 20~\cite{orlando2020refuge}, BraTS 19~\cite{brats19}, CVC~\cite{fan2020pranet}, KVASIR~\cite{jha2020kvasir}                                               & Dice                                                         & Hybrid                & Yes  & Squeeze-and-Expansion transformer where squeeze block regularize the self-attention module and expansion block learns diversified represetations.                                                                           \\ \hline
MBT-Net\cite{zhang2021multi}         & Eye            & Pathology                & 2D            & TM-EM3000 (private), Alisarine~\cite{ruggeri2010system}                                                               & Dice, F1, Sensitivity, Specificity                           & Hybrid                & No   & Body edge branch for precise edge location information for corneal endothelial cells segmentation.                                                                                                     \\ \hline
DS-TransUNet\cite{lin2021ds}    & Multi-organ    & Colonoscopy, Histology   & 2D            & Kvasir~\cite{jha2020kvasir}, Colon DB~\cite{tajbakhsh2015automated} and Clinic DB~\cite{bernal2015wm}, EndoScene~\cite{vazquez2017benchmark}, ETIS~\cite{silva2014toward}, ISIC 18~\cite{codella2019skin}, GLAS~\cite{sirinukunwattana2017gland}, Data Science Bowl 18~\cite{caicedo2019nucleus} & Mean Dice, Mean IoU, Precision, Recall                       & Hybrid                & Yes  & Dual-branch Swin Transformer in encoder and decoder to extract multiscale representation. Transformer Interactive Fusion module to build long-range  dependencies between features of different scales \\ \hline
MCTrans\cite{ji2021multi}         & Multi-organ    & Colonoscopy, Pathology   & 2D            & Pannuke dataset, Colon DB~\cite{tajbakhsh2015automated}, Clinic DB~\cite{bernal2015wm}, ETIS~\cite{silva2014toward}, KVASIR~\cite{jha2020kvasir}, ISIC 2018~\cite{codella2019skin}                     & Dice                                                         & Hybrid                & No   & Transformer self-attention for cross-scale contextual dependencies Transformer cross attention layer for semantic correspondence.                                                                      \\ \hline
Li \textit{et al.}\cite{li2021more}       & Multi-organ    & MRI, CT                  & 2D            & Synapse~\cite{synapse}, MSD Brain~\cite{simpson2019large}                                                                 & Dice, Hausdorff distance                                     & Hybrid                & No   & Investigate the use of transformer decoder for medical image segmentation and its usage in upsampling.                                                                                                 \\ \hline
UTNet\cite{gao2021utnet}           & Heart          & MRI                      & 2D            & MRI Challenge Cohort~\cite{campello2021multi}                                                               & Dice, Hausdorff distance                                     & Hybrid                & No   & Self-attention modules in encoder and decoder. Design relative position encoding to reduce the complexity of self-attention from quadratic to linear.                                                  \\ \hline
TransClaw UNet\cite{chang2021transclaw}  & Multi-organ    & CT                       & 2D            & Synapse~\cite{synapse}                                                                            & Dice, Hausdorff distance                                     & Hybrid                & No   & Integrated transformer layer in the encoder path of Claw-UNet to extract shallow spatial features.                                                                                                     \\ \hline
TransAttUNet\cite{chen2021transattunet}    & Multi-organ    & Xray, CT                 & 2D            & ISIC 2018~\cite{codella2019skin}, JSRT~\cite{shiraishi2000development}, Montgomery~\cite{jaeger2014two}, NIH~\cite{tang2019xlsor}, Clean-CC-CCII~\cite{he2020benchmarking}, Data Science Bowl 18~\cite{caicedo2019nucleus}, GLAS~\cite{sirinukunwattana2017gland}                         & Dice, F1                                                     & Hybrid                & No   & Multi-level guided attention and multi-scale skip connections to mitigate information recession problem.                                                                                               \\ \hline
LeViT-UNet\cite{xu2021levit}      & Multi-organ    & CT, MRI                  & 2D            & Synapse~\cite{synapse}, ACDC~\cite{bernard2018deep}                                                                       & Dice, Hausdorff distance                                     & Hybrid                & Yes  & Integrate multiscale LeViT architecture as the a encoder in UNet.                                                                                                                                      \\ \hline
Polyp-PVT\cite{dong2021polyp}       & Multi-organ    & Colonoscopy              & 2D            & KVASIR~\cite{jha2020kvasir}, Clinic DB~\cite{bernal2015wm}, Colon DB~\cite{tajbakhsh2015automated}, Endoscene~\cite{vazquez2017benchmark}, ETIS~\cite{silva2014toward}                                   & Dice, IoU, MAE, Weighted F-measure, S-measure, E-measure     & Hybrid                & No   & Pyramid vision transformer backbone as encoder to extract robust features. Proposed architectural components to handle noise, occlusions, and capturing global semantic cues.                          \\ \hline
COTRNet\cite{shen2021automated}         & Kidney         & CT                       & 2D            & KITS21 Challenge~\cite{kits2021}                                                                   & Dice, Surface Dice                                           & Hybrid                & Yes  & CNN and transformer based interleaved encoder-decoder. Supervision of decoder's hidden layers.                                                                                                          \\ \hline
nnFormer\cite{zhou2021nnformer}        & Multi-organ    & CT, MRI                  & 3D            & Synapse~\cite{synapse}, ACDC~\cite{bernard2018deep}                                                                       & Dice                                                         & Hybrid                & Yes  & Interleaved convolution and self-attention based encoder-decoder architecture.                                                                                                                          \\ \hline
MISSFormer\cite{huang2021missformer}      & Multi-organ    & CT, MRI                  & 2D            & Synapse~\cite{synapse}, ACDC~\cite{bernard2018deep}                                                                       & Dice, Hausdorff distance                                     & Hybrid                & No   & Hierarchical encoder-decoder network with enhanced transformer block to mitigate the problem of feature inconsistency                                                                                  \\ \hline
TransBridge\cite{deng2021transbridge}     & Heart          & Echocardiography        & 2D            & EchoNet-Dynamic~\cite{ouyang2020video}                                                                     & Dice, Hausdorff distance                                     & Hybrid                & No   & Shuffling layer and group convolution for patch embedding to significantly reduce the number of parameters.                                                                                            \\ \hline
BiTr-UNet\cite{jia2021bitr}       & Brain          & MRI                      & 3D            & BraTS 21~\cite{baid2021rsna}                                                                         & Dice, Hausdorff distance                                     & Hybrid                & No   & Refined version of TransBTS with two sets of ViT layers instead of one.                                                                                                                                \\ \hline
GT UNet\cite{li2021gt}         & Tooth          & X-ray                    & 2D            & Tooth root dataset (private)                                                           & Dice, Accuracy, Sensitivity, Specificity, Jaccard similarity & Hybrid                & No   & Group transformer layers to reduce computational cost. Fourier descriptor based loss function to integrate shape prior.                                                                                \\ \hline
BAT\cite{wang2021boundary}             & --             & Dermoscopy               & 2D            & ISIC 2016+PH2~\cite{gutman2016skin}, ISIC 2018~\cite{codella2019skin}                                                           & Dice, IoU                                                    & Hybrid                & Yes  & Boundary-wise attention gate is added at the end of each transformer encoder layer to tackle challenging cases with ambiguous boundaries.                                                              \\ \hline
AFTer-UNet\cite{yan2021after}      & Multi-organ    & CT                       & 3D            & BCV~\cite{simpson2019large}, Thorax-85~\cite{chen2021deep}, SegTHOR~\cite{lambert2020segthor}                                                            & Dice                                                         & Hybrid                & No   & Axial fusion mechanism to fuse intra-slice and inter-slice contextual information to guide segmentation.                                                                                              \\ \hline
VT-UNet\cite{peiris2021volumetric}      & Multi-organ & CT, MRI                       & 3D            & BraTS 21~\cite{baid2021rsna}, MSD~\cite{simpson2019large}                                                            & Dice, Hausdorff distance                                                         & Hybrid                & Yes   & U-shaped encoder-decoder design. Encoder has two consecutive self-attention layers to encode local and global cues, and our decoder has novel parallel shifted window based self and cross attention blocks to capture fine details.  \\ \hline
Swin UNETR\cite{hatamizadeh2022swin}      & Brain & MRI                       & 3D            & BraTS 21~\cite{baid2021rsna}                                                            & Dice, Hausdorff distance                                                         & Hybrid                & Yes   & Swin UNet based architecture that consists of Swin transformer as the encoder and a CNN-based decoder. Computes self-attention in an efficient shifted window partitioning scheme.                                                         \\ \hlineB{3}
\end{tabular}%
\end{adjustbox}
}
\caption{An overview of ViT-based approaches for medical image segmentation. P.T: pretraining.}
\label{tab:seg_taxonomy}
\end{table*}

\begin{figure*}[t]
\centering
\includegraphics[width = 0.9\textwidth]{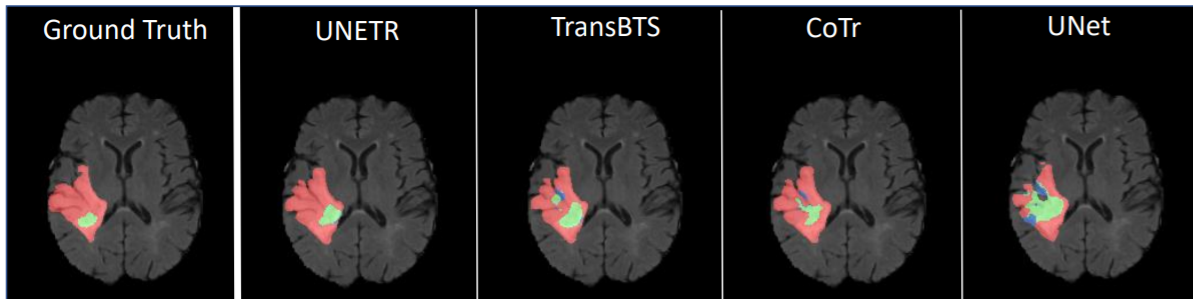}
\caption{Qualitative results of brain tumor segmentation task using transformer. From left to right: Ground truth image, UNETR~\cite{hatamizadeh2021unetr} (ViT-based), TransBTS~\cite{wang2021transbts} (ViT-based), CoTr~\cite{shen2021cotr} (ViT-based), and UNet~\cite{ronneberger2015u} (CNN based). Note that transformer-based approaches demonstrate better performance in capturing the fine-grained details of brain tumors as compared to CNN-based method. Image courtesy \cite{hatamizadeh2021unetr}.}
\label{fig:unetr}
\end{figure*}
\begin{SCfigure*}[][t]
\centering
\caption{Overview of CNN and a Transformer (CoTr) architecture \cite{xie2021cotr} proposed for 3D medical image segmentation. It consists of CNN encoder (left) to extract multi-scale features from the input, followed by  DeTrans-encoder (yellow blocks) to process the flattened multi-scale feature maps. Output features from encoder are fed to the CNN decoder (right) for segmentation mask prediction. Image courtesy \cite{xie2021cotr}.}
\includegraphics[height = 0.3\textwidth,width = 0.70\textwidth]{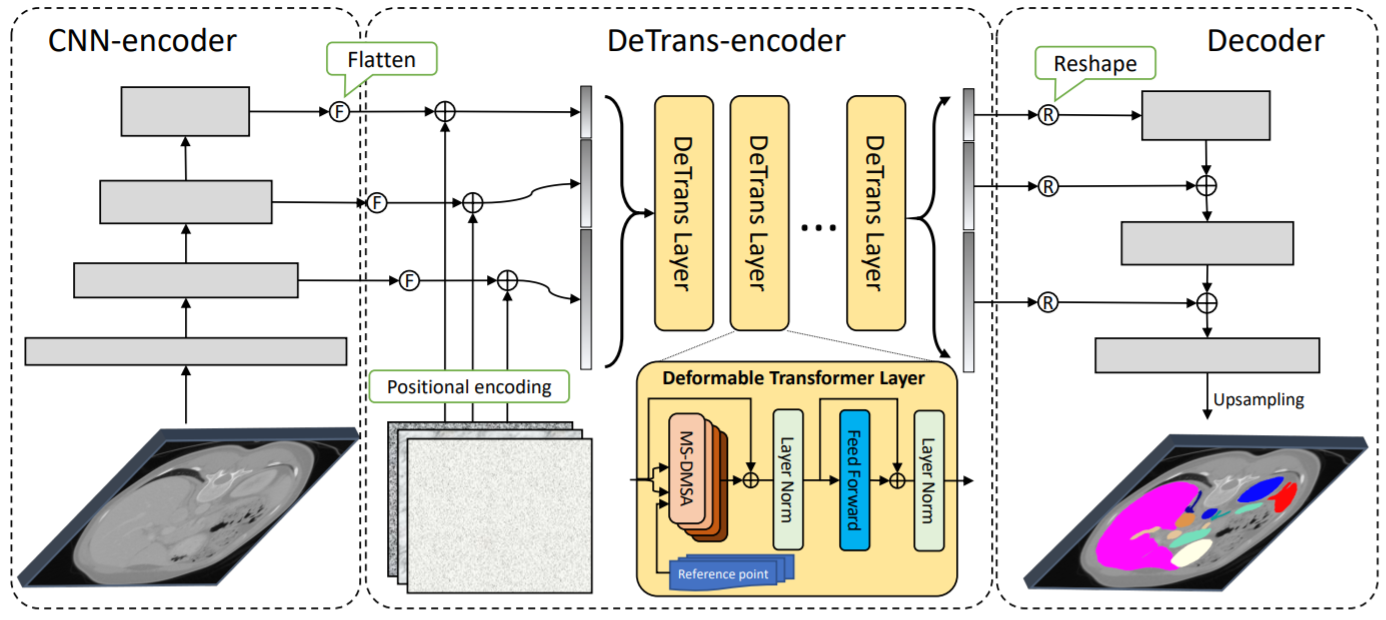}
\label{fig:cotr}
\end{SCfigure*}
\hspace{-1.5em}efficiency. Experiments on MSD Brain and Synapse datasets demonstrate the superiority of their architecture compared to bilinear upsampling.
In another work, 
Li \textit{et. al} \cite{li2021medical} propose SegTran, a Squeeze-and-Expansion Transformer for 2D and 3D medical image segmentation. Specifically, the squeeze block regularizes the attention matrix, and the expansion block learns diversified representations. Furthermore, a learnable sinusoidal positional encoding has been proposed that helps the model to encode spatial relationships. Extensive experiments on Polyp, BraTS19, and REFUGE20 (fundus images) segmentation challenges demonstrate the strong generalization ability of Segtran. 

\begin{figure}[t]
\centering
\includegraphics[width = 0.48\textwidth]{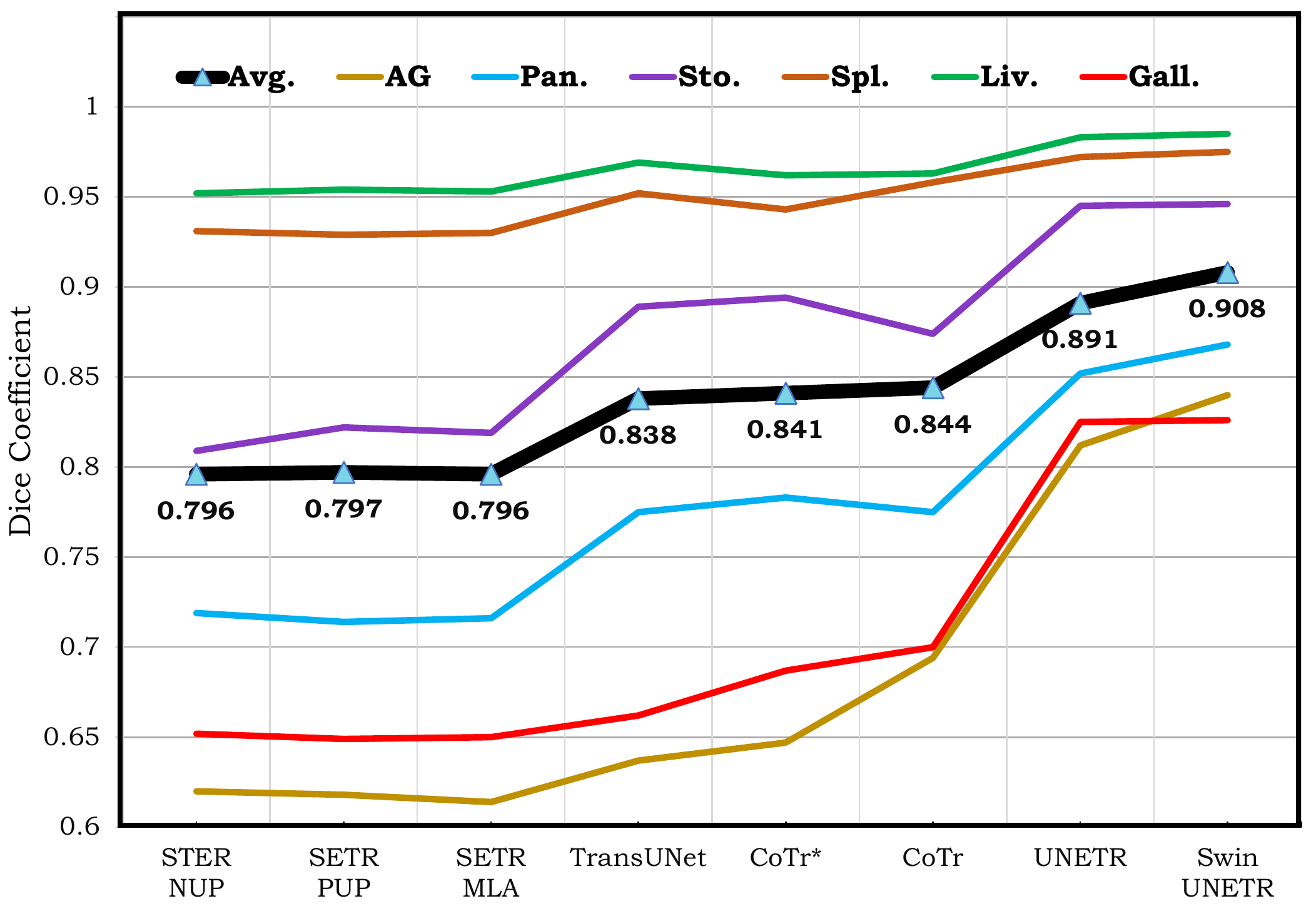} 
\caption{Dice results of BTCV challenge on multi-organ segmentation for various transformer based methods. It can be seen that Swin-UNETR is able to achieve on average 13$\%$ improvement in dice coefficient score compare to SETR method, indicating rapid pace of research in the field. Transformer based approaches used for the comparison include SETR NUP~\cite{zheng2021rethinking}, SETR PUP~\cite{zheng2021rethinking}, SETR MLA~\cite{zheng2021rethinking}, TransUNet~\cite{chen2021transunet}, CoTr*~\cite{xie2021cotr} (small CNN encoder compared to CoTr), CoTr~\cite{xie2021cotr}, UNETR~\cite{hatamizadeh2021unetr}, and Swin UNETR~\cite{tang2021self}.
Note: Avg: Average results (over 12 organs), AG: left and right adrenal glands, Pan: pancreas, Sto: stomach, Spl: spleen, Liv: liver, Gall: gallbladder.}
\label{fig:seg_challenge}
\end{figure}

\paragraph{\underline{\textbf{Multi-Scale Architectures}}}

These architectures process input at multiple scales to effectively segment organs having irregular shapes and different sizes. Here, we highlight various attempts to integrate the multi-scale architectures for medical image segmentation. We further group these approaches into 2D and 3D segmentation categories based on the input image type.

{{\textbf{2D Segmentation}.}}
Most ViT-based multi-organ segmentation approaches struggle to capture information at multiple scales as they partition the input image into fixed-size patches, thereby losing useful information. To address this issue, Zhang \textit{et. al.} \cite{zhang2021pyramid} propose a pyramid medical transformer, PMTrans, that leverage multi-resolution attention to capture correlation at different image scales using a pyramidal architecture~\cite{ghiasi2016laplacian}. PMTrans works on multi-resolution images via an adaptive partitioning scheme of patches to access different receptive fields without changing the overall complexity of self-attention computation. Extensive experiments on three medical imaging datasets of GLAS \cite{sirinukunwattana2017gland}, MoNuSeg~\cite{kumar2017dataset}, and HECKTOR \cite{andrearczyk2020overview} show the effectiveness of exploiting multi-scale information.
In other work, Ji \textit{et al.} \cite{ji2021multi} propose a Multi-Compound transformer (MCTrans) that learns not only feature consistency of the same semantic categories but also capture correlation among different semantic categories for accurate segmentation~\cite{yu2020context}. Specifically, MCTrans captures cross-scale contextual dependencies via the Transformer self-attention module and learned semantic correspondence among different categories via Transformer Cross-Attention module. An auxiliary loss has also been introduced to improve feature correlation of the same semantic category. Extensive experiments on six benchmark segmentation datasets demonstrate the effectiveness of the architectural components of MCTrans.

\textbf{3D Segmentation.}
The majority of multi-scale architectures have been proposed for 2D medical image segmentation. To directly handle volumetric data, Hatamizadeh \textit{et. al.} \cite{hatamizadeh2021unetr} propose a ViT-based architecture (UNETR) for 3D medical image segmentation. UNETR consists of a pure transformer as the encoder to learn sequence representations of the input volume. The encoder is connected to a CNN-based decoder via skip connections to compute the final segmentation output.
UNETR achieves impressive performance on BTCV \cite{landman2015miccai} and MSD~\cite{simpson2019large} segmentation datasets as shown in Fig. \ref{fig:unetr}.
One of the drawbacks of UNETR is its large computational complexity in processing large 3D input volumes. To mitigate this issue, Xie \textit{et. al} \cite{xie2021cotr} propose a computationally efficient 
deformable self-attention module \cite{dai2017deformable}  that casts attention only to a small set using multi-scale features, as shown in Figure \ref{fig:cotr}, to reduce the computational and spatial complexities. Experiments on BTCV~\cite{landman2015miccai} demonstrate the effectiveness of their deformable self-attention module for 3D multi-organ segmentation. 

\begin{tcolorbox}[top=0.5pt,left=0pt,size=minimal,boxrule = 0pt,breakable, enhanced]
\subsection{\textbf{\underline{Discussion}}}
\hspace{0.75em}\textit{From the extensive literature reviewed in this section, we note that the medical image segmentation area is heavily impacted by transformer-based models, with more than 50 publications within one year since the inception of the first ViT model~\cite{dosovitskiy2020image}. We believe such interest is due to the availability of large medical segmentation datasets and challenge competitions associated with them in top conferences compared to other medical imaging applications. As shown in Fig.~\ref{fig:seg_challenge}, a recent transformer-based hybrid architecture is able to achieve 13$\%$ performance gain in terms of dice score compared to the baseline transformer model, indicating rapid progression of the field. In short, ViT-based architectures have achieved impressive results over benchmark medical datasets, competing and most of the time improving over CNN-based segmentation approaches (see Table \ref{tab:seg_taxonomy} for details). Below, we briefly describe some of the challenges associated with ViTs based medical segmentation methods and give possible solutions based on insights from the relevant papers discussed.}
\\
\vspace{0.01em}
\textit{As mentioned before, the high computational cost associated with extracting features at multiple levels hinders the applicability of multi-scale architectures in medical segmentation tasks. These multi-scale architectures exploit processing input image information at multiple levels and achieve superior performance than single-scale architectures. Therefore, designing efficient transformer architectures for multi-scale processing requires more attention.
}
\\
\vspace{0.01em}
\textit{ Most of the proposed ViT-based models are pre-trained on the ImageNet dataset for the downstream task of medical image segmentation. This approach is sub-optimal due to the large domain gap between natural and medical image modalities. Recently, few attempts have been made to investigate the impact of self-supervised pre-training on medical imaging datasets on the ViTs segmentation performance. However, these works have shown that ViT pre-trained on one modality (CT) gives unsatisfactory performance when applied directly to other medical imaging modalities (MRI) due to the large domain gap making it an exciting avenue to explore. We defer detailed discussion related to pre-training ViTs for downstream medical imaging tasks to Sec.~\ref{sec:pretraining}.
} \\
\vspace{0.01em}
\textit{ Moreover, recent ViT-based approaches mainly focus on 2D medical image segmentation. Designing customized architectural components by incorporating temporal information for efficient high-resolution and high-dimensional segmentation of volumetric images has not been extensively explored. Recently, few efforts have been made, e.g.,  UNETR~\cite{hatamizadeh2021unetr} uses Swin Transformer~\cite{liu2021swin} based architectures to avoid quadratic computing complexity; however, it requires further attention from the community. }
\\
\vspace{0.01em}
\textit{In addition to focusing on the scale of datasets,  
with the advent of ViTs, we note there is a need to collect more diverse and challenging medical imaging datasets. Although diverse and challenging  datasets are also crucial to gauge the performance of ViTs in other medical imaging applications, they are particularly relevant for medical image segmentation due to a major influx of ViT-based models in this area. We believe these datasets will play a decisive role in exploring the limits of ViTs for medical image segmentation.}
\end{tcolorbox}

\section{Medical Image Classification} \label{sec:cla}

\begin{figure}[t]
\centering
\includegraphics[trim={0.5cm 8cm 10.5cm 1.2cm},clip,width = 0.48\textwidth]{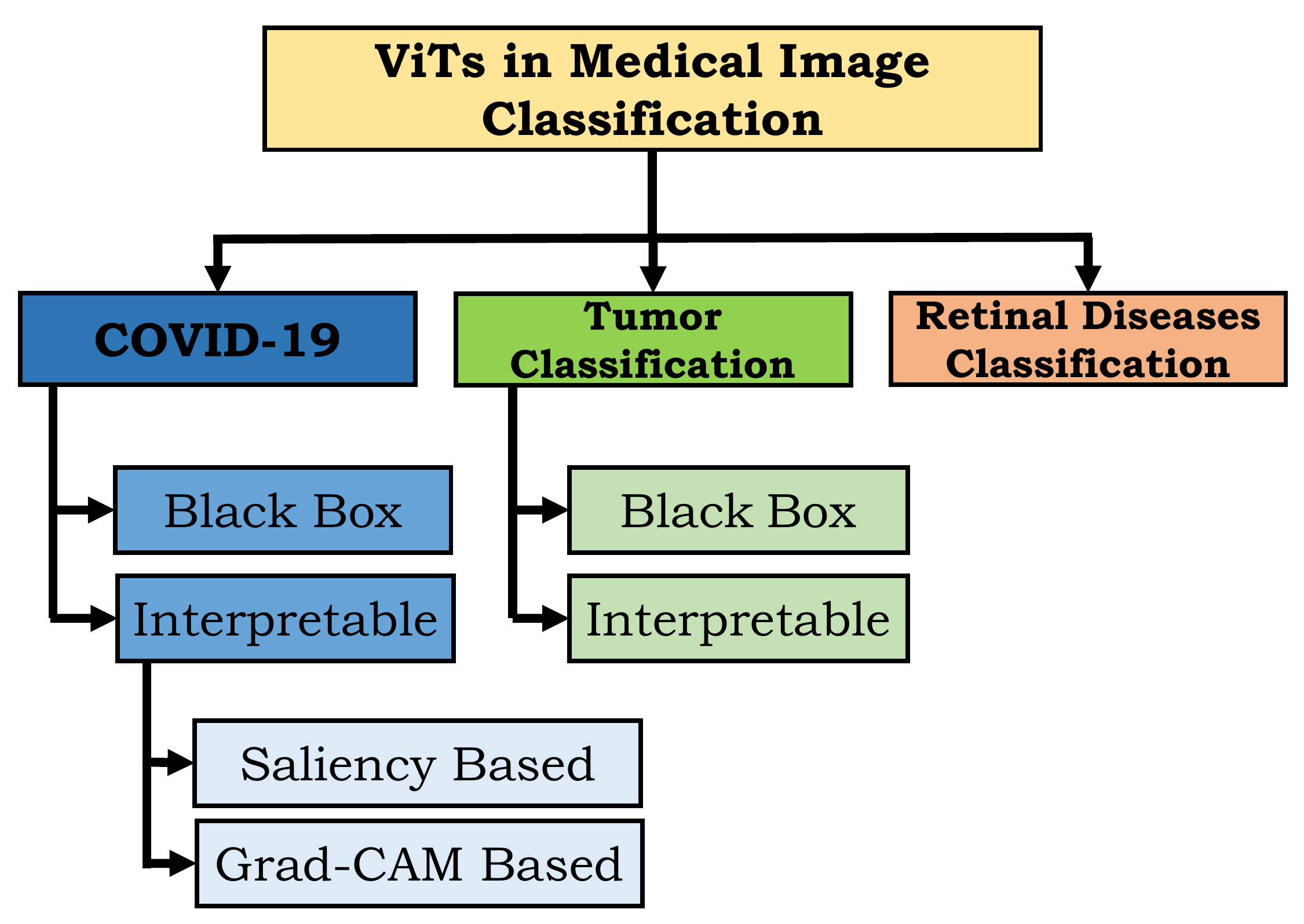}
\caption{Taxonomy of ViT-based medical image classification approaches. The influx of ViT-based COVID-19 classification approaches makes it a dominating category in the taxonomy.
}
\label{fig:classification_taxonomy}
\end{figure}

Accurate classification of medical images plays an essential role in aiding clinical care and treatment. In this section, we comprehensively cover applications of ViTs in medical image classification. 
We have broadly categorized these approaches into COVID-19, tumor, and retinal disease classification based methods due to a different set of challenges associated with these categories as shown in Fig. \ref{fig:classification_taxonomy}. 

\subsection{\textbf{COVID-19 Diagnosis}}

Studies suggest that COVID-19 can potentially be
better diagnosed with radiological imaging as compared to tedious real-time polymerase chain reaction (RT-PCR) test \cite{ai2020correlation,fang2020sensitivity,chen2021can}. Recently, ViTs have been successfully employed for diagnosis and severity prediction of COVID-19, showing SOTA performance. In this section, we briefly describe the impact of ViTs in advancing recent efforts on automated image analysis for the COVID-19 diagnosis process. Most of these works use three modalities, including Computerized tomography (CT), Ultrasound scans (US), and X-ray. We have further categorized ViT-based COVID-19 classification approaches into \textit{Black-box models} and \textit{Interpretable models} according to the level of explainability offered.
\subsubsection{\textbf{Black-Box Models}}

\begin{figure*}[t]
\centering
\includegraphics[width = 0.8\textwidth]{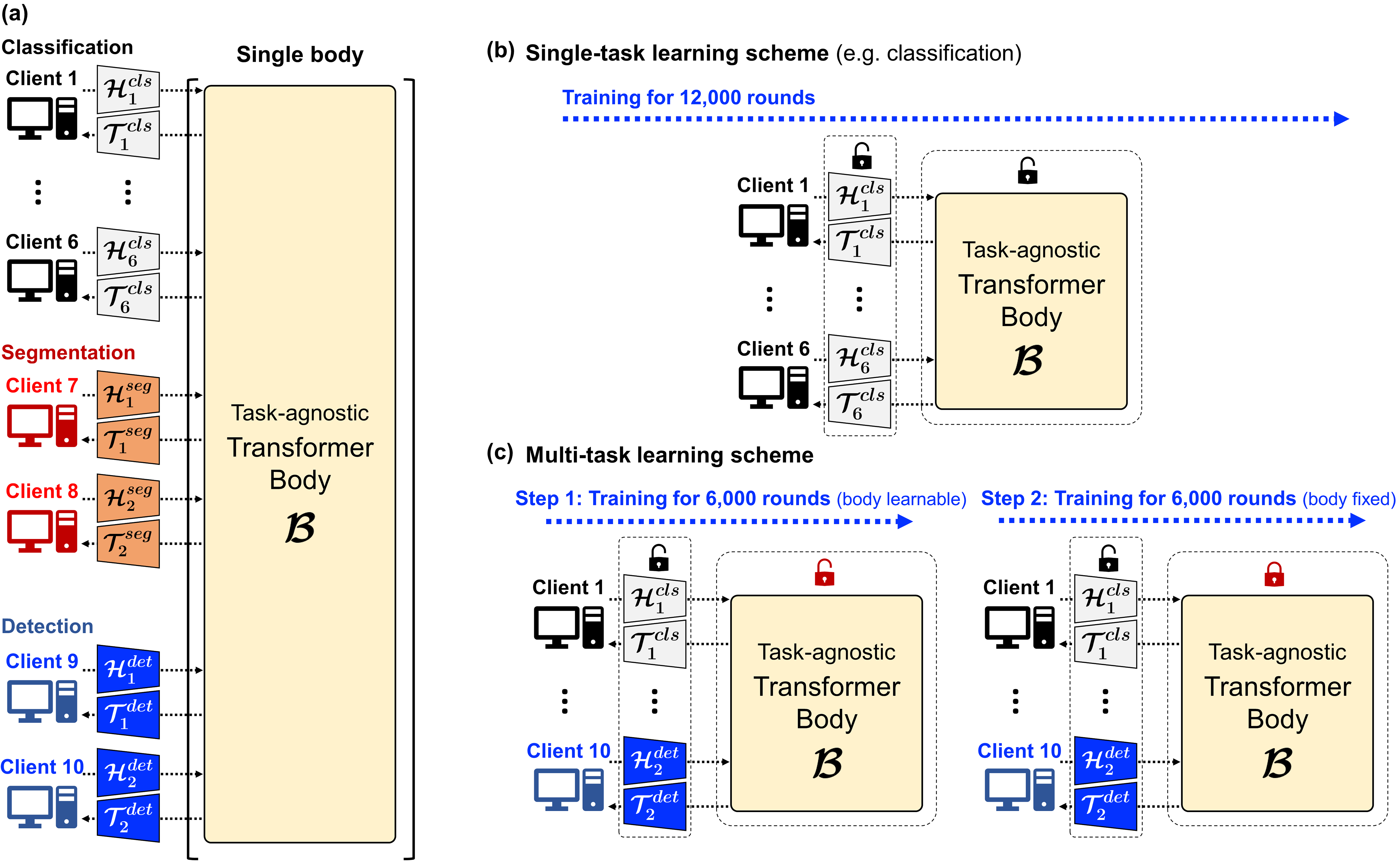}
\caption{Implementation details of Federated Split Task-Agnostic (FESTA) framework \cite{park2021federated} equipped with a Transformer to simultaneously process multiple chest X-ray tasks including diagnosis of COVID-19. (a) Experimental setting for multi-task learning of classification, segmentation, and detection of chest X-ray images. The clients only train the head ($\mathcal{H}$) and tail ($\mathcal{T}$) parts of the network, whereas the transformer body ($\mathcal{B}$) is shared across multiple clients. In the second step, the embedded features in the head are utilized by transformers for processing of individual tasks. (b) shows training scheme for single task. (c) shows training scheme for multi-task learning. Image courtesy \cite{park2021federated}.}
\label{fig:festa}
\end{figure*}

ViT-based Black-box models for COVID-19 imaging classification generally focus on improving accuracy by designing novel and efficient ViT architectures. However, these models are not easily interpretable, making it challenging to gain user-trust.   We have further sub-categorized black-box models into 2D and 3D categories, depending on the input image type. Below, we briefly describe these approaches:

 {\underline{\textbf{2D:}}} The High computational cost of ViTs hinders their deployment on portable devices, thereby limiting their applicability in real-time COVID-19 diagnosis. Perera \textit{et al.}\cite{perera2021pocformer} propose a lightweight \textbf{Point-of-Care Transformer (POCFormer)} to diagnose COVID-19 from lungs images captured via portable devices. Specifically, POCFormer leverages Linformer \cite{wang2020linformer} to reduce the space and time complexity of self-attention from quadratic to linear. POCFormer has two million parameters that are about half of MobileNetv2~\cite{sandler2018mobilenetv2}, thus making it suitable for real-time diagnosis. Experiments on COVID-19 lungs POCUS dataset \cite{born2020pocovid,cohen2020covid} demonstrate the effectiveness of their proposed architecture with above 90$\%$ classification accuracy. 
In other work, Liu \textit{et al.} \cite{liu2021automatic} proposed ViT-based model for COVID-19 diagnosis by exploiting a new attention mechanism named Vision Outlooker (VOLO) \cite{yuan2021volo}. VOLO is effective for \textbf{encoding fine-level features} into ViT token representation, thereby improving classification performance.
Further, they leverage the transfer learning approach to handle the issue of insufficient and generally unbalanced COVID-19 datasets. Experiments on two publicly available COVID-19 CXR datasets \cite{chowdhury2020can,cohen2020covid} demonstrate the effectiveness of their architecture. 
Similarly, Jiang \textit{et al.}~\cite{jiang2021covid} leverage \textbf{Swin Transformer~\cite{liu2021swin} and Transformer-in-Transformer \cite{han2021transformer}} to classify COVID-19 images from Pneumonia and normal images. To further boost the accuracy, they employ model ensembling using a weighted average. Research progress in ViT-based COVID-19 diagnosis approaches is heavily impeded due to the requirement of a large amount of labeled COVID-19 data, thereby demanding collaborations among hospitals. This collaboration is difficult due to limited consent by patients, privacy concerns, and ethical data usage \cite{dou2021federated}. To mitigate this issue, Park \textit{et al.} \cite{park2021federated} proposed a \textbf{Federated Split Task-Agnostic (FESTA) framework that leveraged the merits of Federated and Split Learning}~\cite{yang2019federated,vepakomma2018split} in utilizing ViT to simultaneously process multiple chest X-ray tasks, including the diagnosis in COVID-19 Chest X-ray images on a massive decentralized dataset. Specifically, they split ViT into the shared transformer body and task-specific heads and demonstrate the suitability of ViT body to be shared across relevant tasks by leveraging multitask-learning (MTL) \cite{caruana1997multitask} strategy as shown in Fig. \ref{fig:festa}. They affirm the suitability of ViTs for collaborative learning in medical imaging applications via extensive experiments on the CXR dataset. 

 {\underline{\textbf{3D:}}}
Most of the ViT-based approaches for COVID-19 classification operate on 2D information only. However, as suggested by Kwee \textit{et al.} \cite{kwee2020chest}, the symptoms of COVID-19 might be present at different depths (slices) for different patients. To exploit both 2D and 3D information, Hsu \textit{et al.} \cite{hsu2021visual} propose a hybrid network consisting of transformers and CNNs. Specifically, they determine the importance of slices based on significant symptoms in the CT scan via Wilcoxon signed-rank test \cite{woolson2007wilcoxon} with Swin Transformer~\cite{liu2021swin} as backbone network. To further exploit the intrinsic features in the spatial and temporal dimensions, they propose a Convolutional CT Scan Aware Transformer module to fully capture the context of the 3D scans. Extensive experiments on the COVID-19-CT dataset show the effectiveness of their proposed architectural components. 
Similarly, Zhang \textit{et al.} \cite{zhang2021transformer,zhangmia} also proposed Swin Transformer based two-stage framework for the diagnosis of COVID-19 in the 3D CT scan dataset \cite{kollias2021mia}. Specifically, their framework consists of UNet based lung segmentation model followed by the image classification with Swin Transformer~\cite{liu2021swin} backbone.

\subsubsection{\textbf{Interpretable Models}}

Interpretable models aim to show the features that influence the decision of a model the most, generally via visualization techniques like saliency-based methods, Grad-CAM, etc. Due to their interpretable nature, these models are well suited to gain the trust of physicians and patients and therefore have paved their way for clinical deployment. We have further divided interpretable models into saliency-based~\cite{cong2018review} and Grad-CAM~\cite{selvaraju2017grad} based visualization approaches.
\begin{figure}[ht]
\centering
\includegraphics[width = 0.49\textwidth]{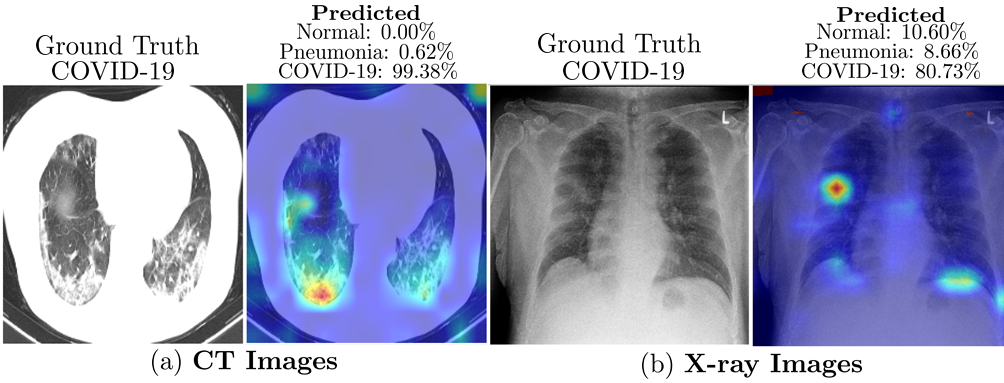}
\caption{ CT scans (a) and X-ray (b) images along with
their ground truth labels (left) and saliency maps (right). For figure (a),  xViTCOS-CT localized suspicious lesion regions exhibiting ground glass opacities, consolidation, reticulations in bilateral postero basal lung. xViTCOS-CT~\cite{mondal2021xvitcos} is able predict these regions correctly. For figure (b), radiologist’s interpretation is: \textit{thick walled cavity in right middle zone with surrounding consolidation}. As shown in last column, xViTCOS-CXR~\cite{mondal2021xvitcos} is able predict it correctly. Figure courtesy \cite{mondal2021xvitcos}.}
\label{fig:saliency}
\end{figure}

 {{\textbf{Saliency Based Visualization.}}} Park \textit{et al.} \cite{park2021vision}, propose a ViT-based method for COVID-19 diagnosis by exploiting the low-level CXR features extracted from the pre-trained backbone network. The backbone network has been trained in a self-supervised manner (using contrastive-learning based SimCLR \cite{chen2020simple} method) to extract abnormal CXR features embeddings from large and well-curated CXR dataset of CheXpert \cite{irvin2019chexpert}. These feature embeddings have been leveraged by ViT model for high-level diagnosis of COVID-19 images. Extensive experiments on three CXR test datasets acquired from  different hospitals demonstrate the superiority of their approach compared to CNN-based models. They also validated the generalization ability of their proposed approach and adopted saliency map visualizations \cite{chefer2021transformer} to provide interpretable results.
Similarly, Gao \textit{et al.} \cite{gao2021covid} propose COVID-ViT  to classify COVID from non-COVID images as part of the MIA-COVID19 challenge \cite{kollias2021mia}. Their experiments on 3D CT lungs images demonstrated the superiority of ViT-based approach over DenseNet \cite{huang2017densely} baseline in terms of F1 score.
In another work, Mondal \textit{et al.} \cite{mondal2021xvitcos} introduce xViTCOS for COVID-19 screening from lungs CT and X-ray images. Specifically, they pre-train xViTCOS on ImageNet to learn generic image representations and fine-tune the pre-trained model on a large chest radiographic dataset. Further, xViTCOS leverage the explainability-driven saliency-based approach \cite{chefer2021transformer} with clinically interpretable visualizations to highlight the role of critical factors in the resulting predictions, as shown in Figure \ref{fig:saliency}. Experiments on COVID CT-2A \cite{gunraj2021covid} and their privately collected Chest X-ray dataset demonstrate the effectiveness of xViTCOS.

 {{\textbf{Grad-CAM Based Visualization.}}} Shome \textit{et al.} \cite{shome2021covid} propose a ViT-based model to diagnose COVID-19 infection at scale. They combine several open-source COVID-19 CXR datasets to form a large-scale multi-class and binary classification dataset. For better visual representation and model interpretability, they further create Grad-CAM based visualization~\cite{selvaraju2017grad}.
\begin{figure*}[t]
\centering
\includegraphics[width = 0.9\textwidth]{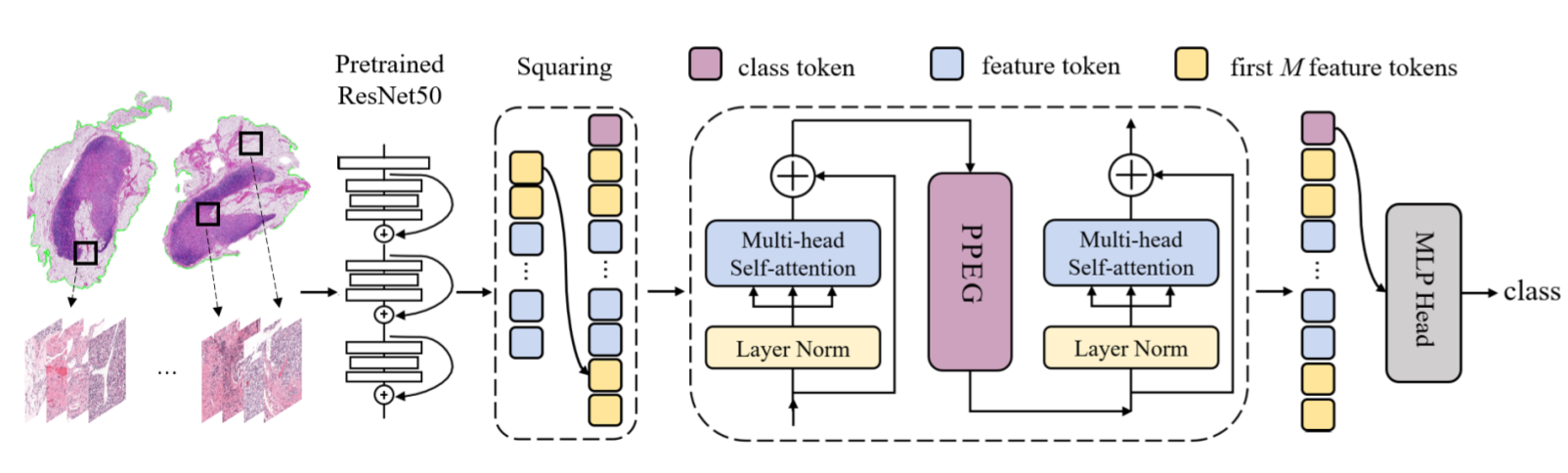}
\caption{Overview of Transformer based Multiple Instance Learning (TransMIL) architecture \cite{shao2021transmil} for whole slide brain tumor classification. Patches of WSI are embedded in the feature space of ResNet-50. The sequence of embedded features are then processed by their proposed pipeline that include: squaring of sequence, Correlation modelling of the sequence, conditional position encoding (via Pyramid Position Encoding Generator (PPEG) module)
and local information fusion, feature aggregation, and mapping from transformer space to label space. Image taken from \cite{shao2021transmil}.}
\label{fig:transmil}
\end{figure*}
\subsection{\textbf{Tumor Classification}}

A tumor is an abnormal growth of body tissues and can be cancerous (malignant) or noncancerous (benign). Early-stage malignant tumor diagnosis is crucial for subsequent treatment planning and can greatly improve the patient's survival rate. In this section, we review
ViT-based models for tumor classification. These models can be mainly categorized into \textit{Black-box models} and \textit{Interpretable models}. We highlight the relevant anatomies in bold.

 {{\textbf{Black-Box Models.}}}
TransMed~\cite{dai2021transmed} is the first work that leverages ViTs for medical image classification. It is a hybrid CNN and transformer-based architecture that is capable of classifying \textbf{parotid} tumors in the multi-modal MRI medical images. 
TransMed also employs a novel image fusion strategy to effectively capture mutual information from images of different modalities, thereby achieving competitive results on their privately collected parotid tumor classification dataset. 
Later, Lu \textit{et al.}~\cite{lu2021smile} propose a two-stage framework that first performs contrastive pre-training on glioma sub-type classification in the \textbf{brain} followed by the feature aggregation via proposed transformer-based sparse attention module. Ablation studies on TCGA-NSCLC~\cite{napel2014nsclc} dataset show the effectiveness of their two-stage framework.
For the task of \textbf{breat} cancer classification, Gheflati \textit{et al.}~\cite{gheflati2021vision} systematically evaluate the performance of pure and hybrid pre-trained ViT models. Experiments on two breast ultrasound datasets provided by Al-Dhabyani \textit{et al.} \cite{al2020dataset} and Yap \textit{et al.} \cite{yap2017automated} shows that Vit-based models provide better results than those of the CNNs for classifying images into benign, malignant, and normal categories.
Similarly, other works employ hybrid Transformer-CNN architectures to solve medical classification problem for different organs. For instance,  Khan \textit{et al.}~\cite{khan2021gene} propose Gene-Transformer to predict the \textbf{lung} cancer subtypes. Experiments on TCGA-NSCLC~\cite{napel2014nsclc} dataset demonstrates the superiority of Gene Transformer over CNN baselines. Chen \textit{et al.}~\cite{chen2021gashis} present a multi-scale GasHis-Transformer to diagnose gastric cancer in the \textbf{stomach}. 
Jiang \textit{et al.}~\cite{jiang2021method} propose a hybrid model 
to diagnose acute lymphocytic leukemia by using symmetric cross-entropy loss function.

\begin{figure}[t]
\centering
\includegraphics[trim={0cm 0cm 2cm 0.1cm},clip,width = 0.49\textwidth]{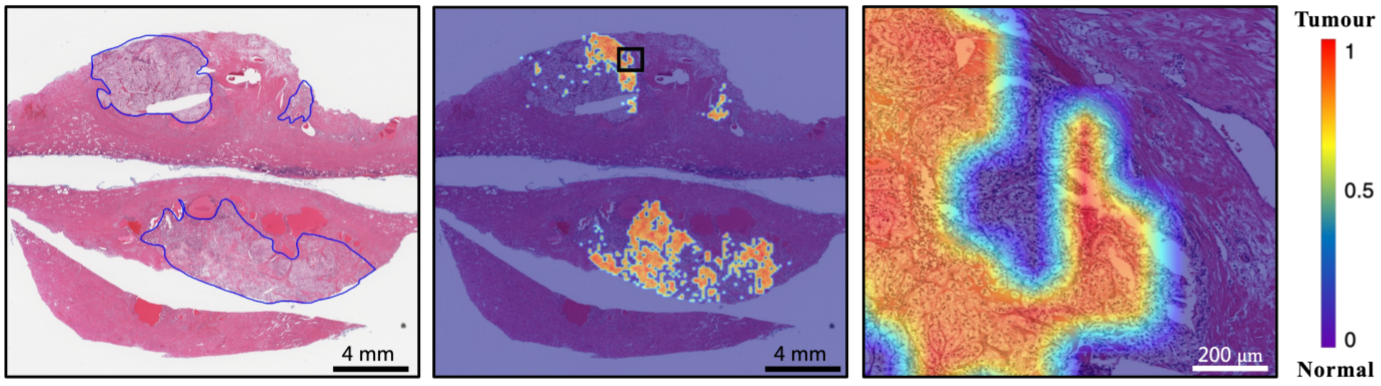}
\caption{ Left: The area within the blue region is the cancer region. Middle: Attention scores from TransMIL are visualised as a heatmap (red for tumor and blue for normal) to interpret the important morphology used for diagnosis. Right: Zoomed-in view of the black square in the middle figure. Figure courtesy \cite{shao2021transmil}.}
\label{fig:saliency_trans}
\end{figure}

 {{\textbf{Interpretable Models.}}} Since the annotation procedure is expensive and laborious, one label is assigned to a set of instances (bag) in whole slide imaging (WSI) based pathology diagnosis. This type of weakly supervised learning is known as Multiple Instance Learning~\cite{fung2007multiple}, where a bag is labeled positive if at least one instance is positive or labeled negative when all instances in a bag are negative. Most of the current MIL methods assume that the instances in each bag are independent and identically distributed, thereby neglecting the correlation among different instances. Shao \textit{et al.}~\cite{shao2021transmil} present TransMIL to explore both morphological and spatial information in weakly supervised WSI classification. Specifically, TransMIL aggregates morphological information with two transformer-based modules and a position encoding layer as shown in Fig. \ref{fig:transmil}. To encode spatial information, a pyramid position encoding generator is proposed. Further, the attention scores from the TransMIL have been visualized to demonstrate interpretability, as shown in Fig. \ref{fig:saliency_trans}. TransMIL shows state-of-the-art performance on three different computational pathology datasets CAMELYON16 \textbf{(breast)}~\cite{bejnordi2017diagnostic}, TCGA-NSCLC \textbf{(lung)}~\cite{napel2014nsclc}, and TCGA-R \textbf{(kidney)}~\cite{tcga}. 
To diagnose \textbf{lung} tumors, Zheng \textit{et al.}~\cite{zheng2021deep} propose graph transformer network (GTN) to leverage the graph-based representation of WSI. GTN consists of a graph convolutional layer~\cite{kipf2016semi}, a transformer layer, and a pooling layer. GTN further employs GraphCAM~\cite{chefer2021transformer} to identify regions that are highly associated with the class label. Extensive evaluations on TCGA dataset~\cite{napel2014nsclc} show the effectiveness of GTN.

\begin{table*}[]
\centering
\resizebox{0.95\textwidth}{!}{%
\begin{tabular}{V{3}l|l|p{2cm}|l|p{2.5cm}|p{2cm}|l|p{7cm}V{3}}
\hlineB{3}
\rowcolor{mygray} \textbf{Method}                                           & \textbf{Organ} & \textbf{Modality}                                                & \textbf{Type}             & \textbf{Datasets}                                                                        & \textbf{Metrics}                                                                                            & \textbf{Arch.} & \textbf{Highlights}                                                                                                                                                                                                 \\ \hlineB{2}
TransMed\cite{dai2021transmed}                                                   & Ear            & MRI (T1,T2)                                                      & 3D                       &  MRI (private)                                                                          &  Accuracy \newline Precision                                 & Pure           & First ViT-based multi-modal medical image classification approach with novel multi-modal fusion strategy                                                                                            \\ \hline
 {\color{teal}TransMIL}~\cite{shao2021transmil}                                                  & Multi-organ    & Pathology                                                        & 2D                        & Camelyon16~\cite{bejnordi2017diagnostic} \newline TCGA-NSCLC~\cite{bakr2018radiogenomic} \newline TCGA-RCC~\cite{tcga}         & Accuracy \newline AuC                  & Hybrid         &  Transformer based architecture to explore morphological and spatial information for Whole Slide Image  classification.                                        \\ \hline

Matsokus \textit{et al.}\cite{matsoukas2021time} & Multi-organ    & Mammograms \newline Dermoscopy & 2D                        & APTOS-2019~\cite{aptos19} \newline ISIC-2019~\cite{isic19} \newline CBIS-DDSM~\cite{lee2017curated}         & Recall \newline AuC                                               & Pure           & Systemetic study of whether one should replace CNNs with ViTs for medical image classification. \\ \hline

Gheflati \textit{et al.}\cite{gheflati2021vision} & Breast         & Ultrasound        & 2D                        &  BUSY~\cite{al2020dataset} \newline Yap \textit{et al.}~\cite{yap2017automated}                             & Accuracy \newline AuC                                                 & Pure           & First application of ViTs to ultrasound images classification.    
\\ \hline
{\color{teal}GTN}~\cite{zheng2021deep} & Lung         & Microscopy        & 2D                        &  TCGA dataset~\cite{napel2014nsclc}                             & Accuracy \newline Precision \newline Sensitivity \newline Specificity \newline Recall                                                 & Hybrid          & Consists of a graph convolutional layer, a transformer module, and a pooling layer for accurate classification of WSI images.    
\\ \hline
MIL-ViT~\cite{yu2021mil} & Eye         & Fundus        & 2D                        &  APTOS-2019~\cite{aptos19} \newline  RFMiD2020~\cite{quellec2020automatic}                             & Accuracy \newline AuC, F1 \newline Precision \newline Recall                                                 & Pure           &  First pretrained on a large fundus image dataset and later fine-tuned
on the downstream task of the retinal disease classification.   \\ \hline
LAT~\cite{sun2021lesion} & Eye         & Fundus        & 2D                        &  Messidor-1 ~~\cite{decenciere2014feedback} \newline  Messidor-2~\cite{decenciere2014feedback} \newline EyePACS~\cite{cuadros2009eyepacs}                             & AuC \newline Kappa                                                & Hybrid           &  Formulate lesion discovery as a weakly supervised lesion localization problem via
a transformer decoder. Jointly solve diabetic retinopathy grading and lesion discovery. 
\\ \hlineB{2}
 \rowcolor{mygray} \multicolumn{8}{V{3}cV{3}}{\textbf{COVID-19}}  \\ \hlineB{2} 

{\color{teal}Park \textit{et al.}}~\cite{park2021vision}                                               & Chest         & X-ray                                                            & 2D                        & CheXpert~\cite{irvin2019chexpert}                      &   AuC                                & Hybrid         & Leveraging a backbone network trained to find low-level abnormal CXR findings in pre-built large-scale dataset to embed feature  corpus suitable for high-level disease classification.     \\ \hline

POCFormer~\cite{perera2021pocformer}  & Chest          & Ultrasound        & 2D                        & POCUS~\cite{born2020pocovid}                                                                             & Recall, F1 \newline Specificity \newline Sensitivity \newline Accuracy & Pure          & Proposed light weight transformer architecture that demonstrates the efficiency and performance improvements.\\ \hline

FESTA~\cite{park2021federated}  & Chest          & X-ray        & 2D                        & CheXpert~\cite{irvin2019chexpert} \newline SIIM-ACR~\cite{siimacr} \newline  RSNA~\cite{rsna}                                                                              & Recall, F1 \newline Specificity \newline Sensitivity \newline AuC & Pure          &   Utilize ViT  to  simultaneously  process  multiple  chest  X-ray  tasks, including the diagnosis in COVID-19 Chest X-ray images on a massive decentralized dataset.\\ \hline

Liu \textit{et al.}~\cite{liu2021automatic}                                                & Chest          & X-ray                                                            & \multicolumn{1}{l|}{2D}                        & COVID-19-1~\cite{chowdhury2020can} \newline COVID-19-2~\cite{cohen2020covid}                      &  Accuracy                                                                                                 & Pure           & Explore VOLO tailored with transfer learning technique to effectively encodes fine-level features into the token representations.                                                            \\ \hline

COVID-VIT~\cite{gao2021covid}                                                & Chest          & CT                                                               & 3D & MIA-COV19~\cite{kollias2021mia}                                                                              & Accuracy, F1                                            & Pure           & Propose ViT based architecture to classifiy COVID-19 CT images in MIA-COV19 competition.                                                                                                     \\ \hline

{\color{teal}xViTCOS}~\cite{mondal2021xvitcos}                                                   & Chest          & X-ray, CT                                                        &  2D                        & xViTCOS-CT~\cite{gunraj2021covid} \newline xViTCOS-CXR~\cite{wang2020covid}                     & Precision \newline Recall, F1 \newline Specificity,  NPV        & Pure           & Propose ViT based multi-stage transfer learning technique to address the issue of COVID-19 data scarcity. The approach is clinically interpretable.                                         \\ \hline

Hsu \textit{et al.}~\cite{hsu2021visual}                                                & Chest          & CT                                                              & 3D                        & COV19 CT DB~\cite{kollias2021mia}                      & Accuracy \newline Recall,F1 \newline Precision                 & Hybrid        & Importance of slices are determined in CT scan via Wilcoxon signed-rank test. Then, spatial and temporal features are exploited via proposed convolutional CT scan Aware Transformer module. \\ \hline

Zhang \textit{et al.}\cite{zhang2021transformer}                                              & Chest          & CT           & 3D                        & MIA-COV19~\cite{kollias2021mia}                                              & F1 score                                                 & Hybrid         & Swin Transformer based two stage framework for diagnosis of COVID-19 in 3D CT scans.                                                                                                          \\ \hline

COViT-GAN~\cite{ambita2021covit}     & Chest          & CT          & 2D                          & COVID-CT~\cite{kollias2021mia} \newline Sars-CoV-2~\cite{angelov2020sars}                                & Accuracy \newline Sensitivity \newline Precision, F1                                                                                                             & Hybrid               &  Generate synthetic images using a self-attention generative adversarial network and use it as a data augmentation method to alleviate the problem of limited data and improve performance.\\ \hline

{\color{teal}COVID-Trans.}\cite{shome2021covid}        & Chest          & X-ray     & 2D                        & El-Shafai \textit{et al.}~\cite{el2020extensive} \newline Sait \textit{et al.}~\cite{sait2020curated} \newline Qi \textit{et al.}~\cite{qi2021chest} & Accuracy \newline Precision \newline AuC, Recall, F1             & Pure           & Propose a ViT model to diagnose COVID-19 at scale. Combines several open source COVID-19 chest X-ray datasets to form large dataset  for binary and multi-class classification.        \\ \hlineB{3}
\end{tabular}%
}
\caption{An overview of ViT-based approaches for medical image classification. {\color{teal} Teal color} indicates that the model is interpretable.}
\label{tab:class_taxonomy}
\end{table*}

\begin{table}[ht!]
\centering
{\renewcommand{\arraystretch}{1.5}
\resizebox{0.48\textwidth}{!}{%
\begin{tabular}{V{3}llccV{3}}
\hlineB{3}
\rowcolor{mygray} \textbf{Initialization} & \textbf{Model} & \textbf{APTOS2019}, $\kappa \uparrow$ & \textbf{ISIC2019}, Recall $\uparrow$ 
\\ \hline
 & ResNet50 & 0.849 $\pm$ 0.022 & 0.662 $\pm$ 0.018 
\\ 
\multirow{-2}{*}{Random} & DeiT-S   & 0.687 $\pm$ 0.017 & 0.579 $\pm$ 0.028 
\\ \hline
 & ResNet50 & 0.893 $\pm$ 0.004 & 0.810 $\pm$ 0.008 
\\ 
\multirow{-2}{*}{ImageNet (supervised)} & DeiT-S   & 0.896 $\pm$ 0.005 & 0.844 $\pm$ 0.021 
\\ \hline
& ResNet50 & 0.894 $\pm$ 0.008 & 0.833 $\pm$ 0.007 
\\
\multirow{-2}{*}{\begin{tabular}[c]{@{}l@{}}ImageNet (supervised) + \\ Self-supervised with DINO \cite{caron2021emerging}\end{tabular}} & DeiT-S   & 0.896 $\pm$ 0.010 & 0.853 $\pm$ 0.009 
\\ \hlineB{3}
\end{tabular}%
}}
\caption{Comparison of vanilla CNNs vs.~ViTs with different initialization strategies on \textit{medical imaging} classification tasks. For APTOS 2019~\cite{aptos19} and ISIC 2019~\cite{tschandl2018ham10000} datasets quadratic Cohen Kappa and recall score have been reported.
\textit{First row}: For randomly initialized networks, CNNs outperform ViTs. \textit{Second row}: ViTs appear to benefit significantly from pre-training on ImageNet dataset. \textit{Third row}: Both ViTs and CNNs perform better with self-supervised pretraining. Table taken from \cite{matsoukas2021time}.}
\label{tab:finetuned}
\end{table}

\subsection{\textbf{Retinal Disease Classification}}

Yu \textit{et al}~\cite{yu2021mil} propose MIL-ViT model which is first pre-trained on a large fundus image dataset and later fine-tuned on the downstream task of the retinal disease classification. MIL-ViT architecture uses MIL-based head that can be used with ViT in a plug-and-play manner. Evaluation performed on APTOS2019~\cite{aptos19} and RFMiD2020~\cite{quellec2020automatic} datasets shows that MIL-ViT is achieving more favorable performance than CNN-based baselines.
Most data-driven approaches treat diabetic retinopathy (DR) grading and lesion discovery as two separate tasks, which may be sub-optimal as the error may propagate from one stage to the other. To jointly handle both these tasks, Sun \textit{et al.}~\cite{sun2021lesion} propose lesion aware transformer (LAT) that consists of a pixel relation based encoder and a lesion-aware transformer decoder. In particular, they leverage transformer decoder to formulate lesion discovery as a weakly supervised lesion localization problem. LAT model sets state-of-the-art on Messidor-1~\cite{decenciere2014feedback}, Messidor-2~\cite{decenciere2014feedback}, and EyePACS~\cite{cuadros2009eyepacs} datasets.
Yang \textit{et al.}~\cite{yang2021fundus} propose a hybrid architecture consisting of convolutional and Transformer layers for fundus disease classification on OIA dataset~\cite{nankaioia}. Similarly, Wu \textit{et al.}\cite{wu2021vision} and Aldahou \textit{et al.} ~\cite{aldahoul2021encoding} also verify that ViT models are more accurate in DR grading than their CNNs counterparts. 
\begin{tcolorbox}[top=0.5pt,left=0pt,size=minimal,boxrule = 0pt,breakable, enhanced]
\subsection{\textbf{\underline{Discussion}}}
\hspace{0.75em}\textit{ In this section, we provide a comprehensive overview of about 25 papers related to applications of ViTs in medical image classification. In particular, we see a surge of Transformer-based architectures for diagnosing COVID-19, compelling us to develop taxonomy accordingly. Below, we briefly highlight some of the challenges associated with this area, identify recent trends, and provide future directions worthy of further exploration.}
 \\
\vspace{0.01em}
\textit{ 
The lack of large COVID-19 datasets hindered the applicability of ViT models to diagnose COVID-19. A recent work by Shome \textit{et al.}~\cite{shome2021covid} attempts to mitigate this issue by combining three open-source COVID-19 datasets to create a large dataset comprising 30,000 images. Still, creating diverse and large COVID-19 datasets is challenging and requires significant effort from the medical community.}
\\
\vspace{0.01em}
\textit{ 
More attention must be given to design \textbf{interpretabile} (to gain end-users trust) and \textbf{efficient} (for point-of-care testing) ViT models for COVID-19 diagnosis to make them a viable alternative of RT-PCR testing in the future. 
} 
\\
\vspace{0.01em}
\textit{ We notice that most works have used the original ViT model~\cite{dosovitskiy2020image} as a plug-and-play manner to boost the medical image classification performance. In this regard, we believe that integrating domain-specific context and accordingly designing architectural components and loss functions can enhance performance and provide more insights in designing effective ViT-based classification models in the future.}
\\
\vspace{0.01em}
\textit{ Finally, let us highlight the exciting work of Matsoukas \textit{et al.}~\cite{matsoukas2021time}
that, for the first time, demonstrates that ViTs pre-trained on ImageNet perform comparably to CNNs for the medical image classification task as shown in Table \ref{tab:finetuned}. This also raises an interesting question ``\textbf{Can ViT models pre-trained on medical imaging datasets perform better than ViT models pre-trained on ImageNet for medical image classification?}."
A recent work by Xie \textit{et al.}~\cite{xie2021unified} attempts to answer this by pre-training the ViT on large-scale 2D and 3D medical images. On the medical image classification problem, their model obtains substantial performance gain over the ViT model pre-trained on ImageNet, indicating that this area is worth exploring further. A brief overview of ViT-based medical image classification approaches has been provided in Table \ref{tab:class_taxonomy}. }
\end{tcolorbox}

\section{Medical Image Detection} \label{sec:det}

In medical image analysis, object detection refers to localization and identification of a region of interest (ROIs) such as lung nodules from X-ray images and is typically an essential aspect of diagnosis. However, it is one of the most time-consuming tasks for clinicians, thereby demanding the accurate computer-aided diagnosis (CAD) system to act as a second observer that may accelerate the process. Following the success of CNNs in medical image detection~\cite{liao2019evaluate,ganatra2021comprehensive}, recently few attempts have been made to improve performance further using Transformer models. These approaches are mainly based on the detection transformer (DETR) framework~\cite{zhu2020deformable}.

Shen \textit{et al.} \cite{shen2021cotr} propose the first hybrid framework COTR, consisting of convolutional and transformer layers for end-to-end polyp detection. Specifically, the encoder of COTR contains six hybrid convolution-in-transformer layers to encode features. Whereas, the decoder consists of six transformer layers for object querying followed by a feed-forward network for object detection. COTR performs better than DETR on two different datasets ETIS-LARIB and CVC-ColonDB. The DETR model~\cite{zhu2020deformable} is also adapted in other works~\cite{liu2021transformer,mathai2021lymph} for the end-to-end polyp detection~\cite{liu2021transformer}, and detecting lymph nodes in T2 MRI scans for the assessment of lymphoproliferative diseases~\cite{mathai2021lymph}. 
\begin{tcolorbox}[top=0.5pt,left=0pt,size=minimal,boxrule = 0pt,breakable, enhanced]
\subsection{\textbf{Discussion}}
 \hspace{0.75em}\textit{Overall, the frequency of new Transformer-based approaches for the problem of medical image detection is lesser than those for the segmentation and classification. This is in contrast to the early years of CNN-based designs that were rapidly developed for the medical image detection, as indicated in Fig. \ref{fig:vit_vs_cnn}. A recent work~\cite{maaz2021multi} shows that generic class-agnostic detection mechanism of multi-modal ViTs (like MDETR~\cite{kamath2021mdetr}) pre-trained on natural images-text pairs performs poorly on medical datasets. Therefore, investigating the performance of multi-modal ViTs by pre-training them on modality-specific medical imaging datasets is a promising future direction to explore. Furthermore, since the recent ViT-based methods yield competitive results on medical image detection problems, we expect to see more contributions in the near future.
}
\end{tcolorbox}


\section{Medical Image Reconstruction} \label{sec:rec}

\begin{figure}[t]
\centering
\includegraphics[trim={1cm 5cm 8.5cm 1.5cm},clip,width = 0.45\textwidth]{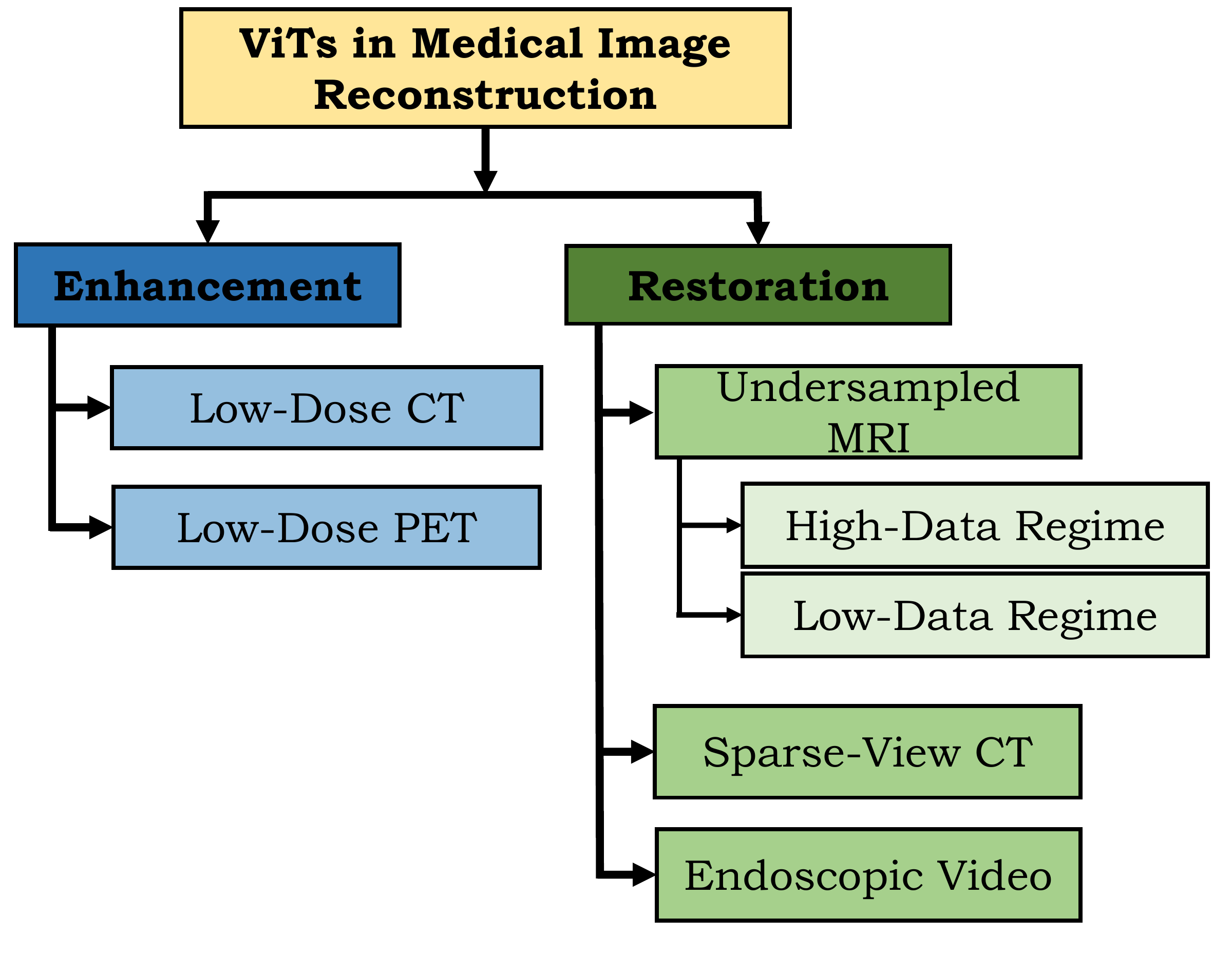}
\caption{Taxonomy of ViT-based medical image reconstruction approaches.}
\label{fig:rec_tax}
\end{figure}

The goal of medical image reconstruction is to obtain a clean image from a degraded input. For example, recovering a high-resolution MRI image from its under-sampled version.
It is a challenging task due to its ill-posed nature.
Moreover, exact analytic inverse transforms in many practical medical imaging scenarios are unknown. Recently, ViTs have been shown to address these challenges effectively. 
We categorize the relevant works into \textit{medical image enhancement} and \textit{medical image restoration} areas, as depicted in Fig. \ref{fig:rec_tax}.

\subsection{Medical Image Enhancement}

ViTs have achieved impressive success in the enhancement of medical images, mostly in the application of Low-Dose Computed Tomography (LDCT)~\cite{gopal2010screening,sadate2020systematic}. In LDCT, the X-ray dose is reduced to prevent patients from being exposed to high radiation. However, this reduction comes at the expense of CT image quality degradation and requires effective enhancement algorithms to improve the image quality and, subsequently, diagnostic accuracy. 

\subsubsection{\textbf{LDCT Enhancement}}
Zhang \textit{et al.}~\cite{zhang2021transct} propose 
an hybrid architecture TransCT that leverages the internal similarity of the LDCT images to effectively enhance them.
TransCT first decomposes the LDCT image into high-frequency (HF) (containing noise) and low-frequency (LF) parts. Next it removes the noise from the HF part with the assistance of latent textures. To reconstruct the final high-quality LDCT images, TransCT further integrates features from the LF part to the output of the transformer decoder. Experiments on Mayo LDCT dataset \cite{mccollough2017low} demonstrate the effectiveness of TransCT over CNN-based approaches.
\begin{figure*}[t]
\centering
\includegraphics[width = 0.95\textwidth]{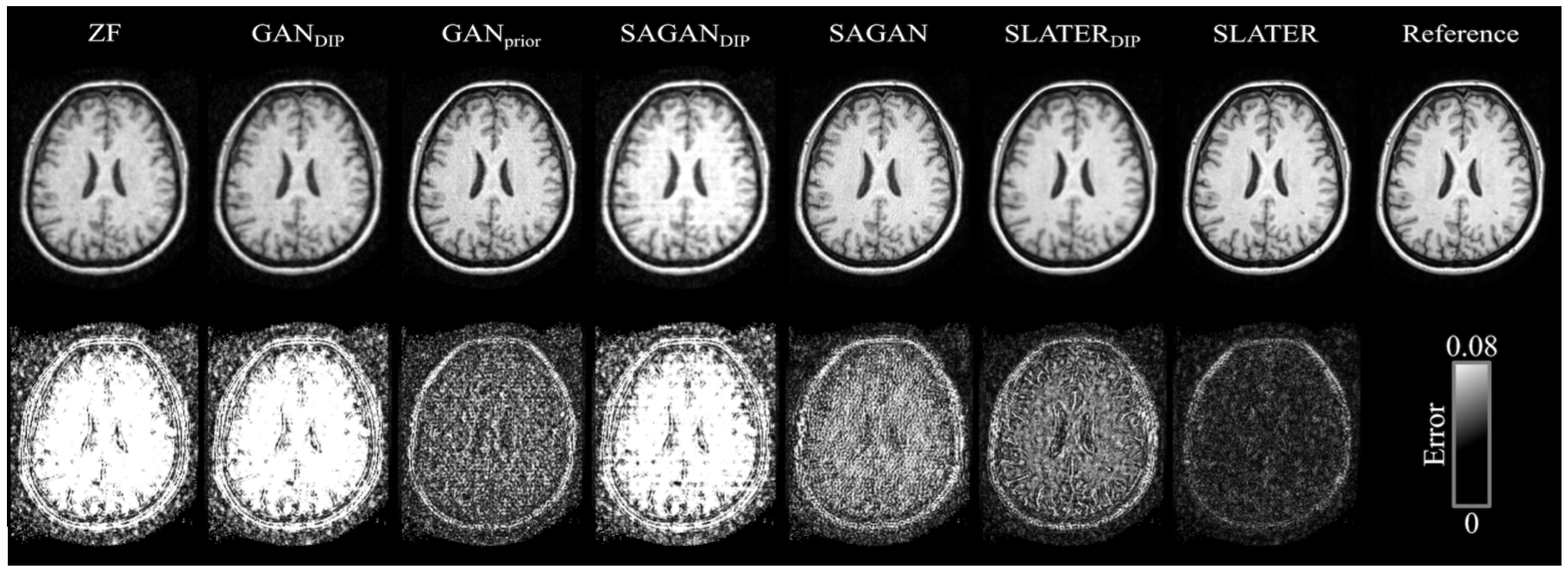}
\caption{Unsupervised under-sampled MRI reconstruction results of CNN based and transformer based approaches. From left to right (top row): Fourier method (ZF) \cite{fessler2010model}, GAN$_{\text{DIP}}$ (unsupervised CNN), GAN$_{\text{Prior}}$ \cite{narnhofer2019inverse} (unsupervised CNN with pre-training), Self-Attention GAN$_{\text{DIP}}$ \cite{ulyanov2018deep} (unsupervised CNN), Self-Attention GAN \cite{narnhofer2019inverse} (unsupervised CNN with pre-training), SLATER$_{\text{DIP}}$ \cite{korkmaz2021unsupervised} (unsupervised transformer), SLATER \cite{korkmaz2021unsupervised} (unsupervised transformer with pre-training), and reference image. Bottom row shows corresponding error maps. It can be seen that SLATTER outperforms all other approaches in term of quality of reconstruction. Results taken from \cite{korkmaz2021unsupervised}. }
\label{fig:slater}
\end{figure*}
To perform LDCT image enhancement, Wang \textit{et al.}~\cite{wang2021ted} propose a convolution-free ViT-based encoder-decoder architecture TED-Net. It employs Token-to-token block \cite{yuan2021tokens} to enrich the image tokenization via a cascaded process. To refine contextual information, TED-Net introduces dilation and cyclic-shift blocks~\cite{cao2021swin} in tokenization. TED-Net shows favorable performance on the Mayo Clinic LDCT dataset ~\cite{mccollough2017low}.
In another work, Luthra \textit{et al.}~\cite{luthra2021eformer} propose Eformer which is Transformer-based residual learning architecture for LDCT images denoising. To focus on edges, Eformer uses the power of Sobel-Feldman operator \cite{liang2020edcnn,irwin1968isotropic} in the proposed edge enhancement block to boost denoising performance. Moreover, to handle the over-smoothness issue, the multi-scale perceptual loss\cite{liang2020edcnn} is used. Eformer achieves impressive image quality gains in terms PSNR, SSIM, and RMSE on the AAPM-Mayo Clinic dataset~\cite{mccollough2017low}.

\subsubsection{\textbf{LDPET Enhancement}} Like LDCT, Low-dose positron emission tomography (LDPET) images reduce the harmful radiation exposure of standard-dose PET (SDPET) at the expense of sacrificing diagnosis accuracy. 
To address this challenge, Luo \textit{et al.}~\cite{luo20213d} propose an end-to-end generative adversarial network (GAN) based method integrated with Transformers, namely Transformer-GAN, to effectively reconstruct SDPET images from the corresponding LDPET images. Specifically, the generator of Transformer-GAN consists of a CNN-based encoder to learn compact feature representation, a transformer network to encode global context, and a CNN-based decoder to restore feature representation. They also introduce adversarial loss to obtain reliable and clinically acceptable images. Extensive experiments on their in-house collected clinical human brain PET dataset show the effectiveness of Transformer-GAN quantitatively and qualitatively.

\subsection{Medical Image Restoration}
Medical image restoration entails transforming signals collected by acquisition hardware (like MRI scanners) into interpretable images that can be used for diagnosis and treatment planning. Recently, ViT-based models have been proposed for multiple medical image restoration tasks, including undersampled MRI restoration, Sparse-View CT image reconstruction, and endoscopic video reconstruction. These models have pushed the boundaries of existing learning-based systems in terms of reconstruction accuracy. Next, we briefly highlight these approaches. 

\subsubsection{\textbf{Undersampled MRI Reconstruction}} Reducing the number of MRI measurements can result in faster scan times and a reduction in artifacts due to patients movement at the expense of aliasing artifacts in the image~\cite{hyun2018deep}. 

 {\textbf{High-Data Regime Approaches.}} Approaches in this category assume the availability of large MRI training datasets to train the ViT model.
Feng \textit{et al.}~\cite{feng2021accelerated} propose Transformer-based architecture, MTrans, for accelerated multi-modal MR imaging.
The main component of MTrans is the cross attention module that extracts and fuses complementary features from the auxiliary modality to the target modality. Experiments on fastMRI and uiMRI datasets for reconstruction and super-resolution tasks show that MTrans achieve good performance gains over previous methods.
However, MTrans requires separate training for MR reconstruction and super-resolution tasks. To jointly reconstruct and super-resolve MRI images, Feng \textit{et al.}~\cite{feng2021task} propose Task-Transformer that leverages the power of multi-task learning to fuse complementary information between the reconstruction branch and the super-resolution branch. Experiments are performed on the public IXI and private MRI brain datasets.
Similarly, Mahapatra \textit{et al.}~\cite{mahapatra2021mr} propose a hybrid architecture to super-resolve MRI images by exploiting the complementary advantages of both CNNs and ViTs. They also propose novel loss functions~\cite{park2020swapping} to preserve semantic and structural information
in the super-resolved images.

\begin{figure}[t]
\centering
\includegraphics[width = 0.48\textwidth]{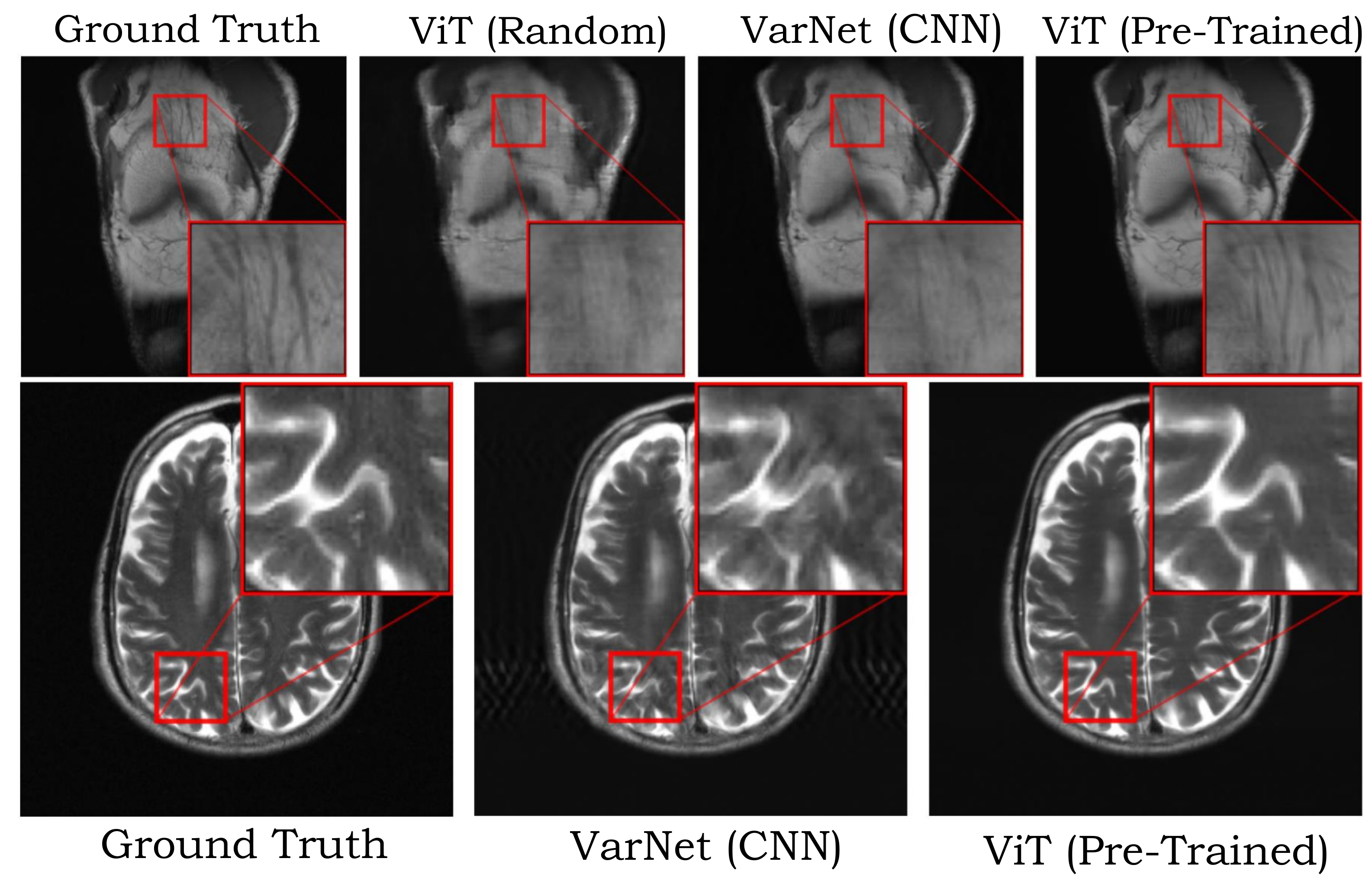}
\caption{\textbf{Top row}: Example reconstructions of models trained on 100 images (low-data regime) of fastMRI dataset. It can be seen that ViT model pre-trained on ImageNet produce sharp results as compared to recent CNN-based model~\cite{sriram2020end} and randomly initialized ViT model. \textbf{Bottom row}: Example reconstructions of Brain images by models pre-trained on ImageNet and fine-tuned on Knee MRI dataset. Results show that pre-trained ViT models are more robust to anatomical shifts. Figures adapted from \cite{lin2021vision}.}
\label{fig:recon_100}
\end{figure}

 {\textbf{Low-Data Regime Approaches.}}
One drawback of the aforementioned approaches is the requirement of a massive paired dataset of undersampled and corresponding fully sampled MRI acquisitions to train ViT models. To alleviate the data requirement issue, Korkmaz \textit{et al.} \cite{korkmaz2021unsupervised,korkmaz2021deep} propose a zero-shot framework, SLATER, that leverages prior induced by randomly initialized neural networks \cite{ulyanov2018deep,qayyum2021untrained} for unsupervised MR image reconstruction. Specifically, during inference, SLATER inverts its transformer-based generative model via iterative optimization over-network weights to minimize the error between the network output and the under-sampled multi-coil MRI acquisitions while satisfying the MRI forward model constraints. SLATER yields quality improvements on single and multi-coil MRI brain datasets over other unsupervised learning-based approaches as shown in Fig.~\ref{fig:slater}.
Similarly, Lin \textit{et al.}~\cite{lin2021vision} show that a ViT model pre-trained on ImageNet, when fine-tuned on only 100 fastMRI images, not only yields sharp reconstructions but is also more robust towards anatomy shifts compared to CNNs as shown in Fig. \ref{fig:recon_100}. Furthermore, their experiments indicate that ViT benefits from higher throughput and less memory consumption than the U-Net baseline.

\begin{table*}[t]
\centering
\resizebox{0.98\textwidth}{!}{%
\begin{tabular}{V{3}l|p{8cm}|c|c|p{2.2cm}|cV{3}}
\hlineB{3}
\rowcolor{mygray}{\textbf{Method}} & \hspace{12em}\textbf{Highlights}                                                                                                                              & \textbf{Modality} & \textbf{Input Type} & \textbf{Datasets}                                         & \textbf{Metric}                                                  \\ \hline
TransCT~\cite{zhang2021transct}                               & Transformer for LDCT enhancement with high and low frequency decomposition.                                 & CT                & 2D                  & NIH-AAPM~\cite{mccollough2017low}                                                & RMSE, SSIM, VIF~\cite{sheikh2006image}  \\ \hline
SLATER~\cite{korkmaz2021unsupervised}                                & Transformer based approach for zero shot MRI  image reconstruction.                                           & MRI               & 3D                  & IXI~\cite{ixi} \newline fastMRI~\cite{zbontar2018fastmri} &  PSNR, SSIM          \\ \hline
TED-Net~\cite{wang2021ted}                               & Pure transformer based encoder decoder dilation architecture for LDCT denoising.                            & CT                & 2D                  & NIH-AAPM~\cite{mccollough2017low}                                                 & RMSE, SSIM         \\ \hline
Eformer~\cite{luthra2021eformer}                               & Transformers based residual image denoising. Incorporate learnable Sobel filters for edge enhancement. & CT                & 2D                  & NIH-AAPM~\cite{mccollough2017low}                                                 & PSNR, SSIM, RMSE \\ \hline
Transformer-GAN\cite{luo20213d}                              & End-to-end 
GAN-based method integrated with Transformers to enhance LDPET images.                           & PET                & 3D                  & Private                                                & PSNR, SSIM, MSE         \\ \hline
MTrans~\cite{feng2021accelerated}                               & Leverage cross-attention module to fuse complementary features
from the auxiliary modality to the target modality for fast multi-modal MRI image reconstruction. & MRI                & 2D                  & fastMRI~\cite{zbontar2018fastmri} \newline uiMRI (private)                                                & PSNR, SSIM, NMSE \\ \hline
Task-Transformer\cite{feng2021task}    & Simultaneously, reconstruct and super-resolve MRI images via multi-task learning.                                           & MRI               & 2D                  & IXI~\cite{ixi} &  PSNR, SSIM, NMSE          \\ \hline
Mahapatra \textit{et al.}~\cite{mahapatra2021mr}                              & Hybrid architecture to super-resolve MRI
images by exploiting the complementary advantages of CNNs and ViTs.                         & MRI                & 2D                  & fastMRI~\cite{zbontar2018fastmri} \newline IXI~\cite{ixi}                                               & PSNR, SSIM, NMSE         \\ \hline
Lin \textit{et al.}~\cite{lin2021vision}                               & ViT pretrained on ImageNet, when fine-tuned on only 100 fastMRI images, yields sharp reconstructions and is robust towards anatomy shifts. & MRI                & 2D                  & fastMRI~\cite{zbontar2018fastmri}                                                & SSIM \\ \hline
DuDoTrans~\cite{wang2021dudotrans}           &  A hybrid CNN-Transformer architecture
that consider the global nature of sinogram’s sampling
process to restore high-quality CT images from sparse views.                             & CT                & 2D                  & NIH-AAPM~\cite{mccollough2017low}                                                & PSNR, SSIM         \\ \hline
MIST-Net~\cite{pan3991087multi}                               & Swin-transformer based projection and image domain two-stage framework to reconstruct high-quality CT images from sparse views.   & CT                & 2D                  & NIH-AAPM~\cite{mccollough2017low}                                                & PSNR, SSIM, RMSE \\ \hline
E-DSSR~\cite{long2021dssr}                               & Leverage lightweight stereo Transformer module to
estimate depth images with high confidence and a segmentor network to accurately predict the surgical tool’s mask.   & Endo.                & 2D                  & Hamlyn~\cite{ye2017self} \newline DaVinci (private)                                                & PSNR, SSIM \\ \hline
TranSMS~\cite{gungor2021transms}                               & ViT-based data consistancy module to super-resolve magnetic particle imaging (MPI) system matrices for accelerated calibration.   & MPI                & 2D, 3D                  & Open MPI~\cite{knopp2020openmpidata} \newline  Private datasets                                                & RMSE \\ \hlineB{3}
\end{tabular}%
}
\caption{An overview of ViT-based approaches for medical image reconstruction.}
\label{tab:recon_taxonomy}
\end{table*}
\subsubsection{\textbf{Sparse-View CT Reconstruction}}
Sparse-view CT~\cite{han2018framing} can effectively reduce the effective radiation dose by acquiring fewer projections. However, a decrease in the number of projections demands sophisticated image processing algorithms to achieve high-quality image reconstruction \cite{kudo2013image}. Wang \textit{et al.} \cite{wang2021dudotrans} present a hybrid CNN-Transformer, named Dual-Domain Transformer (DuDoTrans), by considering the global nature of sinogram's sampling process to better restore high-quality images. In the first step, DuDoTrans reconstructs low-quality reconstructions of sinogram via filtered back projection step and learnable DuDo consistency layer.  In the second step, a residual image reconstruction module performs enhancement to yield high-quality images. Experiments are performed on the NIH-AAPM dataset~\cite{mccollough2017low} to show generalizability, and robustness (against noise and artifacts) of DuDoTrans.

\subsubsection{\textbf{Endoscopic Video Reconstruction}}
Reconstructing surgical scenes from a stereoscopic video is challenging due to surgical tool occlusion and camera viewpoint changes. Long \textit{et al.}~\cite{long2021dssr} propose E-DSSR to reconstruct surgical scenes from stereo endoscopic videos. Specifically, E-DSSR contains a lightweight stereo Transformer module to estimate depth images with high confidence and a segmentor network to accurately predict the surgical tool's mask. Extensive experiments on Hamlyn Centre Endoscopic Video Dataset~\cite{ye2017self} and privately collected DaVinci robotic surgery dataset demonstrate the robustness of E-DSSR against abrupt camera movements and tissue deformations in real-time.

\begin{tcolorbox}[top=0.5pt,left=0pt,size=minimal,boxrule = 0pt,breakable, enhanced]
\subsection{\textbf{ \underline{Discussion}}}
\textit{ In this section, we have reviewed about a dozen papers related to the applications of ViT models in medical image reconstruction, as shown in Table \ref{tab:recon_taxonomy}. Below, we highlight a few challenging problems and provide recent trends in the field.}
 \\
\vspace{0.01em}
\textit{ Recently, an interesting work~\cite{lin2021vision} has investigated the impact of pre-training ViT on the task of MRI image reconstruction. Their results indicate that pre-trained ViT yields sharp reconstructions and is robust towards anatomy shifts (see Fig.~\ref{fig:recon_100}). The robustness of ViTs can be of particular relevance to the pathology image reconstruction as the range of pathology can vary significantly in the anatomy being imaged. Further, it raises an interesting question ``\textbf{Are ViTs pre-trained on medical image dataset able to provide any advantages in terms of reconstruction performance and robustness against anatomy shifts compared to their counterparts pre-trained on ImageNet"}? Extensive and systematic experiments are required to answer this question. Another promising future direction is to investigate the impact on the performance of ViT pre-trained on one image modality (like CT) and fine-tuned on another modality (like MRI) for image reconstruction tasks.  }
 \\
\vspace{0.01em}
\textit{ We notice that most of the Transformer-based approach focus on MRI and CT image reconstruction tasks, and their applicability to other modalities are yet to be explored.
In addition, proposed architectures are mostly generic and have not fully exploited the application-specific aspects. We believe that designing architectural components and formulating loss functions according to the task at hand can significantly boost performance. 
}
 \\
\vspace{0.01em}
\textit{ We want to highlight one particular work that uses the Transformer-layer architecture to regularize the challenging problem of MRI image reconstruction from under-sampled measurements~\cite{korkmaz2021unsupervised}. This work is inspired by the strong prior induced by the structure of untrained neural networks~\cite{ulyanov2018deep,qayyum2021untrained}. These untrained network priors have recently garnered much attention from the medical image community as they do not need labeled training data. Considering advances in the untrained neural network area, we believe this direction requires further attention from medical imaging researchers in the context of Transformers.}
 \\
\vspace{0.01em}
\textit{ We also observe that compared to the early years of CNNs (one paper from 2012 to 2015), Transformers have rapidly gained widespread attention in the medical image reconstruction community (more than a dozen papers in 2021), potentially due to the recent advancement in image-to-image translation frameworks.}
\end{tcolorbox}


\section{Medical Image Synthesis} \label{sec:syn}

\begin{figure}[t]
\centering
\includegraphics[width = 0.48\textwidth]{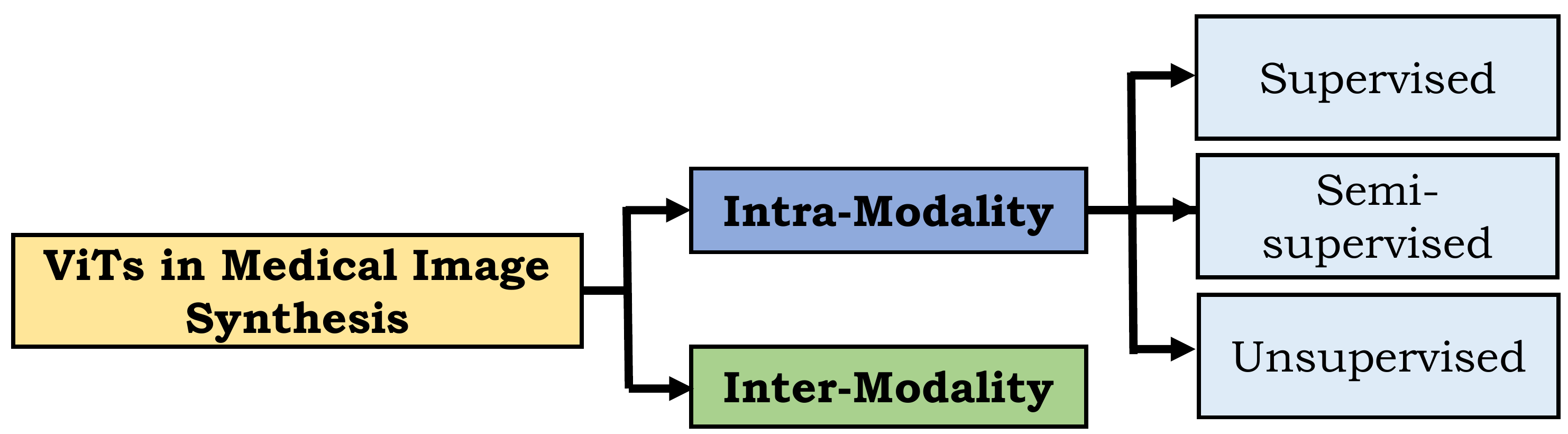}
\caption{Taxonomy of ViT-based medical image synthesis approaches.}
\label{fig:syn_tax}
\end{figure}

In this section, we provide an overview of the applications of ViTs in medical image synthesis. Most of these approaches incorporate adversarial loss to synthesize realistic and high-quality medical images, albeit at the expense of training instability \cite{liu2020loss}. We have further classified these approaches into \textit{intra-modality synthesis} and \textit{inter-modality synthesis} due to a different set of challenges in both categories, as shown in Fig. \ref{fig:syn_tax}.

\subsection{Intra-Modality Approaches}

{The goal of intra-modality synthesis is to generate higher-quality images from the relatively lower quality input images of the same modality.} Next, we describe the details of ViT-based intra-modality medical image synthesis approaches.

\subsubsection{\textbf{\textit{Supervised Methods}}} Supervised image synthesis methods require paired source and target images to train ViT-based models. Paired data is difficult to obtain due to annotation cost and time constraints, thereby generally hindering the applicability of these models in medical imaging applications. Zhang \textit{et al.} \cite{zhang2021ptnet} focus on synthesizing infant brain structural MRIs (T1w and T2w scans) using both transformer and performer (simplified self-attention) layers \cite{choromanski2020rethinking}. Specifically, they design a novel multi-resolution pyramid-like U-Net framework, PTNet, utilizing performer encoder, performer decoder, and transformer bottleneck to synthesize high-quality infant MRI. They demonstrate the superiority of PTNet both qualitatively and quantitatively compared to pix2pix \cite{isola2017image}, and pix2pixHD \cite{wang2018high} on large-scale infant MRI dataset \cite{makropoulos2018developing}. Furthermore, in addition to better synthesis quality, PTNet has a reasonable execution time of around 30 slices per second.

\subsubsection{\textbf{\textit{Semi-Supervised Methods}}} Semi-supervised approaches typically require small amounts of labeled data along with large unlabelled data to train models effectively. Kamran \textit{et al.} \cite{kamran2021vtgan} propose a  multi-scale conditional generative adversarial network (GAN) \cite{isola2017image} using ViT as a discriminator. They train their proposed model in a semi-supervised way to simultaneously synthesize Fluorescein Angiography (FA) images from fundus photographs and predict retinal degeneration. They use softmax activation after MLP head output and a categorical CE loss for classification. Besides adversarial loss, they also use MSE and perceptual losses to train their network. For ViT discriminator, they use embedding feature loss calculated using positional and patch features from the transformer encoder layers by successfully inserting the real and synthesized FA images. Their quantitative results in terms of Frechet inception Distance \cite{heusel2017gans} and Kernel Inception Distance \cite{binkowski2018demystifying} demonstrate the superiority of their approach over baseline methods on diabetic retinopathy dataset provided by Hajeb \textit{et al.} \cite{hajeb2012diabetic}.
\begin{figure*}[t]
\centering
\includegraphics[width = 0.8\textwidth]{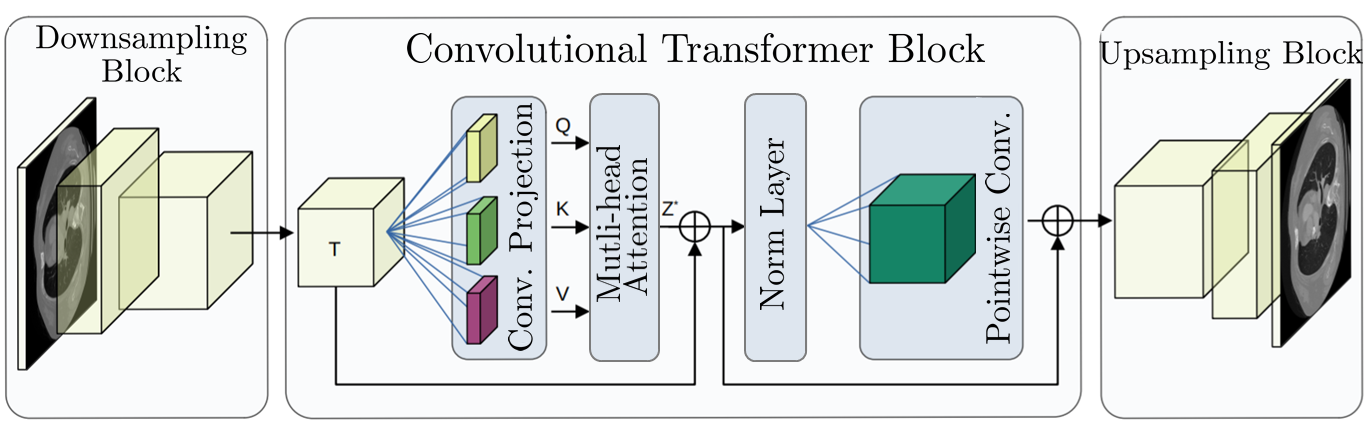}
\caption{Hybrid convolutional-transformer network for CT image generation as proposed in \cite{ristea2021cytran}. It consists of down-sampling convolutional layers to extract features from input images,  a convolutional-transformer block comprising a multi-head self-attention mechanism, and an upsampling block to generate output images.}
\label{fig:cytran}
\end{figure*}
\begin{figure}[t]
\centering
\includegraphics[trim={0cm 0cm 0cm 0cm},clip,width = 0.24\textwidth]{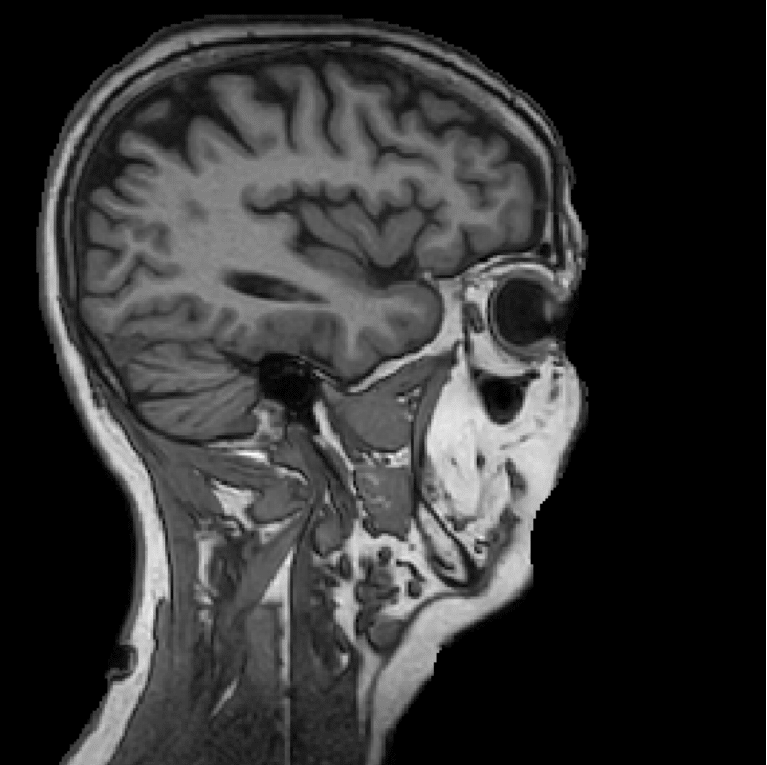} \hspace{-1em}
\includegraphics[trim={0cm 0cm 0cm 0.1cm},clip,width = 0.24\textwidth]{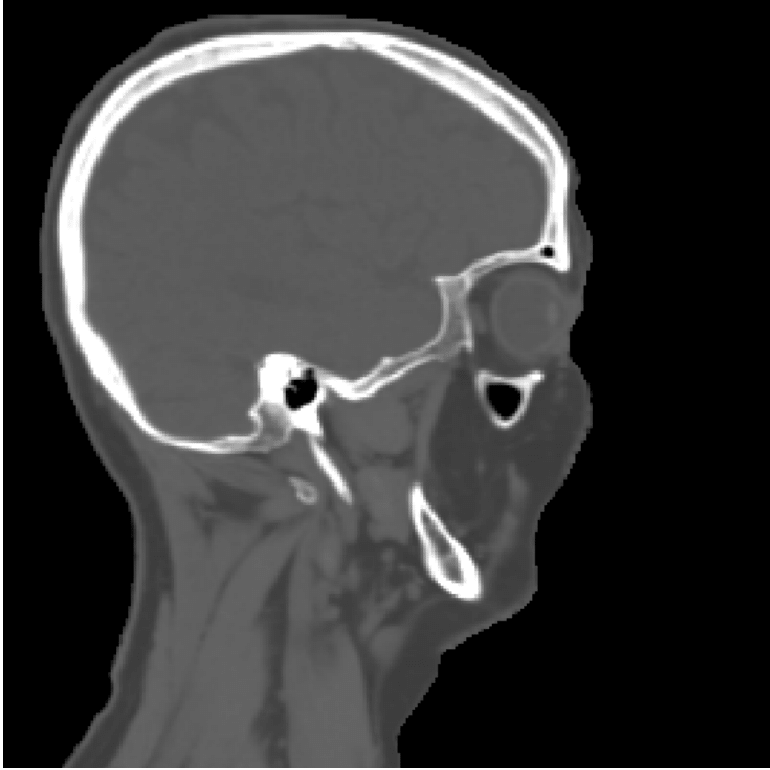} 
\caption{A pair of MRI (left) and CT (right) images of the same subject showing the significant appearance gap between
the two modalities making medical image synthesis from MRI to CT a challenging task. Image is from \cite{wolterink2017deep}.}
\label{fig:mri_ct}
\end{figure}
\subsubsection{\textbf{\textit{Unsupervised Methods}}} These approaches are particularly suitable for medical image synthesis tasks as they do not require paired training datasets. Recently, Ristea \cite{ristea2021cytran} proposed a cycle-consistent generative adversarial transformer (CyTran) to translate unpaired contrast CT scans to non-contrast CT scans and volumetric image registration of contrast CT scans to non-contrast CT scans. To handle high-resolution CT images, they propose hybrid convolution and multi-head attention-based architecture shown in Fig.~\ref{fig:cytran}. CyTran is unsupervised due to the integration of cyclic loss. Morever, they introduce the Coltea-Lung-CT100W dataset formed of 100 3D anonymized triphasic lung CT scans of female patients.

\subsection{Inter-Modality Approaches}

The inter-modality approaches aim to synthesize targets to capture the useful structural information in the source images of different modalities. Examples include CT to MRI translation or vice-versa. Due to challenges associated with inter-modal translation, only supervised approaches have been explored.

Dalmaz \textit{et al.} \cite{dalmaz2021resvit} introduce a novel synthesis approach, ResViT, for the multi-modal imaging based on a conditional deep adversarial network with ViT-based generator. Specifically, ResViT, employs convolutional and transformer branches within a residual bottleneck to preserve both local precision and contextual sensitivity along with the realism of adversarial learning. The bottleneck comprises novel aggregated residual transformer blocks to synergistically preserve local and global context, with a weight-sharing strategy to minimize model complexity. The effectiveness of ResViT model is demonstrated on two multi-contrast brain MRI datasets, BraTS \cite{menze2014multimodal}), and a multi-modal pelvic MRI-CT dataset \cite{nyholm2018mr}.

\begin{tcolorbox}[top=0.5pt,left=0pt,size=minimal,boxrule = 0pt,breakable, enhanced]
\subsection{\textbf{ \underline{Discussion}}}
 \hspace{0.75em}\textit{ In this section, we have reviewed the applications of ViT models in medical image synthesis. Realistic synthesis of medical images is particularly important as, in general, more than one imaging modality is involved in accurate clinical decision-making due to their complementary strengths. Therefore, in many practical applications, a
certain modality is desired but infeasible to acquire due to cost and privacy issues. Recent transformer-based approaches can effectively circumvent these issues due to their ability to generate more realistic images than GAN-based methods.
}
\\
\vspace{0.01em}
\textit{ Furthermore, most Transformer-based medical image synthesis approaches use the adversarial loss to generate realistic images. The adversarial loss can cause mode-collapse, and effective strategies must be employed to mitigate this issue~\cite{wang2021generative}.}
\\
\vspace{0.01em}
\textit{ Lastly, to the best of our knowledge, no work has been done using transformer-based models for inter-modality image synthesis approaches in an unsupervised setting. This can be due to the highly challenging nature of the problem (like CT and MRI images of the same subject have significantly different appearances, as shown in Fig. \ref{fig:mri_ct}), thereby making it a promising direction to explore.}
\end{tcolorbox}

\section{Medical Image Registration} \label{sec:reg}

Medical image registration aims to find dense per-voxel displacement and establish alignment between a pair of fixed and moving images. In medical imaging, registration may be necessary when analyzing a pair of images acquired at different times, from different viewpoints, or using different modalities (like MRI and CT) \cite{haskins2020deep}. Accurate medical image registration is a challenging task due to difficulties in extracting discriminative features from multimodal medical images, complex motion, and lack of robust outlier rejection approaches \cite{alam2018medical}. In this section, we briefly highlight the applications of ViTs in medical image registration.

The first study to investigate the usage of transformers for self-supervised medical volumetric image registration has been proposed by Chen \textit{et al.} \cite{chen2021vit}. Their model, ViT-V-Net, consists of a hybrid architecture composed of convolutional and transformer layers. Specifically, ViTs are applied to the high-level features of fixed and moving images extracted via a series of convolutional and max-pooling layers. The output from ViT is then reshaped and decoded using a V-Net style decoder \cite{milletari2018v}. To efficiently propagate the information, ViT-V-Net uses long skip connections between the encoder and decoder. The output of the ViT-V-Net decoder is a dense displacement field, which is fed to the spatial transformer network for warping. Experiments on in-house MRI dataset show superiority of ViT-V-Net over other competing approaches in terms of Dice score. 
Chen \textit{et al.}~\cite{chen2021transmorph} further extends ViT-V-Net  and propose TransMorph model for volumetric medical image registration. Particularly, TransMorph makes use of Swin Transformer in the encoder to capture the semantic correspondence between input fixed and moving images, followed by long skip connections-based convolutional decoder to predict dense displacement field. For uncertainty estimation, they also introduce Bayesian deep learning by applying variational inference on the parameters of the encoder in TransMorph. Extensive evaluation is performed to compare TransMorph with other approaches for the medical image registration task. Specifically, experiments on inter-patient brain MRI registration provided by John-Hopkin university and XCAT-to-CT registration demonstrate the superiority of TransMorph against twelve different hand-crafted, CNN-based, and transformer-based approaches.
Similarly, Zhang \textit{et al.} \cite{zhang2021learning} present a novel dual transformer architecture (DTN) for volumetric diffeomorphic registration by effectively establishing correspondences between anatomical structures in an unsupervised manner.
The DTN consists of two CNN-based 3D U-Net encoders to extract embeddings of separate and concatenated volumetric MRI images. To further refine and enhance the embeddings, they propose encoder-decoder-based dual transformers to encode the cross-volume dependencies. Given the enhanced embeddings, the CNN decoder infers the deformation fields. Qualitative and quantitative results in terms of Dice similarity coefficient and negative Jacobian determinant on OASIS dataset~\cite{marcus2007open} of MRI scans demonstrate the effectiveness of their proposed architecture.

\begin{tcolorbox}[top=0.5pt,left=0pt,size=minimal,boxrule = 0pt,breakable, enhanced]
\subsection{\textbf{Discussion}}
 \hspace{0.75em}\textit{ Application of transformers to medical image registration problem is still at early stages, and it is difficult to draw any conclusion at this stage. However, seeing the rapid development of Transformer-based registration approaches in generic computer vision, we expect to see the same trend in this field in the near future.
}
\end{tcolorbox}
\section{Clinical report generation} \label{sec:cli}
\begin{figure}[t]
\centering
\includegraphics[trim={1cm 5cm 5.5cm 1.5cm},clip,width = 0.49\textwidth]{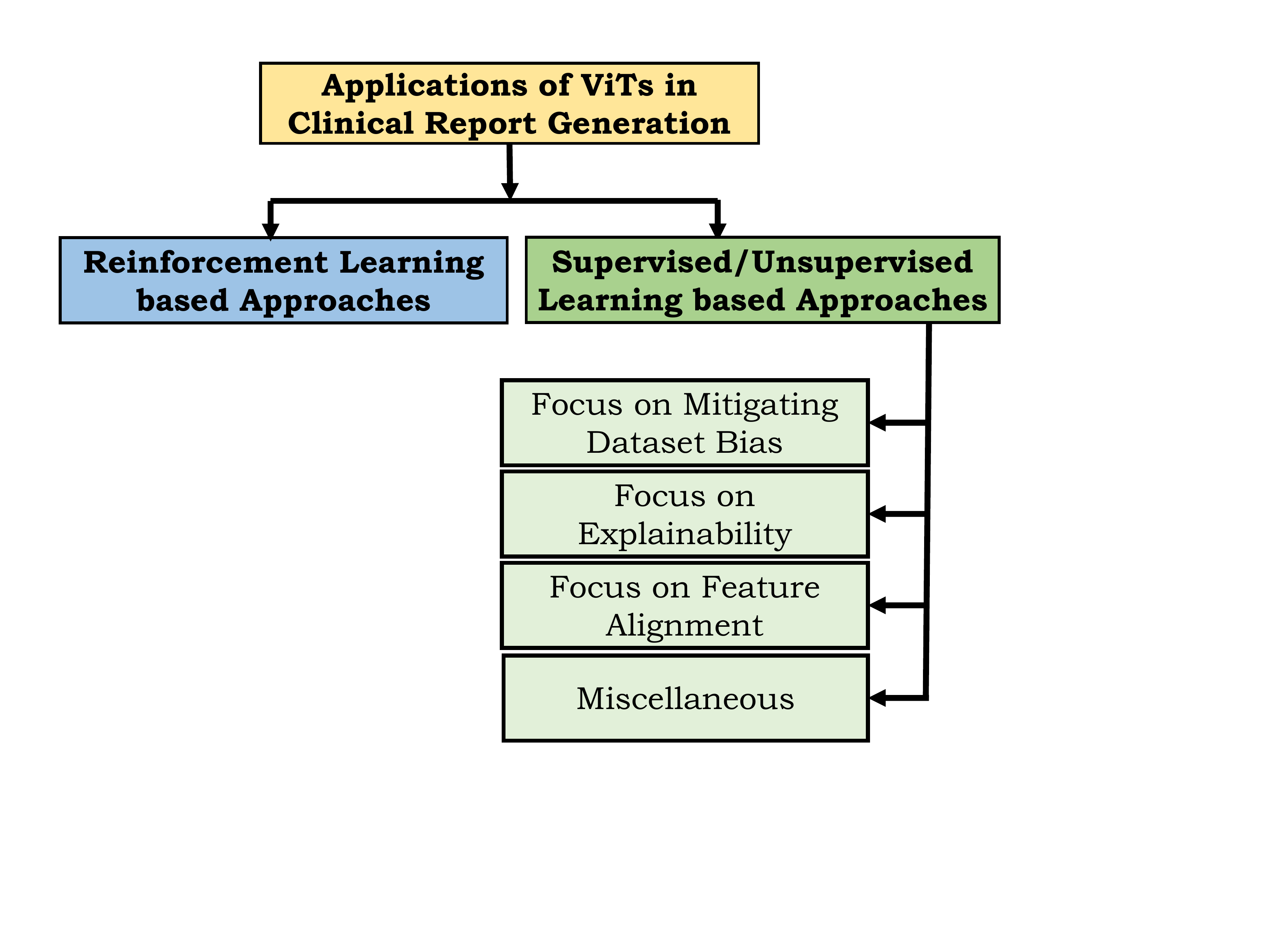}
\caption{Taxonomy of applications of ViTs in clinical report generation.}
\label{fig:report_tax}
\end{figure}
Recently, immense progress has been made to automatically generate clinical reports from medical images using deep learning~\cite{pavlopoulos2021diagnostic,monshi2020deep,kougia2019survey,messina2020survey}. This automatic report generation process can help clinicians in accurate decision-making. However, generating reports (or captions) from the medical imaging data is challenging due to diversity in the reports of different radiologists, long sequence length (unlike natural image captions), and dataset bias (more normal data compared to abnormal). Moreover, an effective medical report generation model is expected to process two key attributes: (1) \textit{language fluency} for human readability and (2) \textit{clinical accuracy} to correctly identify the disease along with related symptoms.
In this section, we briefly describe how transformer models help achieve these desired goals and effectively mitigate the aforementioned challenges associated with medical report generation. Specifically, these transformer-based approaches have achieved state-of-the-art performance both in terms of Natural Language Generation (NLG) and Clinical Efficacy (CE) metrics.
Also note that, unlike previous sections that mainly discuss ViTs, in this section, the focus is on the transformers as powerful language models to exploit the long-range dependencies for sentence generation. We have broadly categorized transformer-based clinical report generation approaches into reinforcement learning (RL) based and supervised/unsupervised learning methods, as shown in Fig.~\ref{fig:report_tax}, due to differences in their underlying training mechanism.

\subsection{Reinforcement Learning Based Approaches}

RL-based medical report generation approaches can directly use the evaluation metrics of interest (like human evaluation, relevant medical terminologies, etc.) as rewards and update the model parameters via policy gradient. All approaches covered in this section use the self-critical RL~\cite{rennie2017self} approach to train models, which is more suitable for the report generation task compared to the conventional RL.

One of the first attempts to integrate transformer in clinical report generation has been made by Xiong \textit{et al.}~\cite{xiong2019reinforced}. They propose Reinforced-Transformer for Medical Image Captioning (RTMIC) that consists of a pre-trained DenseNet~\cite{huang2017densely} to identify the region of interest from the input medical image, followed by a transformer-based encoder to extract visual features.
These features are given as input to the captioning decoder to generate sentences. All these modules are updated via self-critical reinforcement learning method during training on IU Chest X-ray dataset~\cite{demner2016preparing}. 
Similarly, Miura \textit{et al.}~\cite{miura2020improving} show that the high accuracy of automatic radiology reports as measured by natural language generation metrics such as BLEU~\cite{papineni2002bleu} and CIDer~\cite{vedantam2015cider} are often incomplete and inconsistent.
To address these challenges,  Miura \textit{et al.}~\cite{miura2020improving} propose a transformer-based model that directly optimizes the two newly proposed reward functions using self-critical RL. The first reward promotes the coverage of radiology domain entities with corresponding reference reports, and the second reward promotes the consistency of the generated reports with their descriptions in the reference reports. Further, they combine these reward functions with the semantic equivalence metric of BERTScore~\cite{zhang2019bertscore} that results in generated reports with better performance in terms of clinical metrics. 


\textbf{{Surgical Instructions Generation.}}
Inspired by the success of transformers in medical report generation, Zhang \textit{et al.}~\cite{zhang2021surgical} propose a transformer model to generate instructions from the surgical scenes. Lack of a predefined template, as in the case of medical report generation, makes generation of surgical instructions a challenging task. To handle this challenge, Zhang \textit{et al.}~\cite{zhang2021surgical} have proposed an encoder-decoder based architecture back-boned by a transformer model. Specifically, their proposed architecture, optimized via self-critical reinforcement learning~\cite{rennie2017self}, effectively models the dependencies for visual features, textual features, and visual-textural relational features to accurately generate surgical reports on the DAISI dataset~\cite{rojas2020daisi}. 
 
\subsection{Supervised and Unsupervised Approaches}

Supervised/unsupervised approaches use differentiable loss functions to train models for medical report generation and do not interact with the environment via an agent. We have categorized supervised/unsupervised approaches into methods that focus on dataset bias, explainability, feature alignment, and miscellaneous categories based on the challenges these approaches address.


\subsubsection{{{\textbf{Dataset Bias}}}} Dataset bias is a common problem in medical report generation as there are far more sentences describing normalities than abnormalities. 
To mitigate this bias,   Srinivasan~\cite{srinivasan2020hierarchical} propose a hierarchical classification approach using a transformer as a decoder. Specifically, the transformer decoder leverage attention between and across features obtained from reports, images, and tags for effective report generation. The architecture consists of \textit{Abnormality Detection Network} to classify normal and abnormal images, \textit{Tag Classification Net} to generate tags against images, and \textit{Report Generation Net} that takes image features and tags as inputs to generate final reports. Experiments on IU Chest X-ray dataset~\cite{demner2016preparing} demonstrate the effectiveness of the proposed architectural components. 
Similarly, Liu \textit{et al.}~\cite{liu2021exploring} try to imitate the work of radiologists by distilling posterior and prior knowledge to generate accurate radiology reports. Their proposed architecture consists of three modules of Posterior Knowledge Explorer (PoKE), Prior Knowledge Explorer (PrKE), and Multidomain Knowledge Distiller (MKD). Specifically, PoKE identifies the abnormal area in the input images (mitigate image data bias), PrKE explores relevant prior information from the radiological reports and medical knowledge graph (mitigate textual data bias), and MKD (based on transformer decoder) distills the posterior and prior knowledge to generate radiology report.   
In another work, You \textit{et al.} \cite{you2021aligntransformer} propose AlignTransformer to generate a medical report from X-ray images. Specifically, AlignTransformer consists of two modules: align hierarchial attention and multi-grained transformer. Align hierarchial attention module helps to better locate the abnormal region in the input medical images. 
On the other hand, multi-grained transformer leverages multi-grained visual features using adaptive exploiting attention~\cite{cornia2020meshed} to accurately generate long medical reports. AlignTransformer achieves favorable performance on IU-Xray~\cite{demner2016preparing} and MIMIC-CXR~\cite{johnson2019mimic} datasets. 
\begin{figure}[t]
\centering
\includegraphics[trim={0cm 0cm 0cm 0cm},clip,width = 0.48\textwidth]{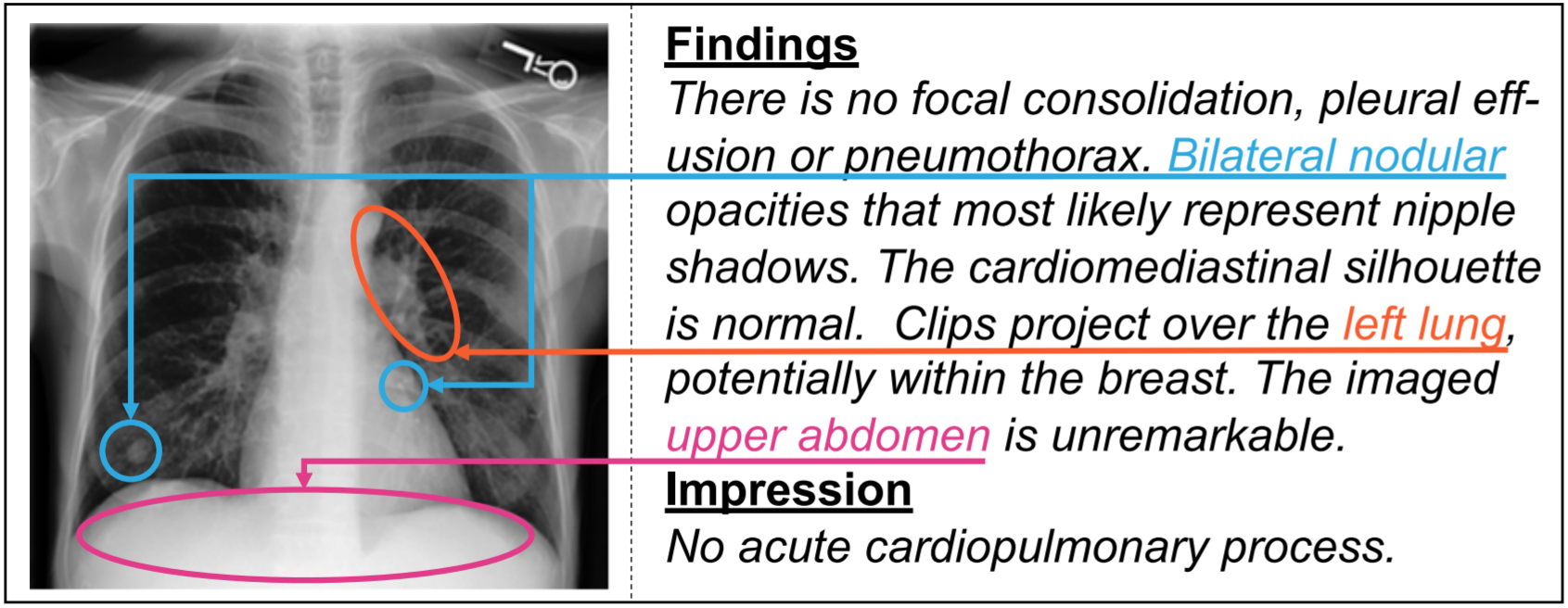}
\caption{A chest X-ray image and its accompanying report, which includes findings and impressions, with aligned visual and textual components highlighted in different colors. Figure taken from \cite{chen2021cross}.}
\label{fig:rep_xray}
\end{figure}

\begin{figure*}[h]
\centering
\includegraphics[width = 0.85\textwidth]{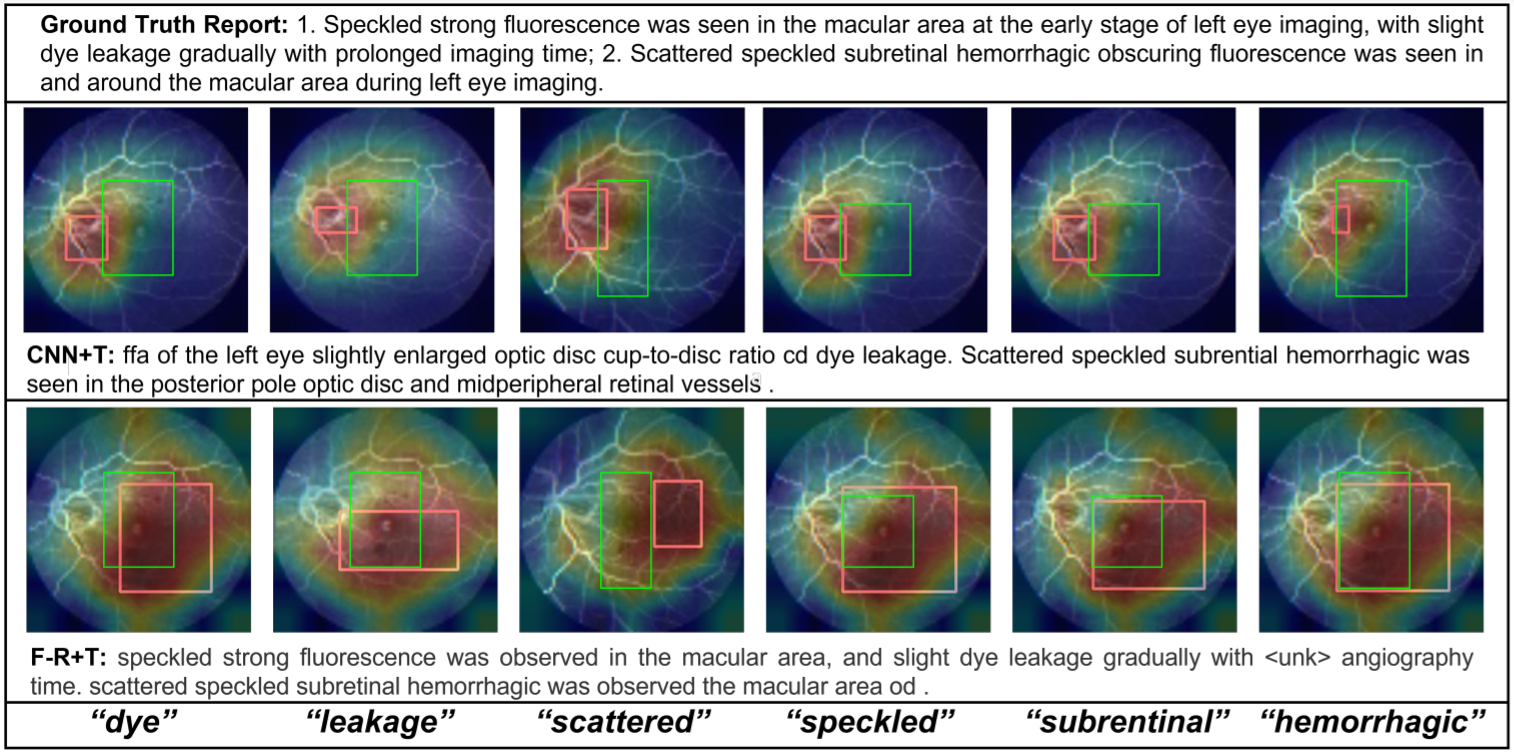}
\caption{The depiction of lesion-image attention mapping areas and ground truth between CNN+Transformer and Faster-RCNN+Transformer samples, where green boxes represent the annotated region for each lesion word and red boxes represent the lesion-image attention mapping regions. Image taken from \cite{li2021ffa}.}
\label{fig:rep_exp}
\end{figure*}
\begin{table}[]
\setlength{\arrayrulewidth}{0.35mm}
\centering
\resizebox{0.48\textwidth}{!}{%
\begin{tabular}{|l|ccc|ccc|}
\hline
\rowcolor{mygray} & \multicolumn{3}{c|}{\textbf{Image}}                                                    & \multicolumn{3}{c|}{\textbf{Report}}                                            \\ 
\rowcolor{mygray}  \multirow{-2}{*}{\textbf{Dataset}}                                 & \multicolumn{1}{l}{Number} & \multicolumn{1}{l}{Modality} & \multicolumn{1}{l|}{View*} & \multicolumn{1}{l}{Length*} & \multicolumn{1}{l}{Language} & \multicolumn{1}{l|}{Cases} \\ \hline
IU X-ray~\cite{demner2016preparing}                         & 7,470                      & X-Ray                        & 2                          & 32.5                        & Eng.                         & 2,955                      \\
MIMIC-CXR~\cite{johnson2019mimic}                         & 377,110                    & X-Ray                        & 1                          & 53.2                        & Eng.                         & 276,778                    \\
PadChest~\cite{bustos2020padchest}                          & 160,868                    & X-Ray                        & 2                          & -                           & Es                           & 22,710                     \\
CX-CHR~\cite{li2019knowledge}                         & 45,598                     & X-Ray                        & 2                          & 66.9                        & Zh                           & 40,410           
                \\
DIARETDB1~\cite{kalviainen2007diaretdb1}                         & 89                         & CFP                          & 1                          & -                           & Eng                          & 89                         \\
MESSIDOR~\cite{decenciere2014feedback}                          & 1,200                      & CFP                          & 2                          & -                           & Fr                           & 587                        \\
FFA-IR\cite{li2021ffa}                            & 1,048,584                  & FFA                          & 87                         & 91.2                        & Eng/Zh                       & 10,790     
\\
COV-CTR~\cite{li2020auxiliary}                           & 728                        & CT-Scans                     & 1                          & 77.3                        & Eng/Zh                       & 728                        \\
DEN~\cite{huang2021deepopht}                               & 15,709                     & CFP+FFA                      & 1                          & 7                           & Eng                          & -                          \\
STARE~\cite{hoover2000locating}                             & 397                        & CFP+FFA                      & 5                          & -                           & Eng                          & 397                        \\ \hline
\end{tabular}%
}
\caption{Statistics of existing medical report generation datasets where * means average number and - for datasets that do not provide relevant information. Most of the transformers based models for medical report generation use Open-IU and MIMIC-CXR for evaluating results. Table adapted from \cite{li2021ffa}.}
\label{tab:rep_datasets}
\end{table}
\subsubsection{{{\textbf{Feature Alignment}}}} Feature alignment based approaches mainly focus on the accurate alignment of encoded representation of the medical images and corresponding text, which is crucial for the interaction and generation across modalities (images and text here) and subsequently for accurate report generation, as indicated in Fig.~\ref{fig:rep_xray}.
To align better, Chen \textit{et al.}~\cite{chen2021cross} propose a cross-modal memory network to augment the transformer-based encoder-decoder model for radiology report generation. They design a shared memory to facilitate the alignment between the features of medical images and texts. Experiments on IU-Xray~\cite{demner2016preparing} and MIMIC-CXR~\cite{johnson2019mimic} datasets demonstrate that the proposed model can better align image and text features as compared to baseline methods. 
Similarly, building on the shared-memory work of Chen \textit{et al.}~\cite{chen2021cross}, Yan \textit{et al.}~\cite{yan2021weakly} introduce a weakly supervised contrastive objective to favor reports that are semantically close to the target, thereby producing more clinically accurate outputs. 
Similarly, Amjoud \textit{et al.}~\cite{amjoud2021automatic} investigate the impact on the report generation performance by modifying different architectural components of the model proposed by Chen \textit{et al.}~ \cite{chen2021cross} including replacing visual extractor and changing the number of layers in transformer-based decoder. 

\subsubsection{{{\textbf{Explainable Models}}}}
Explainability in medical report generation is crucial to improve trustworthiness for deploying models in clinical settings and a mean for extracting bounding boxes for lesion localization.
For model explainability, Hou \textit{et al.}~\cite{hou2021ratchet}  employ attention to identify regions of interest in the input image and demonstrate where the model is focusing for the resulting text. This attention mechanism increases the explainability of black-box models used in clinical settings and provides a method for extracting bounding boxes for disease localization. 
Specifically, they propose RATCHET transformer model to generate reports by using DenseNet-101~\cite{huang2017densely} as an image feature extractor. RATCHET consists of a transformer-based RNN-Decoder for generating chest radiograph reports. 
They assess the model's natural language skills and the medical correctness of generated reports. 
Similarly, despite the immense interest of AI and clinical medicine researchers in the automatic report generation area, benchmark datasets are scarce, and the field lacks reliable evaluation metrics. To address these challenges, Li \textit{et al.}~\cite{li2021ffa} introduce a large-scale Fundus fluorescence in Angiography images and reports dataset containing 10,790 reports describing 1,048,584 images with explainable annotations as shown in Fig. \ref{fig:rep_exp}. The dataset comes with annotated Chinese reports and corresponding translated English reports. Further, they introduce nine reliable metrics based on human evaluation criteria. 

\begin{table*}
\setlength{\arrayrulewidth}{0.4mm}
\centering
\resizebox{\textwidth}{!}{%
\begin{tabular}{|c|l|l|llllll|lll|} 
\hline
\rowcolor{mygray}        &  &  & \multicolumn{6}{c|}{\textbf{\textsc{NLG Metrics}}}                                                                                                                       & \multicolumn{3}{c|}{\textbf{\textsc{CE Metrics}}}                                 \\
 \rowcolor{mygray}   \multirow{-2}{*}{\textbf{\textsc{Dataset}}}                                    & \multicolumn{1}{c|}{\multirow{-2}{*}{\textbf{\textsc{Model}}}}                               &            \multirow{-2}{*}{\textbf{\textsc{Year}}}                    & \multicolumn{1}{c}{\textsc{BL-1}} & \multicolumn{1}{c}{\textsc{BL-2}} & \multicolumn{1}{c}{\textsc{BL-3}} & \multicolumn{1}{c}{\textsc{BL-4}} & \multicolumn{1}{c}{\textsc{MTR}} & \multicolumn{1}{c|}{\textsc{RG-L}} & \multicolumn{1}{c}{\textsc{P}} & \multicolumn{1}{c}{\textsc{R}} & \multicolumn{1}{c|}{\textsc{F1}}  \\ 
\hline
\multirow{10}{*}{\textbf{\textsc{IU X-Ray~\cite{demner2016preparing} }}}   & \textsc{RTMIC~\cite{xiong2019reinforced}}                                                & 2019                           & 0.350                    & 0.234                    & 0.143                    & 0.096                    & -                       & -                         & -                     & -                     & -                        \\
                                      & \textsc{HRG Transformer}                                      & 2019                           & 0.464                    & 0.301                    & 0.212                    & 0.158                    & -                       & -                         & -                     & -                     & -                        \\
                                      & \textsc{KERP~\cite{li2019knowledge}}                                          & 2019                           & 0.482                    & 0.325                    & 0.226                    & 0.162                    & -                       & 0.339                     & -                     & -                     & -                        \\
                                    & \textsc{Hierarchical Transformer~\cite{srinivasan2020hierarchical}}                                    & 2020                           & 0.464                    & 0.301                    & 0.212                    & 0.158                   & -                   & -                     & -                     & -                     & -                        \\ 
                                      & \textsc{Memory Transformer~\cite{chen2020generating}}                                   & 2020                           & 0.470                    & 0.304                    & 0.219                    & 0.165                    & 0.187                   & 0.371                     & -                     & -                     & -                        \\
                                      & \textsc{$\mathcal{M}^2$ TR.~\cite{nooralahzadeh2021progressive} }                                                & 2021                           & 0.475                    & 0.301                    & 0.228                    & 0.180                    & 0.169                   & 0.373                     & -                     & -                     & -                        \\
                                      & \textsc{$\mathcal{M}^2$ TR. Prog.~\cite{nooralahzadeh2021progressive}}                                          & 2021                           & 0.486                    & 0.317                    & 0.232                    & 0.173                    & 0.192                   & 0.390                     & -                     & -                     & -                        \\
                                    & \textsc{Wang et al.~\cite{wang2021confidence}}                                          & 2021                           & 0.481                    & 0.309                    & 0.223                    & 0.169                    & 0.193                   & 0.365                     & -                     & -                     & -                        \\
                                      & \textsc{Nguyen et al~\cite{nguyen2021automated}}                                         & 2021                           & 0.515                    & 0.378                    & 0.293                    & 0.235                    & 0.219                   & 0.436                     & -                     & -                     & -                        \\
                                      & \textsc{PPKED~\cite{liu2021exploring}}                                         & 2021                           & 0.483                    & 0.315                    & 0.224                    & 0.168                    & 0.190                   & 0.376                     & -                     & -                     & -                        \\
                                      & \textsc{Align Transformer~\cite{you2021aligntransformer}}                                    & 2021                           & 0.484                    & 0.313                    & 0.225                    & 0.173                    & 0.204                   & 0.379                     & -                     & -                     & -                        \\ 
                        & \textsc{KGAE Unsupervised~\cite{liu2021auto}}                                    & 2021                           & 0.417                    & 0.263                    & 0.181                    & 0.126                    & 0.149                   & 0.318                     & -                     & -                     & -                        \\ 
                                  & \textsc{KGAE Semi-supervised~\cite{liu2021auto}}                                    & 2021                           & 0.497                    & 0.320                    & 0.232                    & 0.171                    & 0.189                   & 0.379                     & -                     & -                     & -                        \\ 
                                  & \textsc{KGAE Supervised~\cite{liu2021auto}}                                    & 2021                           & 0.512                    & 0.327                    & 0.240                    & 0.179                    & 0.195                   & 0.383                     & -                     & -                     & -                        \\   
\hline
\multirow{12}{*}{\textbf{\textsc{MIMIC-CXR~\cite{johnson2019mimic}} }} & \textsc{Transformers~\cite{vaswani2017attention}}                                         & 2017                           & 0.409                    & 0.268                    & 0.191                    & 0.144                    & 0.157                   & 0.318                     & -                     & -                     & -                        \\

                                      & \textsc{Memory Transformer~\cite{chen2020generating}}                                   & 2020                           & 0.353                    & 0.218                    & 0.145                    & 0.103                    & 0.142                   & 0.277                     & 0.333                 & 0.273                 & 0.276                    \\
                                      & \textsc{Clinical Transformer~\cite{lovelace2020learning}}                                 & 2020                           & 0.415                    & 0.272                    & 0.193                    & 0.146                    & 0.159                   & 0.318                     & 0.411                 & 0.475                 & 0.361                    \\
                                      & \textsc{$\mathcal{M}^2$ TR.~\cite{nooralahzadeh2021progressive}}                                                 & 2021                           & 0.361                    & 0.221                    & 0.146                    & 0.101                    & 0.139                   & 0.266                     & 0.324                 & 0.241                 & 0.276                    \\
                                      & \textsc{$\mathcal{M}^2$ TR. Prog.~\cite{nooralahzadeh2021progressive}}                                            & 2021                           & 0.378                    & 0.232                    & 0.154                    & 0.107                    & 0.145                   & 0.272                     & 0.240                 & 0.428                 & 0.308                    \\
                                      & \textsc{PPKED~\cite{liu2021exploring}}                                         & 2021                           & 0.360                    & 0.224                    & 0.149                    & 0.106                    & 0.149                   & 0.284                     & -                     & -                     & -                        \\
                                      & \textsc{Align Transformer~\cite{you2021aligntransformer}}                                    & 2021                           & 0.378                    & 0.235                    & 0.156                    & 0.112                    & 0.158                   & 0.283                     & -                     & -                     & -                        \\
                                      & \textsc{Ngyuen et al~\cite{nguyen2021automated}}                                         & 2021                           & 0.495                    & 0.360                    & 0.278                    & 0.224                    & 0.222                   & 0.390                     & -                     & -                     & -                        \\
                                      & \textsc{MDT+WCL~\cite{yan2021weakly}}                                              & 2021                           & 0.373                    & -                        & -                        & 0.107                    & 0.144                   & 0.274                     & 0.384                 & 0.274                 & 0.294                    \\
                                      & \textsc{$\mathcal{M}^2$ Trans (CE)~\cite{miura2020improving}}                                        & 2021                           & -                        & -                        & -                        & 0.111                    & -                       & -                         & 0.463                 & 0.732                 & 0.567                    \\
                                      & \textsc{$\mathcal{M}^2$ Trans (EN)~\cite{miura2020improving} }                                       & 2021                           & -                        & -                        & -                        & 0.114                    & -                       & -                         & 0.503                 & 0.651                 & 0.567                    \\
                                            & \textsc{KGAE Unsupervised~\cite{liu2021auto}}                                    & 2021                           & 0.221                    & 0.144                    & 0.096                    & 0.062                    & 0.097                   & 0.208                    & 0.214                     & 0.158                     & 0.156                        \\ 
                                  & \textsc{KGAE Semi-supervised~\cite{liu2021auto}}                                    & 2021                           & 0.352                    & 0.219                    & 0.149                    & 0.108                    & 0.147                   & 0.290                     & 0.360                     & 0.302                     & 0.307                        \\ 
                                  & \textsc{KGAE Supervised~\cite{liu2021auto}}                                    & 2021                           & 0.369                    & 0.231                    & 0.156                   & 0.118                   & 0.153                   & 0.295                     & 0.389                     & 0.362                    & 0.355                       \\  
\hline
\end{tabular}
}
\caption{Quantitative comparison of transformer models for the task of clinical report generation in terms of Natural Language Generation (NLG) and Clinical Efficacy (CE) on two benchmark datasets. The NLG metrics include BLEU (BL)~\cite{papineni2002bleu}, METEOR (MTR)~\cite{denkowski2011meteor} and ROUGE-L (RG-L)~\cite{rouge2004package} and CE metrics include precision, recall and F1 score.}
\label{tab:report_gen}
\end{table*}

\subsubsection{{{\textbf{Miscellaneous}}}}
In this section, we highlight several approaches that try to improve different aspects of clinical report generation from medical images. Examples include a memory-driven transformer to capture similar patterns in reports, uncertainty quantification for reliable report generation, a curriculum learning-based method, and an unsupervised approach to avoid paired training datasets.

Chen~\textit{et al.}~\cite{chen2020generating} propose a \textbf{memory-driven transformer to exploit similar patterns} in the radiology image reports. Specifically, they add a module to each layer of transformer-based decoder by optimizing the original layer normalization with a novel memory-driven conditional layer normalization. Extensive experiments on IU Chest X-ray~\cite{demner2016preparing} and MIMIC-CXR~\cite{johnson2019mimic} datasets demonstrate the superiority of their approach both in terms of Natural Language Generation (NLG) and Clinical Efficacy (CE) metrics. 
Similarly, Lovelace \textit{et al.}~\cite{lovelace2020learning} also leverage the  transformer-based encoder and decoder for accurate medical report generation on  MIMIC-CXR dataset~\cite{johnson2019mimic}. To \textbf{emphasis on clinically relevant report generation}, they design a method to differentiate clinical information from generated reports, which they use to refine the model for clinical coherence. 
In another work, Alfarghaly \textit{et al.}~\cite{alfarghaly2021automated} present a pre-trained transformer-based model to generate a medical report from images. Specifically, the encoder consists of a pre-retrained CheXNet model that can generate semantic features from the input medical images. These semantic features are used to \textbf{condition GPT2 decoder}~\cite{ziegler2019encoder, radford2019language} to generate accurate medical reports. 
Similarly, to judge the reliability of the automatic medical report generating model, uncertainty quantification is the key indicator. To incorporate this measure, Wang \textit{et al.}~\cite{wang2021confidence} propose \textbf{transformer-based confidence guided framework} to quantify both visual and textual uncertainty. These uncertainties are subsequently used to construct an uncertainty-weighted loss to reduce misjudgment risk and improve the overall performance of the generated report. 
In other work, Nguyen \textit{et al.}~\cite{nguyen2021automated} propose differentiable end-to-end framework that consists of \textbf{transformer as generator for report generation}.
Specifically, their proposed framework has three complementary modules: a \textit{classifier} to learn the representation of disease features, a transformer-based \textit{generator} model to generate the medical report, and \textit{interpreter} to make the generated report consistent with the classifier output. They demonstrate the effectiveness of proposed components on IU-Xray~\cite{demner2016preparing} and MIMIC-CXR~\cite{johnson2019mimic} datasets. 
Inspired by \textbf{curriculum learning}~\cite{bengio2009curriculum}, Nooralahzadeh \textit{et al.}~\cite{nooralahzadeh2021progressive} present a two-stage transformer architecture to progressively generate medical reports. Their progressive approach shows better performance over single-stage baselines in generating full-radiology reports. 
In another study, Pahwa \textit{et al.}~\cite{pahwa2021medskip} investigate the \textbf{impact of visual feature extractor} model on the performance of medical report generation. Based on insights, they propose a modified HRNet~\cite{sun2019deep}, MedSkip, to extract visual features for the subsequent processing by the transformer-based decoder to generate an accurate medical report. 
Similarly, Park \textit{et al.}~\cite{park2021medical} \textbf{investigate the expressiveness of features} to discriminate between normal and abnormal images. They demonstrate the superiority of transformer-based decoder without global average pooling over hierarchical LSTM baseline.
Existing transformer-based report generation models are mostly supervised and use paired image-report data during training. The paired data is difficult to obtain due to privacy and cost in the medical domain. To mitigate this issue, Liu \textit{et al.}~\cite{liu2021auto} propose a knowledge graph auto-encoder that works in the share latent domain of images and reports to extract useful information in an \textbf{unsupervised way}. Specifically, they use attention in the encoder to extract the knowledge representation from the knowledge graph and use a three-layer transformer in the decoder to generate reports. Their proposed framework can also be used in a semi-supervised or supervised manner in addition to the unsupervised mode.
Quantitative and qualitative results, as well as evaluation by radiologists, corroborate the effectiveness of their approach.  



\begin{tcolorbox}[top=0.5pt,left=0pt,size=minimal,boxrule = 0pt,breakable, enhanced]
\subsection{\textbf{\underline{Discussion}}}
\hspace{0.75em}\textit{In this section, we have provided a comprehensive overview of the transformer's applications for clinical report generation from X-ray images. In contrast to previous sections that discuss applications of ViTs, this section focuses on transformers as powerful language models. It is also pertinent to note that even though multiple surveys exist covering the applications of deep learning in clinical report generation~\cite{pavlopoulos2021diagnostic,monshi2020deep,kougia2019survey,messina2020survey}, to the best of our knowledge, \textbf{none of these have covered the applications of transformer models} in the area despite having transformers' phenomenal impact since their inception back in 2017.
In this regard, we hope this section will serve as a valuable resource to the research community. Below, we briefly highlight a few challenges associated with transformer-based models for report generation and outline promising future directions to explore.}
\\
\vspace{0.01em}
\textit{ As we have seen, transformer-based report generation models {mostly rely on natural language generation (NLG) evaluation metrics} such as CIDEr and BLEU to assess performance. These NLG metrics often fail to represent clinical efficacy. One recent work by Miura \textit{et al.}~\cite{miura2020improving} addresses this issue by proposing two new reward functions for the transformer model in reinforcement learning framework to better capture disease and anatomical knowledge in the generated reports. Another work by Li \textit{et al.}~\cite{li2021ffa} introduces nine reliable human evaluation criteria to validate the generated reports. Despite these works, we believe that more attention from the research community is required to design reliable clinical evaluation metrics to facilitate the adoption of transformer-based medical report generation models in clinical settings.}
\\
\vspace{0.01em}
\textit{ All transformer-based approaches covered in this section use the X-ray modality for automatic report generation. {Generating reports from other modalities like MRI or PET} have their own challenges associated with them due to their specific nature and distinct characteristics. Further, few medical datasets like ROCO~\cite{pelka2018radiology}, PEIR Gross~\cite{jing2017automatic}, and ImageCLEF~\cite{garcia2018overview} are available that consist of multiple modalities, different body parts, and corresponding captions. These datasets have the potential to become worthy benchmarks to gauge the performance of future multimodal (or unimodal like MRI, PET) transformer-based models for medical report generation. 
We believe that transformer-based models tailored to specific modalities to generate reports must be explored in the future with a focus on creating diverse and challenging datasets of other modalities. Details of a few existing medical report generation datasets are given in Table \ref{tab:rep_datasets}.
}
\\
\vspace{0.01em}
\textit{ Further, we would like to point out the interested researchers toward the recently explored {surgical instructions generation work using transformers}~\cite{zhang2021surgical} that could have a huge impact on surgical robotics, a market that is expected to reach USD 22.27 Billion by 2028. However, only one dataset, DAISI~\cite{rojas2020daisi}, is available to evaluate models in this emerging area, demanding attention from the medical community to create diverse and more challenging datasets. }
\\
\vspace{0.01em}
\textit{ Moreover, datasets for medical report generation like IU X-Ray~\cite{demner2016preparing} {does not contain any standard train-test split} and most of the transformer-based approaches evaluate the performance on different tests data. In this regard, the results in Table \ref{tab:report_gen} are not directly comparable, but they can provide an overall indication about the performance of the models.
We think what seems to be missing is a set of standardized procedures for creating challenging and diverse clinical report generation datasets.} 
\end{tcolorbox}

\section{Other applications} \label{sec:other_app}
In this section, we briefly highlight applications of Transformers in other medical imaging areas, including survival outcome prediction, visual question answering, and medical point cloud analysis. 
\textbf{Survival outcome prediction} is a challenging regression task that seeks to predict the relative risk of cancer death. Recently, transformer models have shown impressive success in predicting survival rates. Chen \textit{et al.}~\cite{chen2021multimodal} propose a Multimodal Co-Attention Transformer (MCAT) for the survival outcome prediction from whole-slide imaging (WSI) in pathology. MCAT learns a co-attention mapping between genomics and WSIs features to discover how histology features attend to genes while predicting patient survival outcomes. Extensive experiments on five cancer datasets demonstrate the superiority of MCAT compared to state-of-the-art CNN-based approaches. Similarly, Kipkogei \textit{et al.}~\cite{kipkogei2021explainable} propose a Transformer-based architecture, Clinical Transformer, to model the relation between clinical and molecular features to predict survival outcomes from cancerous lung dataset~\cite{samstein2019tumor}.
In another work, Eslami~\cite{eslami2021does}  propose \textbf{PubMedCLIP, a fine-tuned version of Contrastive Language-Image Pre-Training (CLIP)}~\cite{radford2021learning} for the medical domain by training it on the image–caption pairs from PubMed articles. Extensive experiments show that PubMedCLIP outperforms the previous state-of-the-art by nearly 3$\%$.
Recently, Liu \textit{et al.}~\cite{liu2019deep} propose \textbf{3D Medical Point Transformer} (3DMPT) to analyse 3D medical data. 3DMPT is tested on 3D medical classification and part segmentation based tasks.
Similarly, Malkiel \textit{et al.}~\cite{malkiel2021pre} propose a \textbf{Transformer-based architecture to analyse fMRI data}. They pre-train the model on 4D fMRI data in a self-supervised manner and fine-tune it on various downstream tasks, including age and gender prediction, as well as diagnosing Schizophrenia.
\begin{figure*}[t]
\centering
\includegraphics[width =1\textwidth]{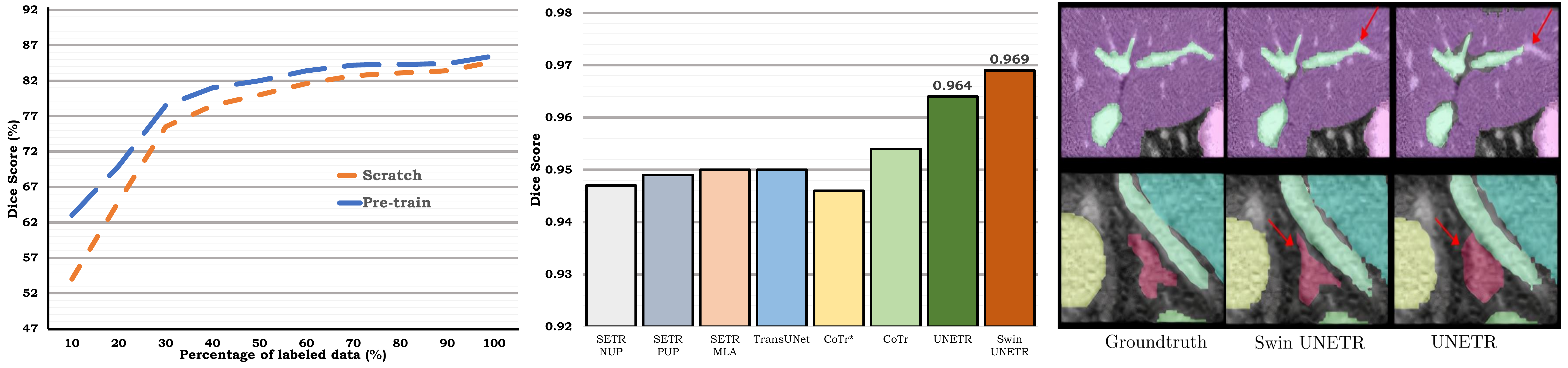}
\caption{Impact of pre-training ViT on domain specific medical imaging dataset. \textit{First column:} By only using 10$\%$ labelled data, Swin UNET pre-trained on CT images ({\color{blue} blue}) is able to achieve 10$\%$ improvement in dice score over Swin UNET trained from scratch ({\color{orange} orange}). \textit{Middle column:} Dice scores of recently proposed transformer models on the spleen segmentation task of MSD dataset. Swin UNETR achieves state-of-the-art performance due to pre-training on the domain specific (CT) medical dataset. \textit{Last column:} Qualitative visualizations of Swin UNETR pre-trained on CT images on BTCV multi-organ segmentation challenge. It can be seen that Swin UNETR predictions are closer to the groundtruth compared to baseline UNETR approach. First and last columns are adapted from \cite{tang2021self}.}
\label{fig:pre_train}
\end{figure*}

\section{Open Challenges and Future Directions}\label{sec:openproblems}

We have reviewed the exciting applications of vision transformers in medical image analysis. Despite their impressive performance, there remain several open research questions. In this section, we outline some of their limitations and highlight promising future research directions. Specifically, we will discuss the challenges of pre-training on large datasets (Sec.~\ref{sec:pretraining}), interpretability of ViT-based medical imaging approaches (Sec.~\ref{sec:interpretability}), robustness against adversarial attacks (Sec.~\ref{sec:adversarialrobustness}), designing efficient ViT architectures for real-time medical applications (Sec.~\ref{sec:edgedevices}), challenges in deploying ViT-based models in distributed settings (Sec.~\ref{sec:decentralized}), and domain adaptation (Sec.~\ref{sec:domainadaptation}). 
Further, wherever possible, we refer interested researchers to relevant CNNs-based medical imaging resources (recent studies, datasets, software libraries, etc.) to explore previously untapped applications by ViT-based models in medical imaging like adversarial robustness.

\subsection{Pre-training} \label{sec:pretraining}

Due to a lack of intrinsic inductive biases in modeling local visual features, ViTs need to figure out the image-specific concepts on their own via pre-training from large-scale training datasets \cite{dosovitskiy2020image}. This may be a barrier to their widespread application in medical imaging, where typically datasets are orders of magnitude smaller compared to natural image datasets due to cost, privacy concerns, and the rarity of certain diseases, thereby making ViTs difficult to train efficiently in the medical domain. {Existing learning-based medical imaging approaches commonly rely on transferring learning via ImageNet
pretraining, which may be sub-optimal due to drastically different image characteristics between medical and natural images.}
Recently, Matsoukas \textit{et al.}~\cite{matsoukas2021time} has studied the impact of pre-training on ViTs performance for image classification and segmentation via a careful set of extensive experiments on several medical imaging datasets. Below, we briefly highlight major findings of their work. 
\begin{itemize}
\item CNNs outperform ViTs for the medical image classification task when initialized with random weights.
\item CNNs and ViTs benefit significantly from ImageNet initialization for medical image classification. ViTs appear to benefit more from transfer learning, as they make up for the gap observed using random initialization, performing on par with their CNN counterparts.
\item CNNs and ViTs perform better with self-supervised pre-training approaches like DINO \cite{caron2021emerging} and BYOL \cite{grill2020bootstrap}. ViTs appear to outperform CNNs in this setting for medical image classification by a small margin.

\end{itemize}
In short, although recent ViT-based data-efficient approaches like DeiT \cite{touvron2021training}, Token-to-Token \cite{yuan2021tokens}, transformer in transformer~\cite{han2021transformer}, etc., report encouraging results in the generic vision applications, the task of learning these transformer models tailored to domain-specific medical imaging applications in a data-efficient manner is challenging. 
Recently, Tang \textit{et al.}~\cite{tang2021self} has made an attempt to handle this issue by investigating the effectiveness of self-supervised learning as a pre-training strategy on domain-specific medical images. Specifically, they propose 3D transformer-based hierarchical encoder, Swin UNETR, and after pre-training on 5,050 CT images, demonstrates its effectiveness via fine-tuning on the downstream task of medical image segmentation. 
The pre-training on the medical imaging dataset also reduces the annotation effort compared to training Swin UNETR from scratch with random initialization. This is shown in Fig.~\ref{fig:pre_train}, where it can be seen that pre-trained Swin UNETR can achieve the same performance by using only 60$\%$ of data as achieved by random initialized Swin UNETR using 100$\%$ of labeled data. This results in 40$\%$reduction of manual annotation effort. Furthermore, as shown in Fig.~\ref{fig:pre_train}, fine-tuning pre-trained Swin UNETR on the downstream medical image segmentation achieves better quantitative and qualitative results as compared to randomly initialized UNETR. Despite these efforts, there still remain several open challenges like Swin UNETR pre-trained on CT dataset gives unsatisfactory performance when applied directly to other medical imaging modalities like MRI due to large domain gap between CT and MRI images. 
Furthermore, the effectiveness of Swin UNETR on other downstream medical imaging tasks like classification and detection requires further investigation. 
Moreover, recent works for CNNs have shown that self-supervised pre-training on both ImageNet and medical datasets can improve the generalization performance (for classification) of the model on distribution shifted medical dataset~\cite{azizi2021big} as compared to pre-training on ImageNet only. We believe such studies for ViT-based models, along with multi-instance contrastive learning to leverage patient meta data~\cite{vu2021medaug}, will provide further insights to the community. Similarly, combining the self-supervised and semi-supervised pre-training in the context of ViTs for medical imaging applications is also an interesting avenue to explore~\cite{chen2020big}.

\subsection{Interpretability} \label{sec:interpretability}
\begin{figure}[t]
\centering
\includegraphics[width = 0.48\textwidth]{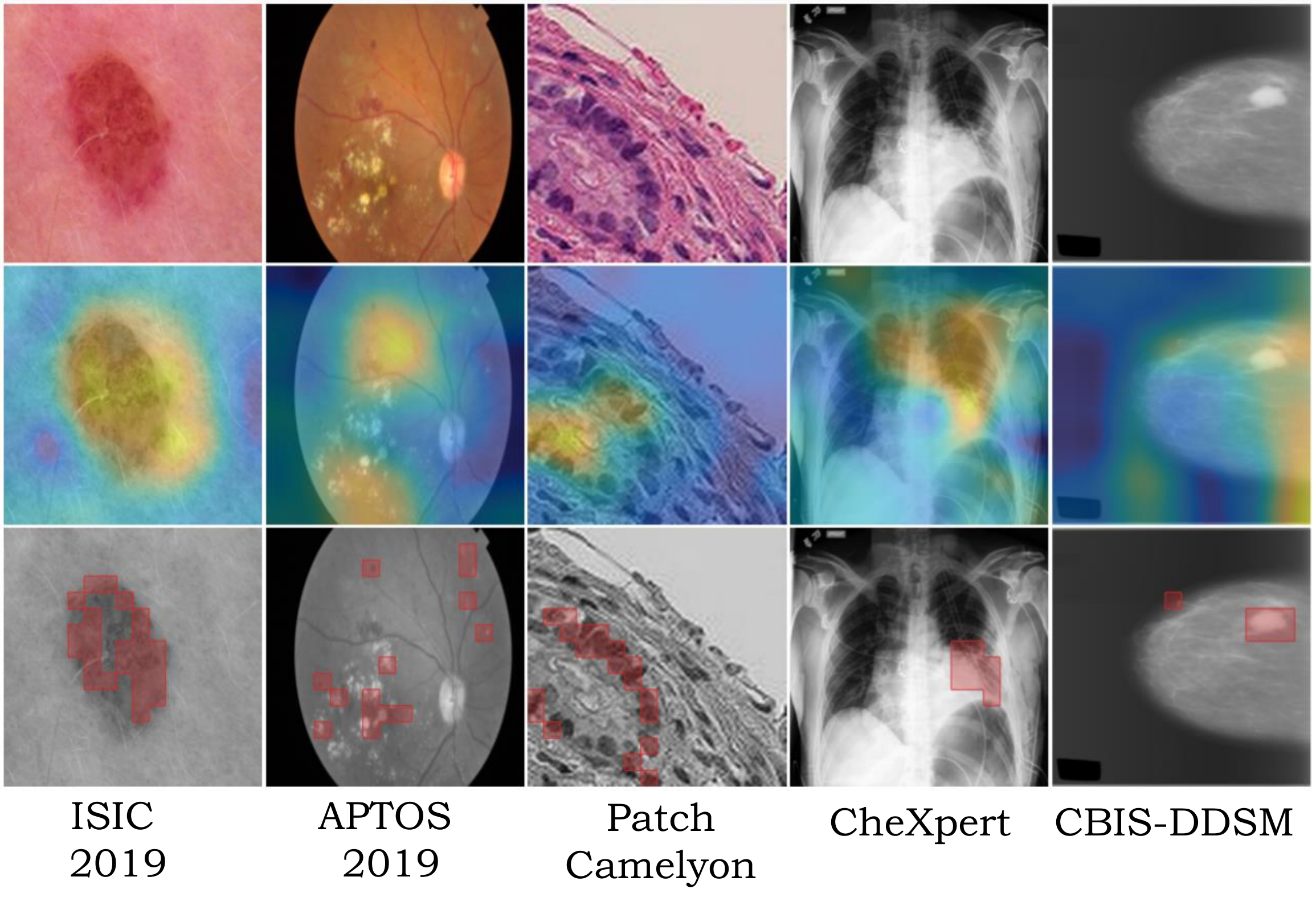} 
\caption{Saliency maps comparison for a CNN-based ResNet-50~\cite{he2016deep} (second row) and ViT-based DEIT-S~\cite{touvron2021training} (third row) on five medical imaging datasets for classification. Each column contains the original image (first row), a visualization of the ResNet-50 Grad-CAM saliency (second row), and a visualization of the DEIT-S’s attention map (third row), respectively. Note that the ViTs provide a clear, localized picture of attention compared to ResNet-50, thus giving insight into how the model makes decisions. Image taken from \cite{matsoukas2021time}.}
\label{fig:gradcam}
\end{figure}

Although the success of transformers has been empirically established in an impressive number of medical imaging applications, it has so far eluded a satisfactory interpretation.
In most medical imaging applications, ViT models have been deployed as block-boxes, thereby failing to provide insights and explain their learning behavior for making predictions. This black-box nature of ViTs has hindered their deployment in clinical practice since, in areas such as medical applications, it is imperative to identify the limitations and potential failure cases of designed systems, where interpretability plays a fundamental role~\cite{reyes2020interpretability}. 
Although several explainable AI-based medical imaging systems have been developed to gain deeper insights into the working of CNNs models for clinical applications~\cite{singh2020explainable,yan2021interpretable,ghoshal2020estimating}, however, the work is still in its infancy for ViT-based medical imaging applications. It is despite the inherent suitability of the self-attention mechanism to interpretability due to its ability to explicitly model interactions between every region in the image a shown in Fig.~\ref{fig:gradcam} \cite{chaudhari2021attentive}. Recent efforts for interpretable ViT-based medical imaging models leverage saliency-based approaches \cite{chefer2021transformer} and Grad-CAM based visualizations \cite{selvaraju2017grad}. Despite these efforts, the development of interpretable and explainable ViT-based approaches, specifically tailored for life-critical medical imaging applications, is a challenging and open research problem. Furthermore, formalisms, challenges, definitions, and evaluation protocols regarding interpretable ViTs based medical imaging systems must also be addressed. We believe that progress in this direction would not only help physicians to decide whether they should follow and trust automatic ViT-based model decisions but could also facilitate the deployment of such systems from a legal perspective.

\subsection{Adversarial Robustness} \label{sec:adversarialrobustness}
\begin{figure}[h]
\centering
\includegraphics[trim={6cm 2.5cm 8cm 1cm},clip,width = 0.48\textwidth]{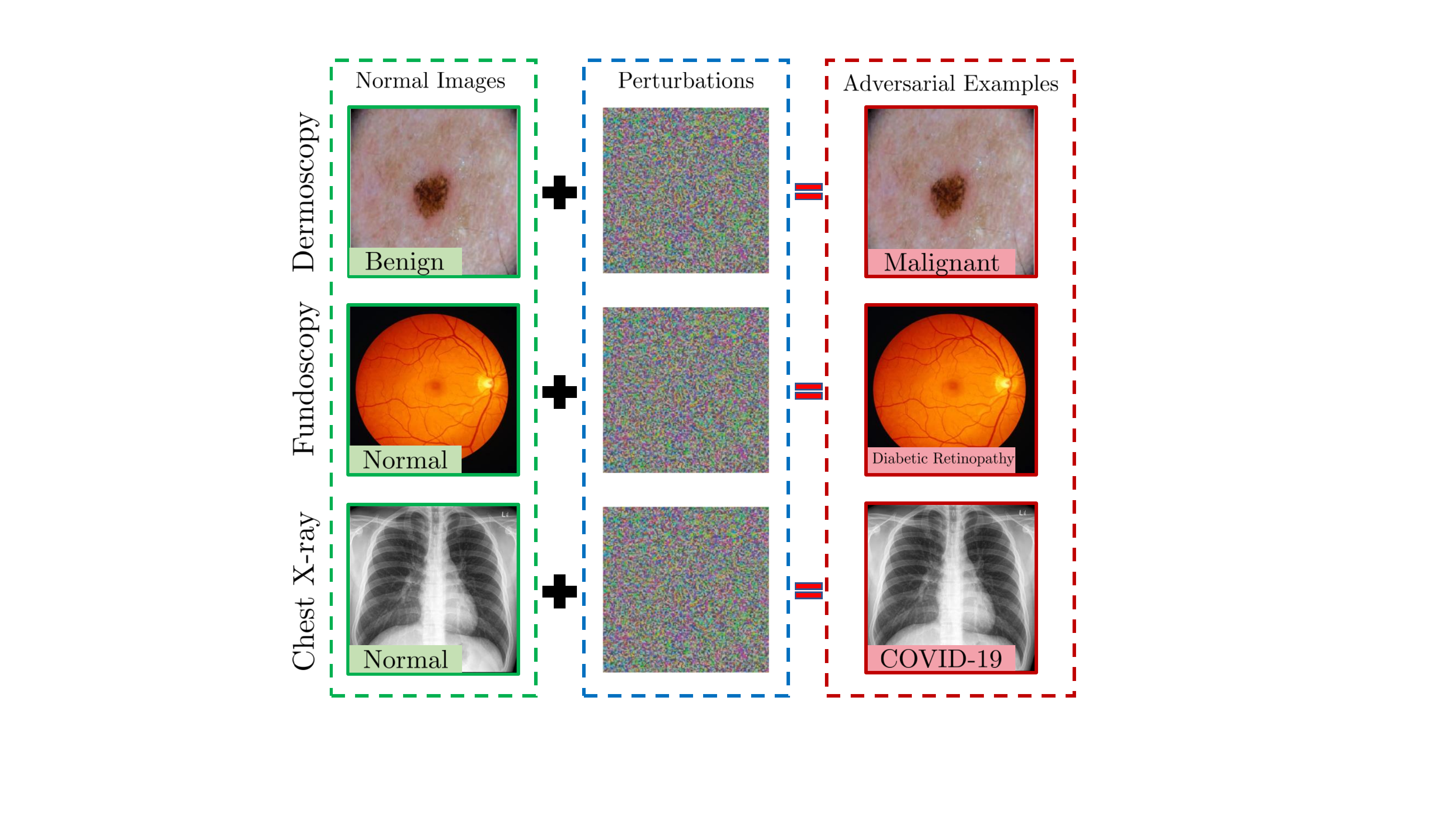} 
\caption{Examples of adversarial attacks to fool learning-based models trained on medical image datasets. Left: normal images, Middle: adversarial perturbations, Right: adversarial images. The bottom tag is the predicted class, and green/red indicates correct/wrong predictions.}
\label{fig:adversarial}
\end{figure}
\begin{table}[t]
\centering
\caption{Description of datasets generally used in medical adversarial deep learning.}
\label{tab:adversarial_datasets}
\resizebox{0.48\textwidth}{!}{%
\begin{tabular}{V{3}c|c|cV{3}}
\hlineB{3}
\rowcolor{mygray} \textbf{Dataset Name} & \textbf{Dataset Size} & \textbf{Modality} \\ \hline
RSNA~\cite{rsna}                 & 29,700                & X-ray             \\ \hline
JSRT~\cite{shiraishi2000development}                 & 247                   & X-ray             \\ \hline
BraTS 2018~\cite{bakas2018identifying}            & 1689                  & MRI               \\ \hline
BraTS 2019~\cite{brats19}           & 1675                  & MRI               \\ \hline
OASIS~\cite{oasis}              & 373-2168              & MRI               \\ \hline
HAM10000~\cite{tschandl2018ham10000}              & 10,000                & Dermatoscopic     \\ \hline
ISIC 18~\cite{codella2019skin}              & 3594                  & Dermatoscopic     \\ \hline
LUNA 16~\cite{luna16}              & 888                   & CT-Scans          \\ \hline
NIH Chest X-ray~\cite{tang2019xlsor}       & 112,000               & X-ray             \\ \hline
APTOS~\cite{aptos19}               & 5590                  & Fundoscopy        \\ \hline
Chest X-ray~\cite{kermany2018identifying}           & 5856                  & X-ray             \\ \hline
NSLT~\cite{nlst}                  & 75,000                & CT-Scans          \\ \hline
Diabetic Retinopathy~\cite{diabeticretino}  & 35,000                & Fundoscopy        \\ \hlineB{3}
\end{tabular}%
}
\end{table}

Advances in adversarial attacks have revealed the vulnerability of existing learning-based medical imaging systems against imperceptible perturbation in the input images \cite{ma2021understanding,papangelou2018toward,finlayson2019adversarial}. Considering the vast amount of money that underpins the medical imaging sector, this inevitably poses a risk whereby potential attackers may seek to profit from manipulation against the healthcare system, as shown in Fig.~\ref{fig:adversarial}. For example, an attacker might try to manipulate the examination reports of patients for insurance fraud or a false medical reimbursement claim, thereby raising safety concerns.
Therefore, ensuring the robustness of ViTs against adversarial attacks in life-critical medical imaging applications is of paramount importance. Although rich literature exists related to the robustness of CNNs in the medical imaging domain, to the best of our knowledge, no such study exists for ViTs, making it an exciting as well challenging direction to explore. Recently, few attempts have been made to evaluate the robustness of ViTs to adversarial attacks for natural images \cite{benz2021adversarial,bhojanapalli2021understanding,wei2021towards,shao2021adversarial,paul2021vision,naseer2021improving,mao2021rethinking,mahmood2021robustness,joshi2021adversarial,aldahdooh2021reveal}. The main conclusions of these attempts, ignoring their nuance difference, can be summarized as \textit{ViTs are more robust to adversarial attacks than CNNs}. However, these robust ViT models cannot be directly deployed for medical imaging applications as the variety and type of patterns and textures in medical images differ significantly from the natural domain. Therefore, a principled approach to evaluate the robustness of ViTs against adversarial attacks in the medical imaging context, which builds the groundwork for resilience, could serve as a critical model to deploy these models in clinical settings. Furthermore, theoretical understanding to provide guarantees about the performance and robustness of ViTs, like CNNs \cite{katz2017towards}, can be of significant interest to medical imaging researchers. In Table \ref{tab:adversarial_datasets}, we provide a description of datasets used in adversarial medical learning to evaluate the robustness of CNNs for interesting researchers to benchmark the robustness of ViT-based models.

\begin{table*}[t]
\centering
\caption{Description of some of the open-source federated learning and privacy-preserving frameworks, including some specifically developed for medical imaging applications.}
\resizebox{0.9\textwidth}{!}{%
\begin{tabular}{V{3}c|c|p{8cm}V{3}}
\hlineB{3}
\hline
\rowcolor{mygray} \textbf{Name}        & \textbf{Framework}         & \hspace{11em}\textbf{Description}                                                                  \\ \hline
TensorFlow Fed\cite{bonawitztensorflow} & TensorFlow                 & Open-source framework for for machine learning and other computations on decentralized data.                           \\ \hline
CrypTen~\cite{knott2021crypten}             & PyTorch                    & Framework to facilitate research in secure and privacy-preserving machine learning.                                                                        \\ \hline
OpenMined~\cite{2017openmined}           & TensorFlow, PyTorch, Keras & Open-source decentralised privacy preserving framework. Includes specialized tools like PySyft, PyGrid, and SyferText. \\ \hline
Opacus~\cite{yousefpour2021opacus}           & PyTorch                    & Enables training PyTorch models with differential privacy. Allows the client to online track the privacy budget.                                                          \\ \hline
Deepee~\cite{kaissis2021end}             & PyTorch                    &  Library for differentially private deep learning for medical imaging in PyTorch.                \\ \hline
PriMIA~\cite{kaissis2021end}             & PyTorch                    & Framework for end-to-end privacy-preserving decentralized deep learning for medical images.                 \\ \hlineB{3}
\end{tabular}%
\label{tab:federated_frameworks}
}
\end{table*}

\subsection{ViTs for Medical Imaging on Edge Devices} \label{sec:edgedevices}
Despite the tremendous success of ViTs in numerous medical imaging applications, the intensive requirements for memory and computation hamper their deployment on resource constraint edge devices \cite{zhai2021scaling,tabani2021improving}.
Due to recent advancements in edge computing, healthcare providers can process, store and analyze complex medical imaging data on-premises, speeding diagnosis, improving clinician workflows, enhancing patient privacy, and saving time—and potentially lives. These edge devices provide extremely fast and highly accurate processing of large amounts of medical imaging data, therefore demanding efficient hardware design to make ViT-based models suitable for edge computing-based medical imaging hardware. Recently few efforts have been made to compress transformer-based models by leveraging enhanced block-circulant matrix-based representation~\cite{li2020ftrans}
and neural architecture search strategies~\cite{wang2020hat}. 
Due to the exceptional performance of ViTs, we believe that there is a dire need for their domain-optimized architectural designs tailored for edge devices. It can have a tremendous impact on medical imaging-based health care applications where on-demand insights help teams make crucial and urgent decisions about patients.

\subsection{Decentralized Medical Imaging Solutions using ViTs} \label{sec:decentralized}

Building robust deep learning-based medical imaging models highly depends on the amount and diversity of the training data. The training data required to train a reliable and robust model may not be available in a single institution due to strict privacy regulations, the low incidence rate of some pathologies,  data-ownership concerns, and limited numbers of patients. Federated Learning (FL) has been proposed to facilitate multi-hospital collaboration while obviating data transfer. Specifically, in FL, a shared model is built using distributed data from multiple devices where each device trains the model using its local data and then shares the model parameters with the central model without sharing its actual data. Although a plethora of approaches exists that address FL for CNNs based medical imaging applications, the work is still in its infancy for ViTs and requires further attention. Recently few research efforts have been made to exploit the inherent structure of ViT in distributed medical imaging applications. Park \textit{et al.} \cite{park2021federated} propose a Federated Split Task-Agnostic (FESTA) framework that integrates the power of Federated and Split Learning \cite{yang2019federated,vepakomma2018split} in utilizing ViT to simultaneously process multiple chest X-ray tasks, including diagnosing COVID-19 CXR images on a large corpus of decentralized data. Specifically, they split ViT into shared body and task-specific heads and demonstrate that ViT body with sufficient capacity can be shared across relevant tasks by leveraging the multitask-learning (MTL) \cite{caruana1997multitask} strategy. 
However, FESTA is just a proof-of-concept study, and its applicability in clinical trials requires further experimentation. Furthermore, challenges like privacy attacks and robustness against communication bottlenecks for ViT-based FL medical imaging systems require in-depth investigation. An interesting future direction is to explore recent privacy enhancement approaches like differential privacy~\cite{abadi2016deep} to prevent gradient inversion attacks~\cite{huang2021evaluating} on FL-based medical imaging systems in the context of ViTs.
In short, we believe that the successful implementation of distributed machine learning frameworks coupled with the strengths of ViTs could hold significant potential for enabling precision medicine at a large scale. This can lead to ViT models that yield unbiased decisions and are sensitive to rare diseases while respecting governance and privacy concerns. In Table~\ref{tab:federated_frameworks}, we highlight various tools and libraries that have been developed to implement distributed and secure deep learning. This can provide useful information for researchers who wish to rapidly prototype their ViT-based models for medical imaging in distributed settings.

\subsection{Domain Adaptation and Out-of-Distribution Detection} \label{sec:domainadaptation}
\begin{figure*}[t]
\centering
\includegraphics[width=0.32\linewidth]{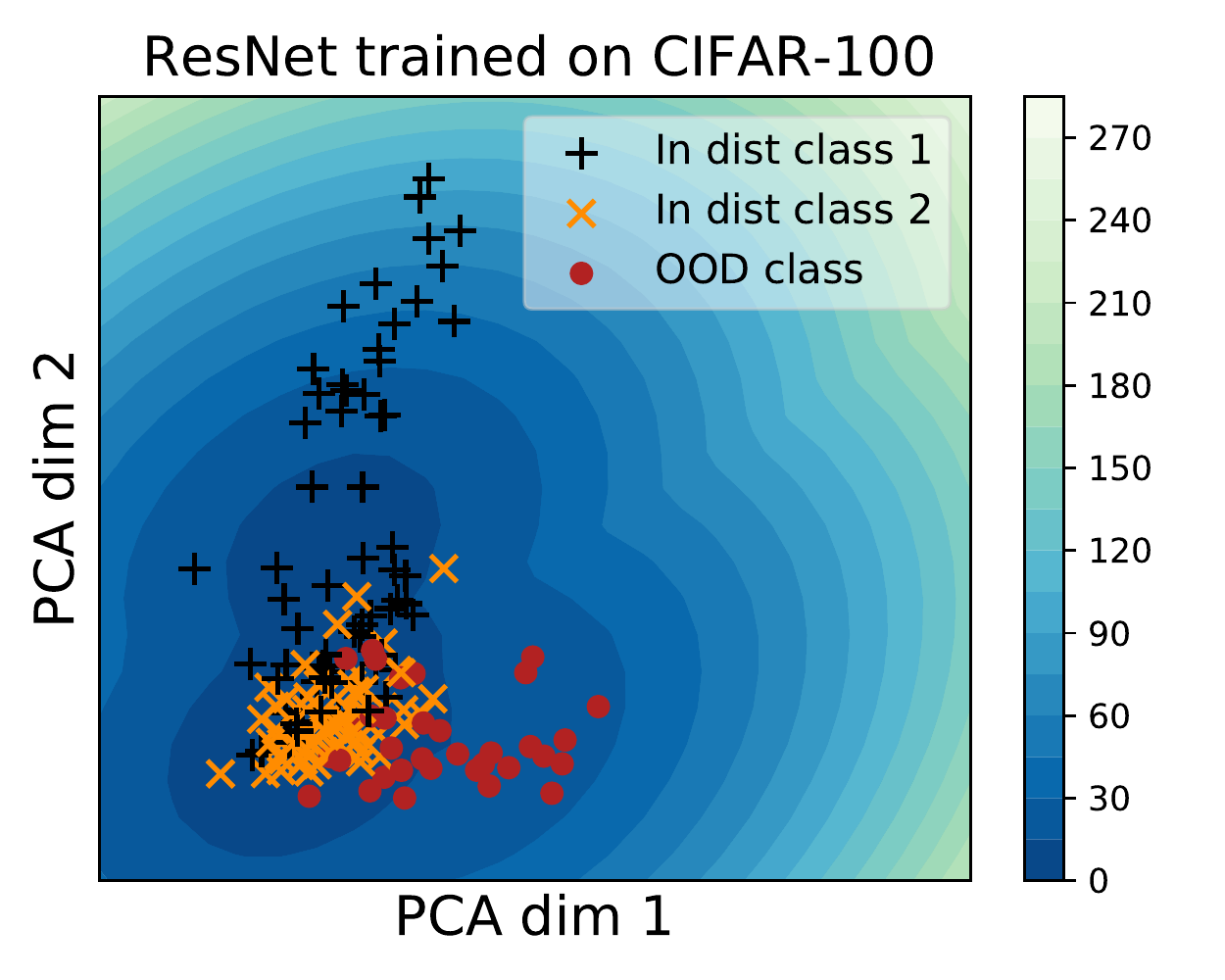}
\includegraphics[width=0.32\linewidth]{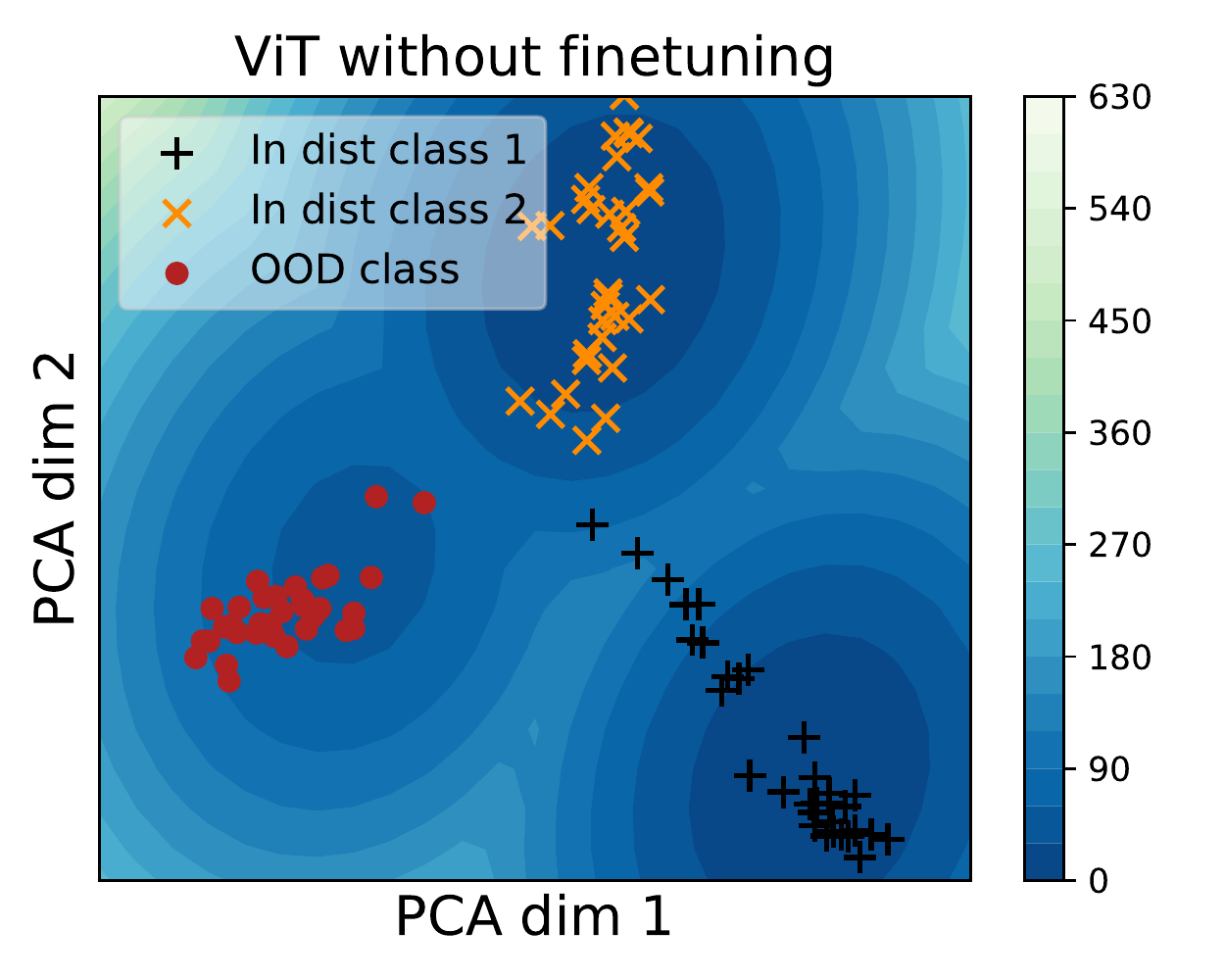}
\includegraphics[width=0.32\linewidth]{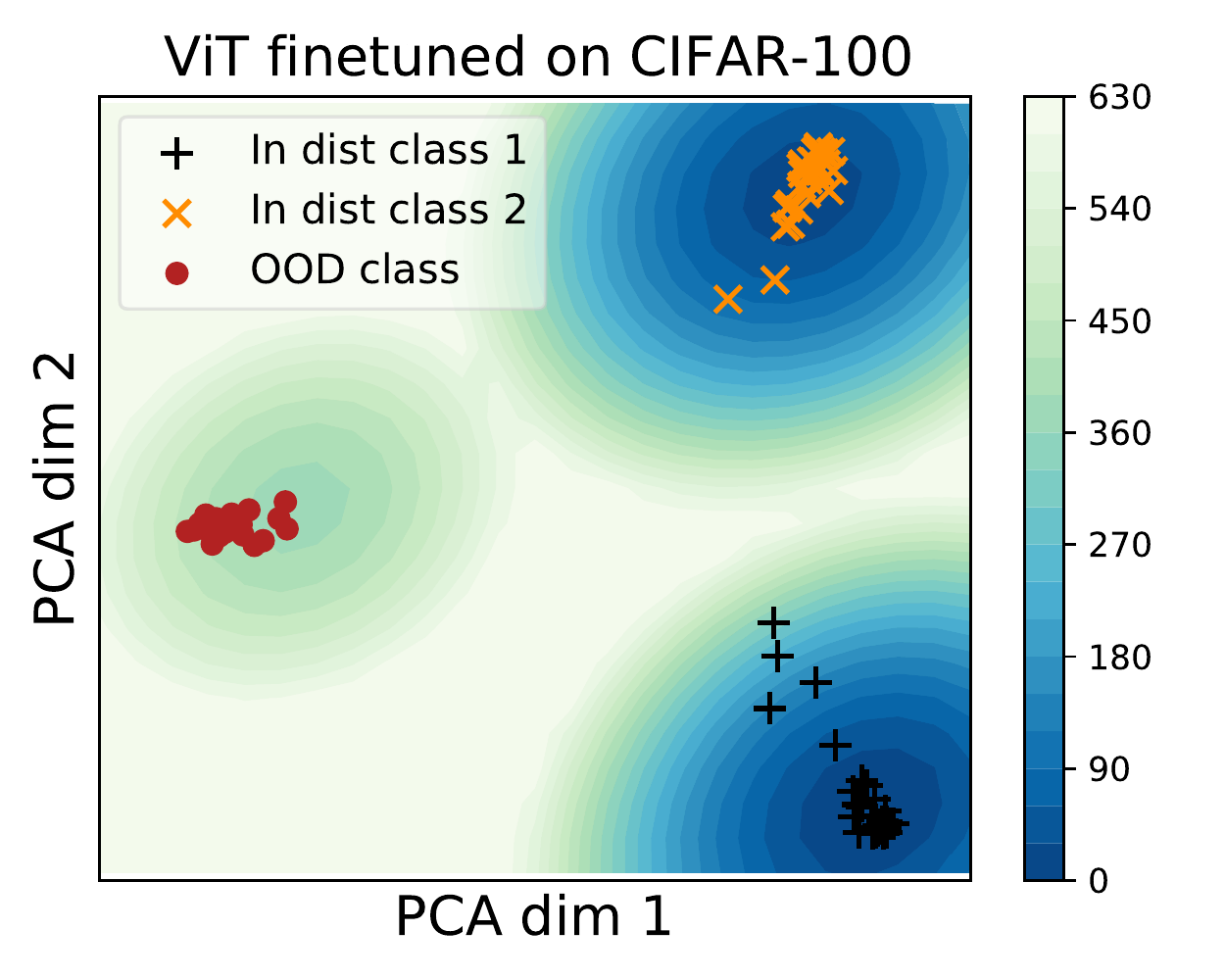}
\caption{A 2D PCA projection of the space of embedding vectors for three models, having two in-distribution and one out-of-distribution class. The points are projections of embeddings of the categories of in-distribution classes ({\color{yellow}yellow} and black) and out-of-distribution classes (red points). The color-coding shows the Mahalanobis outlier score \cite{lee2018simple}. ResNet-20 plot (left panel) leads to overlapping clusters indicating that classes are not well separated. ViT pre-trained on ImageNet-21k (middle panel) can distinguish classes from each other well but does not lead to well-separated outlier scores. ViT fine-tuned on the in-distribution dataset (right panel) is excellent at clustering embeddings based on classes and assigning high Mahalanobis distance to out-of-distribution inputs ({\color{red}red}). Image courtesy \cite{fort2021exploring}.
}
\label{fig:ood}
\end{figure*}
 \begin{figure}[t]
\centering
\includegraphics[trim={0cm 0cm 1.5cm 0cm},clip,width = 0.49\textwidth]{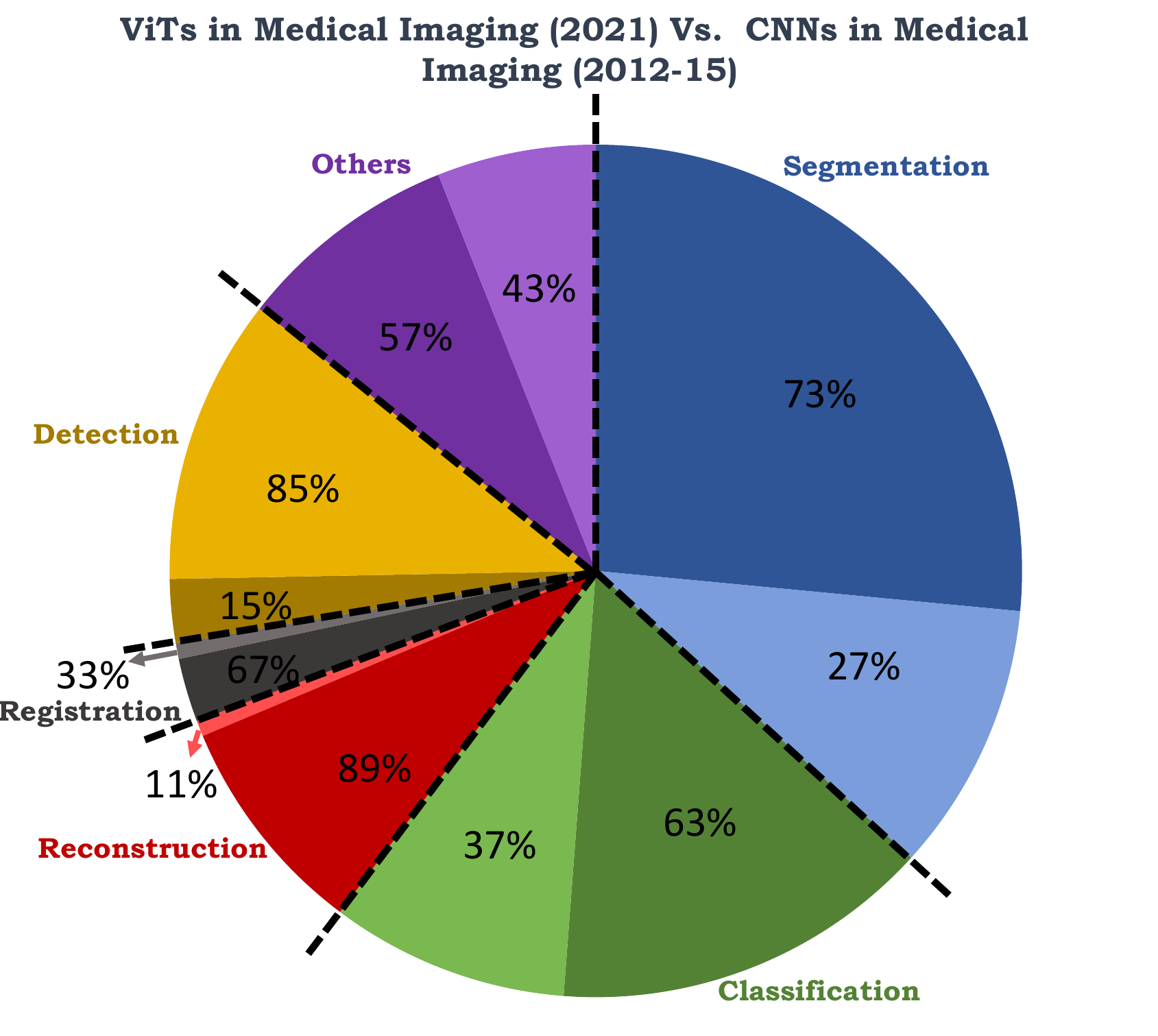}
\caption{ViT-based papers in medical imaging in 2021 (dark shade) vs pure CNN-based papers in medical imaging from 2012 to 2015 (light shade). It can be seen that ViTs have rapidly pervaded into almost all areas of medical imaging in a single year, with segmentation and classification being the most impacted areas. Statistics of CNNs-based papers are taken from \cite{litjens2017survey}.}
\label{fig:vit_vs_cnn}
\end{figure}

Recent efforts for ViT-based medical imaging systems have primarily focused on improving the accuracy and generally lacking a principled mechanism to evaluate their generalization ability under  
different distribution/domain shifts.
Recent studies have shown that test error
generally increases in proportion to the distribution difference between training and test datasets, thereby making it a crucial issue to investigate in the context of ViTs.
In medical imaging applications, these distribution shifts in data arise due to several factors that include: images acquired with a different device model at a different hospital, images of some unseen disease not in the training dataset, images that are incorrectly prepared, e.g., poor contrast, blurry images, etc. Extensive research exists on CNN-based out-of-distribution detection approaches in medical imaging~\cite{yang2021generalized,zhang2021out,zhang2021delving,linmans2020efficient,hendrycks2016baseline}. 
Recently, few attempts have been made to show that large-scale pre-trained ViTs, due to their high-quality representations, can significantly improve the state-of-the-art on a range of out-of-distribution tasks across different data modalities  \cite{fort2021exploring,koner2021oodformer,radford2021learning}. However, investigation in these works has been mostly carried out on toy datasets such as CIFAR-10 and CIFAR-100, therefore not necessarily reflecting out-of-distribution detection performance on medical images with complex textures and patterns, high variance in feature scale (like in X-ray images), and local specific features. This demands further research to design ViT-based medical imaging systems that should be accurate for classes seen during training while providing calibrated estimates of uncertainty for abnormalities and unseen classes. We believe that research in this direction using techniques from transfer learning and domain adaptation will be of interest to the practitioners working in medical imaging based life-critical applications to envision potential practical deployment. In Fig.~\ref{fig:ood}, we highlight the performance gain of ViTs as compared to CNNs for out of distribution detection to inspire medical imaging researchers who wish to explore this area.
Another possible direction is to explore the recent advancements in continual learning~\cite{qu2021recent} to effectively mitigate the issue of domains shift using ViTs. Few preliminary efforts have been made to explore this direction~\cite{lenga2020continual}; however, the work is still in its infancy and requires further attention from the community.
Further, standardized and rigorous evaluation protocols also need to be established for domain adaptation in the medical imaging applications, similar to \textsc{DomainBed}~\cite{gulrajani2020search} framework in the natural image domain. Such a framework will also help in advocating models reproducibility.

\section{Discussion and Conclusion} \label{sec:conc}
From the papers reviewed in this survey, it is evident that ViTs have pervaded every area of medical imaging (see Fig. \ref{fig:vit_vs_cnn}). To keep pace with this rapid development, we recommend organizing the relevant workshops in top computer vision and medical imaging conferences and arranging special issues in prestigious journals to quickly disseminate the relevant research to the medical imaging community.

In conclusion, we present the first comprehensive review of the applications of Transformers in medical imaging. We briefly cover the core concepts behind the success of Transformer models and then provide a comprehensive literature review of Transformers in a broad range of medical imaging tasks. Specifically, we survey the applications of Transformers in medical image segmentation, detection, classification, reconstruction, synthesis, registration, clinical report generation, and other tasks. In particular, for each of these applications, we develop taxonomy, identify application-specific challenges as well as give insights to solve them and specify recent trends. 
Despite their impressive performance, we anticipate there is still much exploration left to be done with Transformers in medical imaging, and we hope this survey provides a roadmap to researchers to progress this field further.

\ifCLASSOPTIONcompsoc
  \section*{Acknowledgments}
\else
  \section*{Acknowledgment}
\fi
{\footnotesize The authors would like to thank Maryam Sultana (MBZ University of Artificial Intelligence) for her help with a few figures.}

\bibliographystyle{unsrt}
\bibliography{Reference}

\end{document}